\documentclass[a4paper,11pt]{article}
\pdfoutput=1
\usepackage{silence}
\usepackage{jheppub}
\WarningsOn[hyperref]
\makeatletter
\renewcommand{\@fpheader}{}
\makeatother
\usepackage{graphicx}
\usepackage{color}
\usepackage{mathtools}
\usepackage{mathrsfs}
\usepackage{amsfonts}
\usepackage{varioref}
\usepackage{textcomp}
\usepackage{cancel}
\usepackage{float}
\usepackage{subcaption}
\usepackage[T1]{fontenc}
\usepackage{bm} 
\title{\boldmath Holographic Timelike Entanglement and Subregion Complexity in
	Localized $\mathrm{AdS}_3\times S^3\times T^4$ Black Holes}

\author[a,b]{Jitendra Pal}
\author[a,b]{Yu Shi}

\affiliation[a]{Wilczek Quantum Center, Shanghai Institute for Advanced Studies, Shanghai 201315, China}
\affiliation[b]{University of Science and Technology of China, Hefei 230026, China}

\emailAdd{jeetupal007@gmail.com}
\emailAdd{yu\_shi@ustc.edu.cn}

\abstract{We study timelike entanglement entropy and timelike subregion complexity in
	localized black holes with asymptotic AdS$_3\times S^3\times T^4$ geometry,
	focusing on the black-pole solution. Unlike the BTZ solution, the black pole
	exhibits a nontrivial dependence on the internal sphere through the functions
	$K_y(r,\theta)$ and $G(r,\theta)$. Both observables are constructed from
	spacelike and timelike Lorentzian branches, but they probe the geometry in
	different ways: timelike entanglement yields a complex lifted area, while
	timelike complexity gives a real, finite renormalized volume. We employ a
	localized timelike prescription in which the branch profile is built at an
	angular label $\theta_0$ and subsequently lifted over the physical internal
	angle $\theta$. In the large-$r$ regime, the leading angular dependence
	drops out, recovering the expected short-interval behaviour. In the exact
	black-pole geometry, the temporal families become non-monotonic, making a
	fixed-boundary-interval selection essential. As the boundary interval
	increases, the selected branches move inward and become sensitive to the
	localized cap–horizon transition region. These results demonstrate that
	timelike Lorentzian observables probe localized-geometry effects that are
	absent in BTZ and in the leading large-$r$ description.}

\begin{document} 
\maketitle
\raggedbottom

\section{Introduction}

Black holes with AdS$_3\times S^3\times T^4$ asymptotics admit various
ten-dimensional descriptions. The prototypical example is the uplift of the
BTZ black hole, where the geometry is independent of the internal
$S^3\times T^4$ coordinates, so the horizon is spread uniformly over the
compact space. This configuration is the standard black-hole saddle of the
D1-D5 system and is well described by the three-dimensional theory obtained
by reduction on $S^3\times T^4$
\cite{Strominger:1996sh,Maldacena:1997de,David:2002wn,Banados:1992wn,Banados:1992gq}.
At sufficiently low energies, type IIB supergravity also admits black-hole
solutions with the same asymptotics but with horizons localized on the
internal $S^3$ \cite{Bena:2024lbh,Dias:2025ads3localised,Aharony:2026d1d5phase}.
Such solutions are related to the fact that horizons extended along compact
directions can become non-uniform or localized, as seen in the
Gregory–Laflamme instability and its AdS generalizations
\cite{Gregory:1993vy,Gregory:1994bj,Gubser:2001ac,Hubeny:2002xn,Peet:1998cr,Dias:2015pda,Dias:2016eto,Buchel:2015gxa}.
Because these localized solutions vary over the internal sphere, they encode
information that is lost in the purely three-dimensional BTZ description.

Holography \cite{Maldacena:1997re,Gubser:1998bc,Witten:1998qj} provides a
principled framework for asking how ten-dimensional bulk structure is
reflected in the boundary theory. The most prominent example is holographic
entanglement entropy, where the entropy of a spatial boundary region is
computed as the area of a bulk extremal surface, as in the Ryu–Takayanagi
prescription and its covariant extension
\cite{Ryu:2006bv,Ryu:2006ef,Hubeny:2007xt,Headrick:2007km,
	Wall:2012uf,Nishioka:2009un,Rangamani:2016dms}. Its gravitational
derivation and quantum generalizations have revealed that boundary
entanglement is intimately connected to semiclassical geometry, bulk
reconstruction, and quantum extremal surfaces
\cite{Lewkowycz:2013nqa,Faulkner:2013ana,Engelhardt:2014gca,
	Jafferis:2015del,Almheiri:2014lwa,Dong:2016eik,Harlow:2016vwg}.
Another useful geometric probe is holographic complexity, which investigates
how the structure and growth of a boundary quantum state are encoded in
bulk volumes or gravitational actions
\cite{Susskind:2014rva,Stanford:2014jda,Brown:2015bva,Brown:2015lvg,
	Carmi:2017jqz,Chapman:2021jbh}. These observables are typically
formulated for spatial boundary regions. A natural Lorentzian question is
what they become when the boundary region is extended in time. This leads
to timelike entanglement entropy \cite{Doi:2022iyj,Doi:2023zaf} and
timelike subregion complexity \cite{Alishahiha:2025timelike}, whose bulk
descriptions are built from spacelike and timelike Lorentzian branches
rather than from a single spatial extremal surface
\cite{Doi:2022iyj,Doi:2023zaf,Alishahiha:2025timelike,
	Afrasiar:2025timelike,Prihadi:2026scalarhair}.

Timelike entanglement entropy defines a geometric observable for a boundary
interval extended in time. The bulk surface is no longer an ordinary
spatial RT surface; it is a Lorentzian configuration made of spacelike and
timelike branches. Already in AdS$_3$ and BTZ, this branch structure leads
to a complex area, whose real and imaginary parts are determined by how the
surface is continued through the Lorentzian geometry
\cite{Doi:2022iyj,Doi:2023zaf}. This has made timelike entanglement a
useful tool for studying holographic aspects inaccessible to purely spatial
extremal surfaces. It is closely related to pseudo-entropy and transition
matrices, and has been applied to Lorentzian extremal surfaces in deformed
CFTs, non-relativistic theories, black-hole and wormhole geometries, RG
flows, higher-curvature theories, scalar-hair backgrounds, and evaporating
black-hole setups
\cite{Doi:2022iyj,Doi:2023zaf,Prihadi:2026scalarhair,Anegawa:2024timelike,Li:2023HolographicTLEE,Afrasiar:2024NonConformalTEE,Afrasiar:2025NonRelativisticTEE,
	Heller:2025GeometricTEE,Heller:2025TemporalHEE,Jena:2025LifshitzTEE,Jiang:2023TTbarTEE,Basu:2024ReflectedTEE,
	Fujiki:2025dSCFTEinstein,Li:2025TEEFirstLaw,Kawamoto:2025Wormhole,
	Harper:2025WormholeTEE,Jiang:2023dS3CFT2,Chu:2023AdSBCFT,
	Grieninger:2024TemporalRG,Narayan:2024TimePseudoEntropy,
	He:2024RindlerTEE,Liu:2024FreeFermionsTimelike,Nunez:2025TopDownTEE,
	Xu:2025ImaginaryTEE,Guo:2025TimeSpacelike,Chu:2025AnomaliesTEE,
	Giataganas:2025RGTEE,Katoch:2025VaidyaTEE,Jiang:2025RevisitedTEE,
	Zhao:2025HigherCurvatureTEE,Nunez:2025InterpolatingTEE,
	Goki:2026HolographicTEE,Li:2026BHInteriorTEE,Ladghami:2026HawkingTEE,Dai:2026RotatingBTZTEE,Li:2026bof}.
Two works are particularly close to our setup. Ref.~\cite{Anegawa:2024timelike}
used timelike and ordinary entanglement to study black-hole singularities
and complex saddles, while Ref.~\cite{Afrasiar:2025timelike} developed the
Lorentzian branch construction directly in black-hole geometries, including
BTZ and higher-dimensional AdS–Schwarzschild backgrounds. We use this
branch-based black-hole prescription for a different purpose: to ask
whether timelike Lorentzian observables can detect the internal angular
structure of the localized black pole, which is absent in the ordinary BTZ
uplift.

Timelike subregion complexity \cite{Alishahiha:2025timelike} is the volume
counterpart of the timelike entanglement construction. In ordinary
subregion complexity, one associates a bulk volume to the region selected
by a boundary subsystem
\cite{Alishahiha:2015rta,Carmi:2016wjl,Ben-Ami:2016qex,Chapman:2018hou,
	Chen:2018mcc,Ling:2019ien}. For a boundary interval separated in time,
the relevant bulk region is no longer bounded by a single spatial extremal
surface; instead, it is determined by the same Lorentzian branch geometry
that appears in timelike entanglement: a spacelike branch reaching the
asymptotic boundary and a timelike branch ending at a turning point
\cite{Afrasiar:2025timelike,Prihadi:2026scalarhair}. The two observables
therefore share the same branch kinematics, but they measure different bulk
data. Timelike entanglement yields a lifted area that can be complex,
whereas timelike complexity gives a real, finite renormalized volume. This
difference is particularly useful in the localized black-pole geometry:
both the area and the volume probe the same Lorentzian branches, but they
respond differently to the angular dependence of $K_y(r,\theta)$ and
$G(r,\theta)$.

Applying timelike observables to the localized black pole requires the
localized lifting prescription of Ref.~\cite{Bena:2024lbh}. In BTZ or
AdS–Schwarzschild, the branch problem is essentially radial. In the
black-pole geometry this is no longer true, because the AdS$_3$ metric
components depend on the internal angle through $K_y(r,\theta)$ and
$G(r,\theta)$. We therefore first construct the reduced Lorentzian branches
at an angular label $\theta_0$, and only afterwards lift the result to the
full ten-dimensional geometry by integrating over the physical internal
angle $\theta$. The two angles play distinct roles: $\theta_0$ labels the
branch family, while $\theta$ is the internal angle in the lifted area or
volume. This is the key localized ingredient in both the timelike
entanglement and timelike complexity calculations.

Our analysis is organized around the relation between the BTZ benchmark and
the localized black-pole geometry. If we set $K_y=G=1$, the branch
equations reduce to their BTZ form, providing a useful check of the
prescription. In the localized coordinates, however, this limit describes
the BTZ outside-horizon patch. To recover the usual two-sided BTZ
description, with both exterior and interior regions, one must perform the
standard BTZ coordinate transformation. This distinction is especially
important for timelike subregion complexity. In the BTZ benchmark, after
the appropriate subtraction, the finite contribution comes from the
interior branch, while the outside-horizon branch gives no finite
complexity. The localized black-pole solution is naturally written in the
corresponding outside-horizon patch of the ten-dimensional geometry. For
this reason, our localized complexity should be interpreted as a finite
renormalized volume associated with the Lorentzian branch construction in
this localized patch, not as the full late-time interior complexity of an
eternal BTZ black hole.

The large-$r$ regime of the black pole is not exactly BTZ, but it
asymptotically approaches the BTZ geometry. The leading terms reproduce
the BTZ-like branch structure, while the first localized corrections enter
at order $1/r^2$. This regime thus provides an analytic check of the
prescription and fixes the short-boundary-interval behaviour. We then
restore the exact black-pole functions $K_y(r,\theta)$ and $G(r,\theta)$.
In the exact geometry, the internal angular dependence becomes part of the
branch problem: the boundary interval can become non-monotonic as a
function of the turning point, and more than one radial branch can
represent the same boundary time interval. The physical comparison must
therefore be made only after fixing the boundary interval and evaluating
the lifted area or finite volume on all admissible branches.

The exact black-pole geometry produces effects that are absent in the BTZ
benchmark and in the large-$r$ regime. For timelike entanglement entropy,
the lifted area becomes sensitive to the internal angular structure after
the ten-dimensional lift. The real part is selected by the fixed-boundary-
interval minimization, while the imaginary part is evaluated on the same
selected surface and depends on where the lifted area samples Lorentzian
regions of the geometry. For timelike subregion complexity, the same
selected Lorentzian branches define a real finite volume. The selected
complexity is therefore controlled not by a complex area, but by the
weighted spacelike and timelike volume contributions in the localized
geometry. In both observables, as the boundary interval increases, the
selected branches move away from the asymptotic region and become
sensitive to the cap/horizon transition region of the internal sphere.

A central point in both calculations is that the saddle comparison must be
performed at fixed boundary interval. In the exact black-pole geometry,
the boundary time is not a simple monotonic function of the turning point.
For a given angular label, more than one radial branch can reach the same
boundary interval, and different angular labels may cease to be available
when the target interval is increased. Therefore, the time equation is
solved first for all admissible branches. Only after this step do we
compare the corresponding lifted areas or finite volumes. For timelike
entanglement entropy, the selected surface is determined by the real part
of the renormalized lifted area, while the imaginary part is evaluated on
the same surface. For timelike subregion complexity, the selected saddle
is determined by the real finite renormalized volume. This fixed-boundary-
interval prescription is essential for obtaining a well-defined boundary
observable in the localized geometry.

The selected saddles reveal how the localized geometry enters the two
observables. At short boundary intervals, the branches remain close to the
asymptotic region, where the black-pole geometry is BTZ-like and the
internal angular dependence is suppressed. As the boundary interval grows,
the selected branches move inward and become sensitive to the cap/horizon
transition region of the internal sphere. This transition region controls
which angular families can support larger boundary intervals, while the
final saddle is fixed by the relevant variational quantity. In timelike
entanglement entropy, this quantity is the real part of the renormalized
lifted area; in timelike subregion complexity, it is the finite
renormalized lifted volume. The same Lorentzian branch geometry therefore
leads to two different probes of the black pole: one complex and
area-based, the other real and volume-based.

The main outcome is that the localized black-pole geometry leaves a clear
imprint on timelike Lorentzian observables. The large-$r$ regime gives a
BTZ-like short-interval check, but the exact geometry produces branch
structures absent in the BTZ benchmark. The boundary-interval map becomes
non-monotonic, the allowed angular range depends on the target interval,
and the selected saddles move toward the cap/horizon transition region as
the boundary interval grows. Timelike entanglement records these features
through the real and imaginary parts of a lifted complex area, whereas
timelike subregion complexity records them through a real finite volume.
In this sense, the two observables provide complementary probes of the
same localized ten-dimensional geometry.

The paper is organized as follows. In section~\ref{sec:geometry} we review
the localized black-pole geometry, emphasizing the functions
$K_y(r,\theta)$ and $G(r,\theta)$ and the cap/horizon angular structure.
Section~\ref{sec:timelike-entanglement} develops the timelike-entanglement
calculation. We first derive the Lorentzian branch equations, then study
the large-$r$ regime as an analytic reference, and finally restore the exact
black-pole functions. The fixed-boundary-interval selection is then imposed
to obtain the selected real and imaginary parts of the lifted area.
Section~\ref{sec:timelike-complexity-prescription} applies the same Lorentzian branch
geometry to timelike subregion complexity. After reviewing the BTZ
benchmark and the large-$r$ limit, we compute the finite renormalized lifted
volume in the exact black-pole geometry and perform the corresponding
fixed-boundary-interval minimisation. We conclude in section~\ref{sec:discussion-conclusion} with a comparison of the two observables and
their sensitivity to the localized cap/horizon transition region.

\section{Localized black hole geometry}
\label{sec:geometry}

In this section we collect the geometric ingredients needed for the
timelike observables. The background is a type IIB supergravity solution with asymptotic
AdS$_3\times S^3\times T^4$ structure. The localized family contains
several possible horizon distributions on the internal $S^3$, including the
black pole, black belt and black bi-pole solutions. In this work we focus
on the black pole, because it is the simplest localized geometry that already
contains the cap-side sector, the horizon-side sector and the transition
angle between them. Unlike the BTZ uplift, which is homogeneous on the
internal sphere, the black pole depends nontrivially on the internal angle
$\theta$. This angular dependence is the geometric structure probed by the
timelike observables below.

The ten-dimensional metric is \cite{Bena:2024lbh}
\begin{equation}
	\begin{split}
		ds_{10}^2
		=&\,
		\frac{1}{\sqrt{Q_1Q_5}}
		\left[
		-r^2K_y(r,\theta)\,dt^2
		+
		\frac{r^2+\ell^2}{K_y(r,\theta)}\,dy^2
		\right]
		+
		\sqrt{\frac{Q_1}{Q_5}}\,ds^2(T^4)
		\\
		&\,
		+
		\sqrt{Q_1Q_5}
		\left[
		G(r,\theta)
		\left(
		\frac{dr^2}{r^2+\ell^2}
		+
		d\theta^2
		\right)
		+
		\cos^2\theta\,d\phi_1^2
		+
		\sin^2\theta\,d\phi_2^2
		\right].
	\end{split}
	\label{eq:ten-dimensional-metric}
\end{equation}
Here $Q_1$ and $Q_5$ are the D1- and D5-brane charges, $y$ is the
AdS$_3$ circle, and $\theta$ is the Hopf coordinate on the internal $S^3$. We use the angular range
$0\leq \theta\leq \pi/2$.
The functions $K_y(r,\theta)$ and $G(r,\theta)$ carry the information about
the localization of the black-hole region on the internal sphere. We also
define $Q= \sqrt{Q_1Q_5}$.

The localized solutions are described by a set of source scales $\ell_i$,
with total scale $\ell^2=\sum_i\ell_i^2$. For each source it is useful to introduce local variables $(r_i,\theta_i)$.
Following the notation of the localized solution, these variables are defined
through
\begin{align}
	r_i^2
	=
	\rho_i+\rho_{i-1}-\frac{\ell_i^2}{2},\qquad
	\ell_i^2\cos^2\theta_i
	=
	\rho_{i-1}-\rho_i+\frac{\ell_i^2}{2},
	\label{eq:thetai-def}
\end{align}
where
\begin{equation}
	\rho_i
	=
	\frac{1}{4}
	\sqrt{
		\left[
		(2r^2+\ell^2)\cos 2\theta
		+
		\ell^2
		-
		2\sum_{j=1}^{i}\ell_j^2
		\right]^2
		+
		4r^2(r^2+\ell^2)\sin^2 2\theta
	}.
	\label{eq:rhoi-def}
\end{equation}
These source-centered variables make the angular structure of the localized
solution transparent. At $r=0$, one of the $r_i$ vanishes depending on the
value of the internal angle. The corresponding degeneration may be either
horizon-side or cap-side.

The localized family is specified by two sets of labels. A source in $U_t$
corresponds to a degeneration of the time circle, while a source in $U_y$
corresponds to a smooth degeneration of the AdS$_3$ spatial circle. For the
black pole,
\begin{equation}
	n=2,
	\qquad
	U_t=\{1\},
	\qquad
	U_y=\{2\}.
	\label{eq:black-pole-sets}
\end{equation}
Thus the first source describes the horizon-side sector and the second source
describes the cap-side sector. This is the geometric origin of the black-pole
localization. For the black pole the warp factor $K_y(r,\theta)$ and $G(r,\theta)$ takes the following form 
\begin{equation}
	K_y(r,\theta)
	=
	1+\frac{\ell_2^2}{r_2^2},\qquad G(r,\theta)
	=
	\frac{
		\left(
		(r_2^2+\ell_2^2)\cos^2\theta_2
		+
		r_1^2\sin^2\theta_1
		\right)
		\left(
		r_2^2\cos^2\theta_2
		+
		(r_1^2+\ell_1^2)\sin^2\theta_1
		\right)
	}{
		\left(
		(r_2^2+\ell_2^2)\cos^2\theta_2
		+
		(r_1^2+\ell_1^2)\sin^2\theta_1
		\right)
		\left(
		r_2^2\cos^2\theta_2
		+
		r_1^2\sin^2\theta_1
		\right)
	}.
	\label{eq:Ky-black-pole}
\end{equation}

Equations~\eqref{eq:Ky-black-pole} are the
central inputs for the exact black-pole calculation. They encode the fact
that the AdS$_3$ part of the metric changes as one moves along the internal
sphere.

For the black-pole solution, the regularity conditions fix the geometry in
terms of a single temperature parameter. We denote the black-hole temperature
by ${\cal T}$ and introduce $\tau \equiv 4\pi^2{\cal T}^2$. This notation keeps the symbol $T$ free for the boundary time interval used in
the timelike observables. In terms of $\tau$, the total length scale of the
black pole is
\begin{equation}
	\ell^2
	=
	\frac{Q_1Q_5}{R_y^2}\,
	\frac{\tau}{1+\tau}.
	\label{eq:ell-total}
\end{equation}
The two length scales which resolve the $r=0$ locus are
\begin{equation}
	\ell_1^2
	=
	\frac{\ell^2}{1+\tau},
	\qquad
	\ell_2^2
	=
	\frac{\tau\,\ell^2}{1+\tau}.
	\label{eq:ell12}
\end{equation}
With these definitions they satisfy
\begin{equation}
	\ell^2=\ell_1^2+\ell_2^2.
	\label{eq:length-sum}
\end{equation}
Thus $\ell$ sets the total size of the two-source geometry, while $\ell_1$
and $\ell_2$ describe how this size is divided between the two possible
endings of the spacetime. The scale $\ell_1$ controls the horizon side,
where the time circle degenerates, whereas $\ell_2$ controls the cap side,
where the $y$-circle closes off smoothly. The black pole therefore differs
from the BTZ uplift in an essential way: the BTZ horizon is homogeneous on the
internal $S^3$, while the black pole has a localized horizon on part of the
sphere and a smooth cap on the remaining part.

The same information can be expressed through the ratios
\begin{equation}
	\frac{\ell_1^2}{\ell^2}
	=
	\frac{1}{1+\tau},
	\qquad
	\frac{\ell_2^2}{\ell^2}
	=
	\frac{\tau}{1+\tau}.
	\label{eq:length-ratios}
\end{equation}
These ratios determine how the $r=0$ surface is split between the cap side and
the horizon side. They also fix the angular position at which the geometry
changes its local endpoint. The transition between the two regions occurs at an angle $\theta_\star$ on
the internal $S^3$. It is fixed by
\begin{equation}
	\cos^2\theta_\star
	=
	\frac{\ell_1^2}{\ell^2}
	=
	\frac{1}{1+\tau},
	\qquad
	\sin^2\theta_\star
	=
	\frac{\ell_2^2}{\ell^2}
	=
	\frac{\tau}{1+\tau}.
	\label{eq:theta-star}
\end{equation}
Equivalently,
\begin{equation}
	\ell_1^2\sin^2\theta_\star
	-
	\ell_2^2\cos^2\theta_\star
	=
	0.
	\label{eq:theta-star-equivalent}
\end{equation}
The $r=0$ locus is therefore separated into
\begin{equation}
	0\leq \theta < \theta_\star
	\quad \text{cap side},
	\qquad
	\theta_\star < \theta \leq \frac{\pi}{2}
	\quad \text{horizon side}.
	\label{eq:angular-sectors}
\end{equation}
On the cap side the spacetime ends smoothly through the degeneration of the
$y$-circle. On the horizon side the time circle degenerates and produces the
localized black-pole horizon. This angular split is the main
ten-dimensional feature that the exact timelike probes are sensitive to.

It is useful to describe the same family at fixed energy. We define
\begin{equation}
	x_E\equiv \frac{E}{E_{\max}},
	\qquad
	E_{\max}=\frac{c}{96}.
	\label{eq:xE-def}
\end{equation}
Here $c$ is the central charge of the dual D1-D5 CFT. In the normalization
used here,
\begin{equation}
	c=6N_1N_5
	=
	\frac{3\pi Q_1Q_5}{2G_5R_y}.
	\label{eq:central-charge}
\end{equation}
With the standard energy convention, global ${\rm AdS}_3$ has energy
$E=-c/12$. The black-pole solutions exist in the range
\begin{equation}
	-\frac{c}{12}<E\leq E_{\max}.
	\label{eq:black-pole-energy-range}
\end{equation}
At fixed energy, the temperature parameter can be written as \cite{Bena:2024lbh}
\begin{equation}
	\tau_\sigma(x_E)
	=
	\frac{4}{3+\sigma\sqrt{1-x_E}}-1,
	\qquad
	\sigma=\pm1.
	\label{eq:tau-branch}
\end{equation}
The branch with $\sigma=-1$ is the minus branch, while the branch with
$\sigma=+1$ is the plus branch. For positive energy both branches can occur,
whereas for negative energy only the minus branch is physical. In all
numerical calculations in this paper we work on the minus branch $\sigma=-1$.

For the numerical analysis we use
\begin{equation}
	Q_1=Q_5=R_y=1.
	\label{eq:numerical-units}
\end{equation}
This is a choice of units. In the general solution, $\ell$, $\ell_1$ and
$\ell_2$ are length scales. In the plots, their values are measured in the
unit system specified by eq.~\eqref{eq:numerical-units}. With this choice,
$Q_1Q_5/R_y^2=1$, and the regularity relations become
\begin{equation}
	\ell^2
	=
	\frac{\tau}{1+\tau},
	\qquad
	\ell_1^2
	=
	\frac{\ell^2}{1+\tau},
	\qquad
	\ell_2^2
	=
	\frac{\tau\,\ell^2}{1+\tau}.
	\label{eq:numerical-lengths}
\end{equation}
All numerical results below should be understood in this unit convention.

We now introduce the effective three-dimensional metric used in the reduced
problem. In the localized RT prescription, the internal deformation can be
absorbed into an effective AdS$_3$ metric. In the present notation this
metric is \cite{Bena:2024lbh}
\begin{equation}
	d\widetilde{s}_3^2
	=
	Q\,G(r,\theta)
	\left[
	-r^2K_y(r,\theta)\,dt^2
	+
	R_y^2\frac{r^2+\ell^2}{K_y(r,\theta)}\,d\varphi^2
	+
	Q\,G(r,\theta)\frac{dr^2}{r^2+\ell^2}
	\right],
	\label{eq:effective-three-metric}
\end{equation}
where $\varphi=y/R_y$ is the angular coordinate of the boundary spatial
circle. For the timelike observables studied in this paper, the boundary
interval extends in the time direction at fixed spatial position. Hence
\begin{equation}
	d\varphi=0.
	\label{eq:fixed-spatial-coordinate}
\end{equation}
The relevant two-dimensional part of the effective metric is therefore
\begin{equation}
	ds_{2,\theta}^2
	=
	-Q\,G(r,\theta)r^2K_y(r,\theta)\,dt^2
	+
	\frac{Q^2G(r,\theta)^2}{r^2+\ell^2}\,dr^2.
	\label{eq:two-dimensional-metric-theta}
\end{equation}
The reduced Lorentzian profile is first evaluated at an angular label
$\theta_0$, with $0\leq\theta_0\leq\pi/2$. We write
\begin{equation}
	ds_{2,\theta_0}^2
	=
	-F(r;\theta_0)\,dt^2
	+
	H(r;\theta_0)\,dr^2.
	\label{eq:reduced-two-dimensional-metric}
\end{equation}
The two metric functions are
\begin{align}
	F(r;\theta_0)
	=
	Q\,r^2K_y(r,\theta_0)G(r,\theta_0),\qquad
	H(r;\theta_0)
	=
	\frac{Q^2G(r,\theta_0)^2}{r^2+\ell^2}.
	\label{eq:H-definition}
\end{align}
The function $F(r;\theta_0)$ gives the time-direction weight in the reduced
Lorentzian problem, while $H(r;\theta_0)$ gives the radial weight. Since
both depend on the angular label $\theta_0$, different choices of
$\theta_0$ define different effective branch profiles.

The distinction between $\theta_0$ and $\theta$ is essential. The angular
label $\theta_0$ fixes the reduced profile in the effective $(t,r)$ plane.
The physical internal angle $\theta$ is restored only when the area or volume
is lifted to ten dimensions. Thus the final timelike entanglement entropy
and timelike subregion complexity are not obtained by simply setting
$\theta=\theta_0$ everywhere. The reduced profile fixes the radial and
temporal shape, while the ten-dimensional lift restores the full internal
angular dependence of the localized geometry.

Finally, we write the asymptotic limit. For $r$ much larger than
the intrinsic black-pole scales $\ell_1$ and $\ell_2$, the exact functions
have the expansion
\begin{align}
	K_y(r,\theta)
	=
	1+\frac{\ell_2^2}{r^2}
	+
	O(r^{-4}),\qquad
	G(r,\theta)
	=
	1+O(r^{-4}).
	\label{eq:G-asymptotic}
\end{align}
Consequently,
\begin{equation}
	F(r;\theta_0)
	=
	Q(r^2+\ell_2^2)+O(r^{-2}).
	\label{eq:F-asymptotic}
\end{equation}
At this order the internal angular dependence is suppressed. The large-$r$
limit is therefore useful as an analytic consistency check and for
understanding the asymptotic tail of the UV subtraction. It cannot
reproduce the cap/horizon distinction or the inward motion of the selected
fixed-boundary-interval saddles. Those effects require the exact localized
functions $K_y(r,\theta)$ and $G(r,\theta)$.

\section{Timelike entanglement}
\label{sec:timelike-entanglement}

\subsection{Lorentzian branch equations}
\label{sec:tee-branch-equations}

We first derive the Lorentzian branch equations that will be used throughout
the timelike-entanglement calculation. The boundary subsystem is a temporal
interval at fixed spatial position,
\begin{equation}
	-\frac{T}{2}\leq t\leq \frac{T}{2},
	\qquad
	\varphi=\mathrm{constant}.
	\label{eq:boundary-timelike-interval}
\end{equation}
The bulk profile lies in the effective $(t,r)$ plane and is written as
\begin{equation}
	t=t(r).
	\label{eq:tr-profile}
\end{equation}
Because the localized black-pole geometry depends on the internal sphere,
the reduced Lorentzian problem is first specified at a fixed angular label
$\theta_0$. We use the effective two-dimensional metric introduced in
eqs.~\eqref{eq:reduced-two-dimensional-metric}--\eqref{eq:H-definition},
\begin{equation}
	ds^2_{2,\theta_0}
	=
	-F(r;\theta_0)\,dt^2
	+
	H(r;\theta_0)\,dr^2.
	\label{eq:tee-effective-metric-repeat}
\end{equation}
The angular label $\theta_0$ fixes the branch profile in the reduced
$(t,r)$ geometry. The physical internal angle $\theta$ is restored only
when the surface is lifted to ten dimensions.

Along the curve $t=t(r)$, the induced line element is
\begin{equation}
	ds_{\rm ind}^2
	=
	\left[
	H(r;\theta_0)-F(r;\theta_0)t'(r)^2
	\right]dr^2.
	\label{eq:induced-line-element-tee}
\end{equation}
The sign of this expression separates the reduced Lorentzian surface into a
spacelike branch and a timelike branch. This separation determines the
first-order branch equations in the effective geometry. The final real and
imaginary parts of the timelike-entanglement area are determined only after
the ten-dimensional lift, where the same branch profiles are weighted by the
full internal angular dependence of the localized geometry.

The branch equations follow from the conserved momentum conjugate to $t(r)$.
For the spacelike branch, the real length functional is
\begin{equation}
	{\cal L}_{\rm sp}
	=
	\sqrt{
		H(r;\theta_0)-F(r;\theta_0)t_{\rm sp}'(r)^2
	}.
	\label{eq:Lsp}
\end{equation}
Since $t$ is a cyclic variable, the corresponding conserved quantity is
\begin{equation}
	\Pi_{\rm sp}
	=
	-
	\frac{
		F(r;\theta_0)t_{\rm sp}'(r)
	}{
		\sqrt{
			H(r;\theta_0)-F(r;\theta_0)t_{\rm sp}'(r)^2
		}
	}.
	\label{eq:Pi-sp}
\end{equation}
Solving for the slope gives
\begin{equation}
	t_{\rm sp}'(r)^2
	=
	\frac{
		\Pi_{\rm sp}^2 H(r;\theta_0)
	}{
		F(r;\theta_0)
		\left[
		F(r;\theta_0)+\Pi_{\rm sp}^2
		\right]
	}.
	\label{eq:tsp-prime-general}
\end{equation}
Only $\Pi_{\rm sp}^2$ enters the profile, so the sign of $\Pi_{\rm sp}$ does
not affect the branch geometry.

For the timelike branch, the induced line element has the timelike sign in
the reduced $(t,r)$ geometry. The profile is obtained by extremizing the
corresponding real branch functional,
\begin{equation}
	{\cal L}_{\rm tim}
	=
	\sqrt{
		F(r;\theta_0)t_{\rm tim}'(r)^2-H(r;\theta_0)
	}.
	\label{eq:Ltim}
\end{equation}
The conserved quantity is
\begin{equation}
	\Pi_{\rm tim}
	=
	\frac{
		F(r;\theta_0)t_{\rm tim}'(r)
	}{
		\sqrt{
			F(r;\theta_0)t_{\rm tim}'(r)^2-H(r;\theta_0)
		}
	}.
	\label{eq:Pi-tim}
\end{equation}
Solving for the timelike slope gives
\begin{equation}
	t_{\rm tim}'(r)^2
	=
	\frac{
		\Pi_{\rm tim}^2 H(r;\theta_0)
	}{
		F(r;\theta_0)
		\left[
		\Pi_{\rm tim}^2-F(r;\theta_0)
		\right]
	}.
	\label{eq:ttim-prime-general}
\end{equation}

We parametrize the Lorentzian branch family by the radial turning point
$r_0$ and define
\begin{equation}
	F_0\equiv F(r_0;\theta_0).
	\label{eq:F0-definition}
\end{equation}
The timelike branch closes at $r=r_0$. At this point $dr/dt=0$, or
equivalently $t'(r)\to\infty$. From
eq.~\eqref{eq:ttim-prime-general}, this gives
\begin{equation}
	\Pi_{\rm tim}^2=F_0.
	\label{eq:tim-turning-condition}
\end{equation}
We use the same positive scale $F_0$ to label the associated spacelike
branch,
\begin{equation}
	\Pi_{\rm sp}^2=F_0.
	\label{eq:sp-parameter-choice}
\end{equation}
The two branch slopes are therefore
\begin{subequations}
	\begin{align}
		t_{\rm sp}'(r)
		&=
		\left[
		\frac{
			F_0 H(r;\theta_0)
		}{
			F(r;\theta_0)
			\left[
			F(r;\theta_0)+F_0
			\right]
		}
		\right]^{1/2},
		\label{eq:tsp-prime}
		\\
		t_{\rm tim}'(r)
		&=
		\left[
		\frac{
			F_0 H(r;\theta_0)
		}{
			F(r;\theta_0)
			\left[
			F_0-F(r;\theta_0)
			\right]
		}
		\right]^{1/2}.
		\label{eq:ttim-prime}
	\end{align}
\end{subequations}

The spacelike branch reaches the asymptotic boundary and contains the
UV-sensitive part of the surface. The timelike branch has finite radial
extent and exists only where
\begin{equation}
	F_0>F(r;\theta_0).
	\label{eq:tim-domain-condition}
\end{equation}
This condition determines the radial interval covered by the timelike
branch.

The boundary interval generated by a branch configuration is fixed by the
difference between the timelike and spacelike time integrals. We denote the
lower endpoint of the reduced radial problem by $r_{\rm L}$. In the
large-$r$ calculation, $r_{\rm L}=r_\ast$ marks the lower edge of the
validity region of the expansion. In the exact black-pole calculation,
$r_{\rm L}=\epsilon$ regulates the small-$r$ endpoint of the localized core.
With this notation,
\begin{subequations}
	\begin{align}
		I_{\rm tim}(r_0,\theta_0)
		&=
		\int_{r_{\rm L}}^{r_0}
		dr\,t_{\rm tim}'(r),
		\label{eq:Itim-definition}
		\\
		I_{\rm sp}(r_0,\theta_0)
		&=
		\int_{r_{\rm L}}^{\infty}
		dr\,t_{\rm sp}'(r).
		\label{eq:Isp-definition}
	\end{align}
\end{subequations}

The boundary interval associated with the Lorentzian surface is
\begin{equation}
	\frac{T(r_0,\theta_0)}{2}
	=
	I_{\rm tim}(r_0,\theta_0)
	-
	I_{\rm sp}(r_0,\theta_0).
	\label{eq:time-map-tee}
\end{equation}
Equivalently,
\begin{align}
	\frac{T(r_0,\theta_0)}{2}
	=&
	\int_{r_{\rm L}}^{r_0}
	dr\,
	\left[
	\frac{
		F_0 H(r;\theta_0)
	}{
		F(r;\theta_0)
		\left[
		F_0-F(r;\theta_0)
		\right]
	}
	\right]^{1/2}
	\nonumber
	\\
	&-
	\int_{r_{\rm L}}^{\infty}
	dr\,
	\left[
	\frac{
		F_0 H(r;\theta_0)
	}{
		F(r;\theta_0)
		\left[
		F(r;\theta_0)+F_0
		\right]
	}
	\right]^{1/2}.
	\label{eq:time-map-tee-expanded}
\end{align}

This map is central to the saddle problem. A branch configuration is first
specified by $(r_0,\theta_0)$, and the corresponding boundary interval is
then computed from the geometry:
\begin{equation}
	(r_0,\theta_0)
	\longrightarrow
	T(r_0,\theta_0).
	\label{eq:surface-to-time-map}
\end{equation}
In the exact black-pole geometry, this time map can be non-monotonic. The
same boundary interval can therefore be reached by more than one radial
branch, and different angular labels may support different ranges of
boundary intervals. Thus $(r_0,\theta_0)$ is a branch label, not yet the
selected boundary saddle. The physical comparison is made only after the
branch family has been restricted to a common boundary interval. This
fixed-boundary-interval selection is carried out in
section~\ref{sec:tee-fixed-boundary-interval}, after the lifted area and its
UV subtraction have been defined.

For later use, we specify the time embedding of the two branches on the
upper half of the Lorentzian surface. The timelike branch is measured from
its turning point,
\begin{equation}
	t_{\rm tim}^{+}(r)
	=
	\int_{r}^{r_0}
	d\tilde r\,
	t_{\rm tim}'(\tilde r),
	\qquad
	t_{\rm tim}^{+}(r_0)=0.
	\label{eq:ttim-positive-profile}
\end{equation}
The spacelike branch is measured from the asymptotic endpoint,
\begin{equation}
	t_{\rm sp}^{+}(r)
	=
	\frac{T}{2}
	+
	\int_{r}^{\infty}
	d\tilde r\,
	t_{\rm sp}'(\tilde r),
	\qquad
	t_{\rm sp}^{+}(\infty)=\frac{T}{2}.
	\label{eq:tsp-positive-profile}
\end{equation}
The time map in eq.~\eqref{eq:time-map-tee} ensures that these two profiles
join at the lower radial endpoint,
\begin{equation}
	t_{\rm tim}^{+}(r_{\rm L})
	=
	t_{\rm sp}^{+}(r_{\rm L}).
	\label{eq:profile-meeting}
\end{equation}
Thus the upper half of the surface is a continuous Lorentzian curve made
from a timelike branch ending at $r_0$ and a spacelike branch reaching the
asymptotic boundary. The lower half is obtained by time reflection,
\begin{equation}
	t_{\rm tim}^{-}(r)=-t_{\rm tim}^{+}(r),
	\qquad
	t_{\rm sp}^{-}(r)=-t_{\rm sp}^{+}(r).
	\label{eq:profile-reflection-tee}
\end{equation}
The complete surface is therefore anchored at $t=\pm T/2$ on the
asymptotic boundary.

The equations derived in this subsection will first be applied in the
large-$r$ regime, where the leading angular dependence drops out and the
calculation can be checked analytically. We then restore the exact
black-pole functions $K_y(r,\theta)$ and $G(r,\theta)$ to study how the
localized geometry modifies the Lorentzian branches.

\subsection{Asymptotic limit for timelike entanglement}
\label{sec:tee-asymptotic-limit}

We first consider the large-$r$ regime of the black-pole geometry, where the
surface remains far from the localized core. In this region the exact
functions $K_y(r,\theta)$ and $G(r,\theta)$ take the expansion in
eqs.~\eqref{eq:G-asymptotic}--\eqref{eq:F-asymptotic}, and the leading
dependence on the internal angle drops out of the branch equations. The
calculation is therefore analytically tractable: the time map, the area
integrals, the UV subtraction and the short-boundary-interval limit can all
be obtained explicitly. This regime captures the universal near-boundary
behaviour of the timelike surface, but it does not distinguish the cap-side
and horizon-side sectors. The cap/horizon transition region belongs to the
exact localized geometry and appears only after the full functions
$K_y(r,\theta)$ and $G(r,\theta)$ are restored.

We introduce
\begin{equation}
	a=\ell_2^2,
	\qquad
	b=\ell^2,
	\qquad
	c=r_0^2+2\ell_2^2.
	\label{eq:abc-limit-tee}
\end{equation}
The validity window of the expansion is
\begin{equation}
	r_0>r_\ast\gg \ell,\ell_1,\ell_2.
	\label{eq:validity-window-limit-tee}
\end{equation}
Here $r_0$ is the radial turning point of the timelike branch, while
$r_\ast$ marks the lower edge of the large-$r$ regime. The region
$r<r_\ast$ is part of the localized core and is not described by the present
approximation.

Using the branch equations derived in
section~\ref{sec:tee-branch-equations}, the large-$r$ slopes are
\begin{subequations}
	\begin{align}
		t_{\rm tim}'(r)
		&=
		\left[
		\frac{
			Q(r_0^2+a)
		}{
			(r^2+b)(r^2+a)(r_0^2-r^2)
		}
		\right]^{1/2},
		\qquad
		r_\ast\leq r\leq r_0,
		\label{eq:ttim-limit}
		\\
		t_{\rm sp}'(r)
		&=
		\left[
		\frac{
			Q(r_0^2+a)
		}{
			(r^2+b)(r^2+a)(r^2+c)
		}
		\right]^{1/2},
		\qquad
		r_\ast\leq r<\infty.
		\label{eq:tsp-limit}
	\end{align}
\end{subequations}

The timelike branch has finite radial extent and ends at $r=r_0$. The
spacelike branch reaches the asymptotic boundary and therefore contains the
UV-sensitive part of the surface. The boundary interval is fixed by the
difference between the two time integrals,
\begin{equation}
	\frac{T(r_0,r_\ast)}{2}
	=
	I_{\rm tim}(r_0,r_\ast)
	-
	I_{\rm sp}(r_0,r_\ast),
	\label{eq:T-limit-definition}
\end{equation}
where
\begin{subequations}
	\begin{align}
		I_{\rm tim}
		&=
		\sqrt{Q(r_0^2+a)}
		\int_{r_\ast}^{r_0}
		\frac{dr}{
			\sqrt{(r^2+b)(r^2+a)(r_0^2-r^2)}
		},
		\label{eq:Itim-limit}
		\\
		I_{\rm sp}
		&=
		\sqrt{Q(r_0^2+a)}
		\int_{r_\ast}^{\infty}
		\frac{dr}{
			\sqrt{(r^2+b)(r^2+a)(r^2+c)}
		}.
		\label{eq:Isp-limit}
	\end{align}
\end{subequations}

This expression makes the branch construction explicit. The timelike
integral measures the inward part of the Lorentzian surface, while the
spacelike integral measures the boundary-reaching part. Their difference
sets the temporal size of the boundary interval.

Before evaluating the area, it is useful to display the corresponding
$t$-$r$ profile. The profile is reconstructed from
eqs.~\eqref{eq:ttim-positive-profile}--\eqref{eq:profile-reflection-tee}
using the large-$r$ slopes in eqs.~\eqref{eq:ttim-limit} and
\eqref{eq:tsp-limit}. For the representative plot we use
\begin{equation}
	x_E=0.2,
	\qquad
	\sigma=-1,
	\qquad
	Q_1=Q_5=R_y=1,
	\qquad
	r_\ast=10,
	\qquad
	r_0=100.
	\label{eq:limit-profile-parameters}
\end{equation}

\begin{figure}[H]
	\centering
	\includegraphics[width=0.70\textwidth]{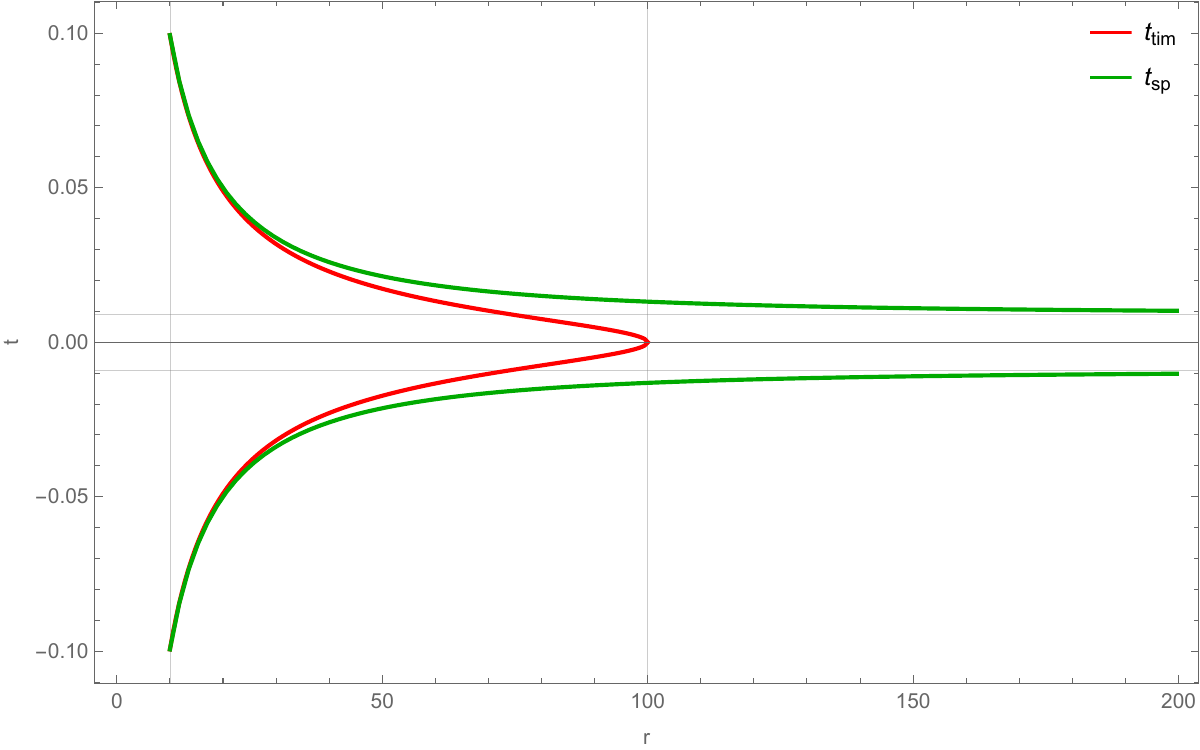}
	\caption{
		Lorentzian branch profile in the large-$r$ regime. The timelike branch
		closes at the radial turning point $r_0$, while the spacelike branch
		reaches the asymptotic boundary and is anchored at $t=\pm T/2$. The
		vertical line marks $r=r_\ast$, the lower edge of the validity window of
		the asymptotic expansion.
	}
	\label{fig:tee-limit-tr-profiles}
\end{figure}

Figure~\ref{fig:tee-limit-tr-profiles} provides a direct check of the branch
gluing in the large-$r$ approximation. The two upper branches meet at
$r=r_\ast$, and the lower half follows by time reflection. Thus the figure
shows the same Lorentzian construction used later in the exact black-pole
geometry, but in a regime where the angular dependence has dropped out.

The time integrals can also be written in closed form. For the timelike
integral we set
\begin{equation}
	u=\frac{r}{\sqrt{r_0^2-r^2}},
	\qquad
	u_\ast=\frac{r_\ast}{\sqrt{r_0^2-r_\ast^2}},
	\label{eq:u-def-limit}
\end{equation}
and define
\begin{equation}
	A_a=\frac{r_0^2+a}{a},
	\qquad
	A_b=\frac{r_0^2+b}{b}.
	\label{eq:Aa-Ab-limit}
\end{equation}
Then
\begin{equation}
	I_{\rm tim}
	=
	\frac{\sqrt{Q(r_0^2+a)}}{\sqrt{ab}}
	\left[
	{\cal F}_{\rm tim}(\infty)
	-
	{\cal F}_{\rm tim}(u_\ast)
	\right],
	\label{eq:Itim-analytic-limit}
\end{equation}
where
\begin{equation}
	{\cal F}_{\rm tim}(u)
	=
	u\,F_1
	\left(
	\frac{1}{2};
	\frac{1}{2},\frac{1}{2};
	\frac{3}{2};
	-A_bu^2,-A_au^2
	\right).
	\label{eq:Ftim-Appell-limit}
\end{equation}
For the spacelike integral we set
\begin{equation}
	z=\frac{r}{\sqrt{r^2+a}},
	\qquad
	z_\ast=\frac{r_\ast}{\sqrt{r_\ast^2+a}},
	\label{eq:z-def-limit}
\end{equation}
and
\begin{equation}
	B_b=\frac{b-a}{b},
	\qquad
	B_c=\frac{c-a}{c}.
	\label{eq:Bb-Bc-limit}
\end{equation}
The integral becomes
\begin{equation}
	I_{\rm sp}
	=
	\frac{\sqrt{Q(r_0^2+a)}}{\sqrt{bc}}
	\left[
	{\cal F}_{\rm sp}(1)
	-
	{\cal F}_{\rm sp}(z_\ast)
	\right],
	\label{eq:Isp-analytic-limit}
\end{equation}
with
\begin{equation}
	{\cal F}_{\rm sp}(z)
	=
	z\,F_1
	\left(
	\frac{1}{2};
	\frac{1}{2},\frac{1}{2};
	\frac{3}{2};
	B_bz^2,B_cz^2
	\right).
	\label{eq:Fsp-Appell-limit}
\end{equation}
Equations~\eqref{eq:T-limit-definition},
\eqref{eq:Itim-analytic-limit}, and \eqref{eq:Isp-analytic-limit} give an
analytic representation of the time map.

\begin{figure}[H]
	\centering
	\includegraphics[width=0.70\textwidth]{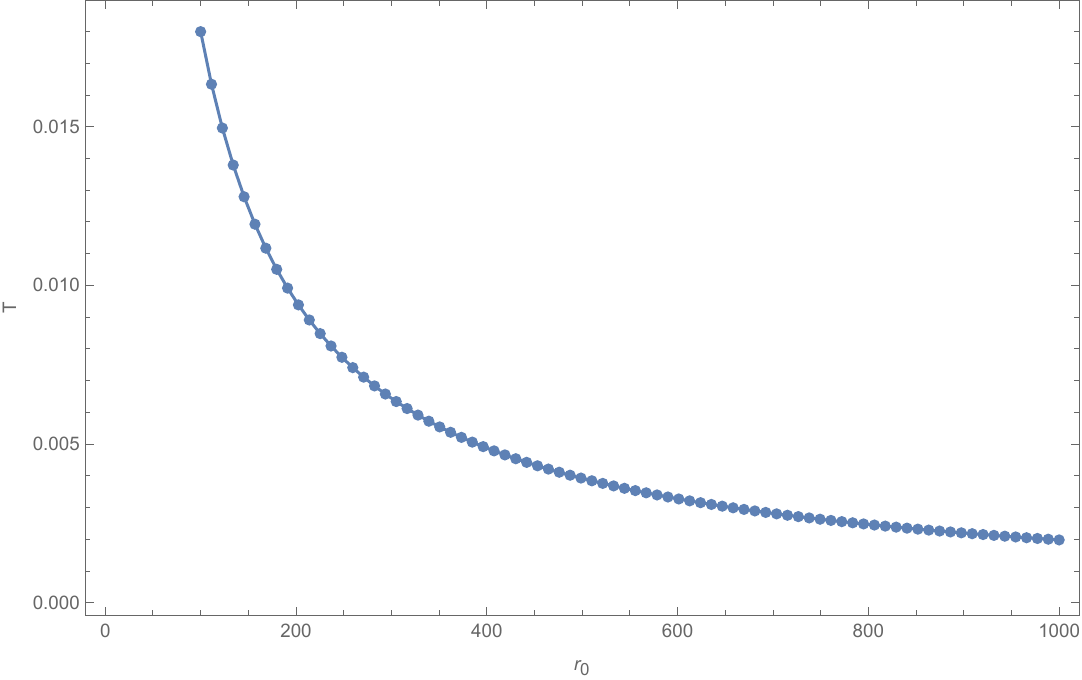}
	\caption{
		Boundary interval $T$ as a function of the radial turning point $r_0$ in
		the large-$r$ regime. The map is single-valued inside the validity window
		of the asymptotic expansion.
	}
	\label{fig:tee-limit-time-map}
\end{figure}

Figure~\ref{fig:tee-limit-time-map} shows that the large-$r$ time map is
single-valued. Therefore the multibranch fixed-boundary-interval structure
found later cannot be attributed to the universal near-boundary region. It
is generated only when the exact localized functions are restored and the
surface becomes sensitive to the black-pole core.

We now turn to the area. Let
\begin{equation}
	{\cal N}=(2\pi)^6V_4
	\label{eq:N-limit-def}
\end{equation}
be the compact-space factor. The spacelike branch gives the regulated real
contribution
\begin{equation}
	{\cal A}_{\rm sp}(R)
	=
	{\cal N}Q
	\int_{r_\ast}^{R}
	dr
	\left[
	\frac{r^2+a}{(r^2+b)(r^2+c)}
	\right]^{1/2}.
	\label{eq:Asp-regulated-limit}
\end{equation}
At large $r$ the integrand has the universal asymptotic tail
\begin{equation}
	\left[
	\frac{r^2+a}{(r^2+b)(r^2+c)}
	\right]^{1/2}
	=
	\frac{1}{r}
	+
	O(r^{-3}).
	\label{eq:Asp-tail-limit}
\end{equation}
Hence
\begin{equation}
	{\cal A}_{\rm sp}(R)
	=
	{\cal N}Q\log R+\text{finite}.
	\label{eq:Asp-div-limit}
\end{equation}
The renormalized real contribution is defined by subtracting the logarithmic
UV divergence,
\begin{equation}
	{\cal A}_{\rm sp,ren}
	=
	{\cal N}Q
	\lim_{R\to\infty}
	\left[
	\int_{r_\ast}^{R}
	dr
	\left[
	\frac{r^2+a}{(r^2+b)(r^2+c)}
	\right]^{1/2}
	-
	\log R
	\right].
	\label{eq:Asp-ren-limit}
\end{equation}
This subtraction removes the divergence associated with the boundary-reaching
branch.

The timelike branch gives
\begin{equation}
	{\cal A}_{\rm tim}
	=
	i\,{\cal N}Q
	\int_{r_\ast}^{r_0}
	dr
	\left[
	\frac{r^2+a}{(r^2+b)(r_0^2-r^2)}
	\right]^{1/2}.
	\label{eq:Atim-limit}
\end{equation}
This contribution is finite. The endpoint at $r=r_0$ gives an integrable
square-root behaviour, and the lower endpoint $r=r_\ast$ lies inside the
large-$r$ validity window. No UV subtraction is required because this
branch does not reach the asymptotic boundary.

The renormalized area in the asymptotic regime is
\begin{equation}
	{\cal A}_{\rm tEE,ren}
	=
	{\cal A}_{\rm sp,ren}
	+
	{\cal A}_{\rm tim},
	\qquad
	{\cal S}_{\rm tEE}
	=
	\frac{{\cal A}_{\rm tEE,ren}}{4G_{10}}.
	\label{eq:Stee-limit}
\end{equation}
The real part is the UV-renormalized contribution from the spacelike branch.
The imaginary part is finite and comes from the timelike branch.

The area integrals can be written in analytic form. Define
\begin{equation}
	{\cal A}_{\rm sp,ren}
	=
	{\cal N}Q\, A_{\rm Re},
	\qquad
	{\cal A}_{\rm tim}
	=
	i\,{\cal N}Q\, A_{\rm Im}.
	\label{eq:dimensionless-area-limit}
\end{equation}
For the imaginary part,
\begin{equation}
	A_{\rm Im}
	=
	\frac{\sqrt{a}}{\sqrt{b}}
	\left[
	{\cal J}_{\rm tim}(\infty)
	-
	{\cal J}_{\rm tim}(u_\ast)
	\right],
	\label{eq:AIm-analytic-limit}
\end{equation}
where
\begin{equation}
	{\cal J}_{\rm tim}(u)
	=
	u\,F_D^{(3)}
	\left(
	\frac{1}{2};
	1,\frac{1}{2},-\frac{1}{2};
	\frac{3}{2};
	-u^2,-A_bu^2,-A_au^2
	\right).
	\label{eq:Jtim-Lauricella-limit}
\end{equation}
For the real part, let
\begin{equation}
	z_R=\frac{R}{\sqrt{R^2+a}}.
	\label{eq:zR-limit}
\end{equation}
Then
\begin{equation}
	A_{\rm Re}
	=
	\lim_{R\to\infty}
	\left[
	\frac{a}{\sqrt{bc}}
	\left(
	{\cal J}_{\rm sp}(z_R)
	-
	{\cal J}_{\rm sp}(z_\ast)
	\right)
	-
	\log R
	\right],
	\label{eq:ARe-analytic-limit}
\end{equation}
with
\begin{equation}
	{\cal J}_{\rm sp}(z)
	=
	z\,F_D^{(3)}
	\left(
	\frac{1}{2};
	1,\frac{1}{2},\frac{1}{2};
	\frac{3}{2};
	z^2,B_bz^2,B_cz^2
	\right).
	\label{eq:Jsp-Lauricella-limit}
\end{equation}
The special-function expressions are useful checks of the radial integrals.
The integral form, however, makes the physics immediate: the
boundary-reaching branch carries the UV divergence, while the finite
timelike branch gives the imaginary contribution.

\begin{figure}[H]
	\centering
	\includegraphics[width=0.48\textwidth]{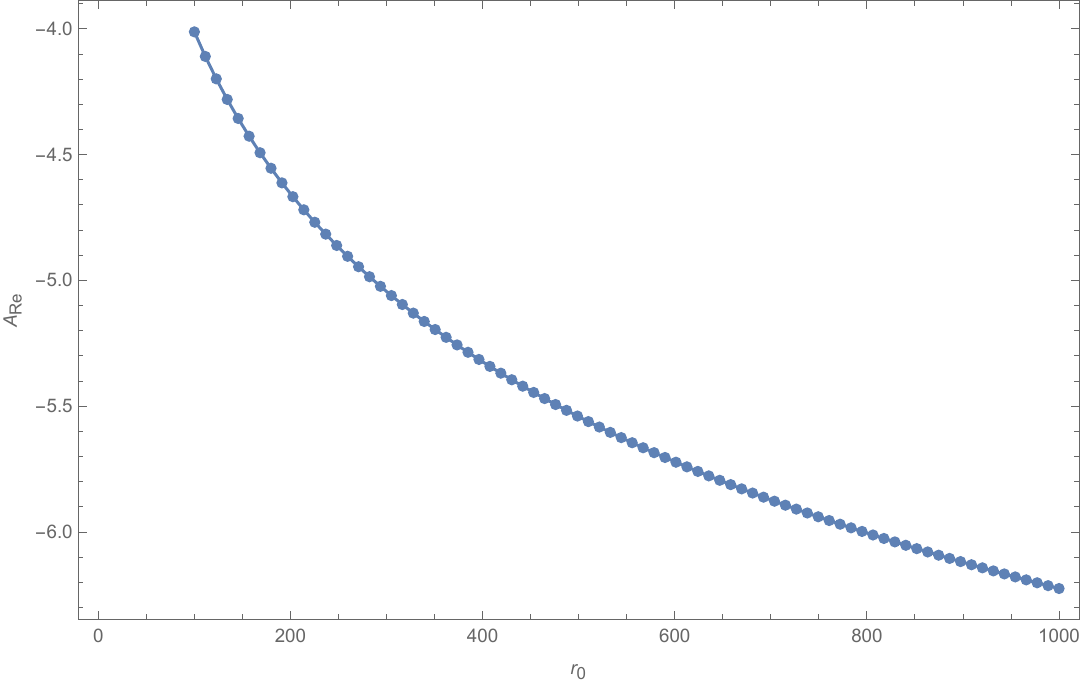}
	\hfill
	\includegraphics[width=0.48\textwidth]{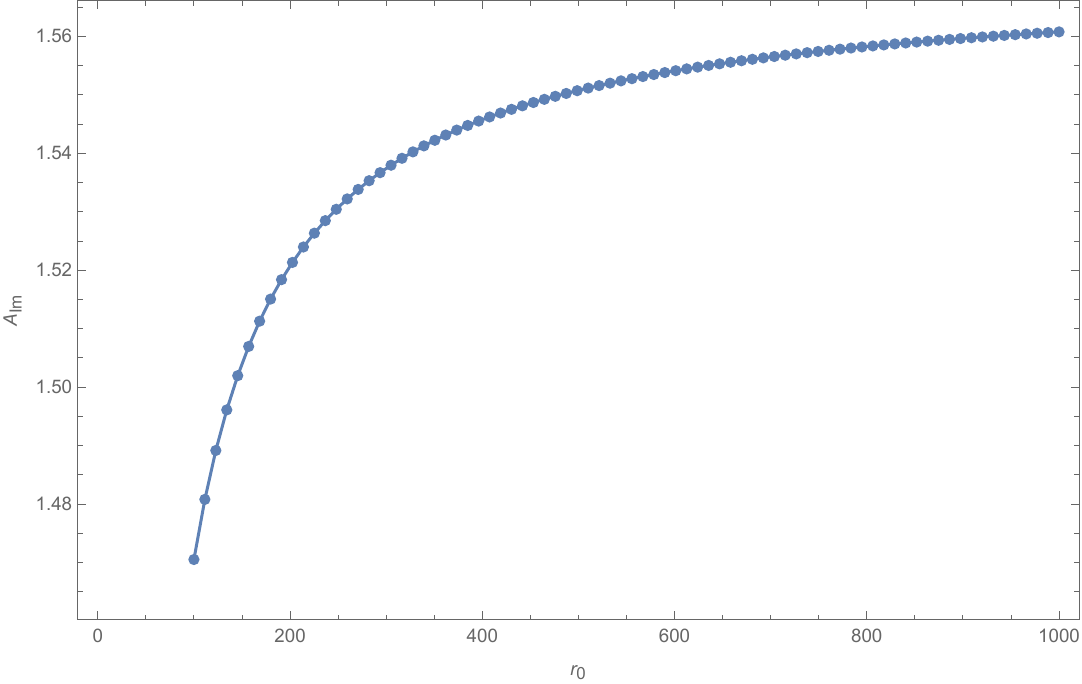}
	\caption{
		Timelike-entanglement area in the asymptotic regime as a function of the
		radial turning point $r_0$. The left panel shows the renormalized real
		part after the logarithmic UV subtraction. The right panel shows the
		finite imaginary part from the timelike branch.
	}
	\label{fig:tee-limit-area-r0}
\end{figure}

Figure~\ref{fig:tee-limit-area-r0} shows the clean separation between the
two parts of the asymptotic area. Varying $r_0$ changes the size of the
Lorentzian surface, but the origin of the two contributions remains fixed:
the real part is tied to the boundary-reaching spacelike branch, while the
imaginary part is tied to the finite timelike branch.

Because the time map in figure~\ref{fig:tee-limit-time-map} is single-valued,
the same area data can be re-expressed as functions of the boundary interval
$T$.

\begin{figure}[H]
	\centering
	\includegraphics[width=0.48\textwidth]{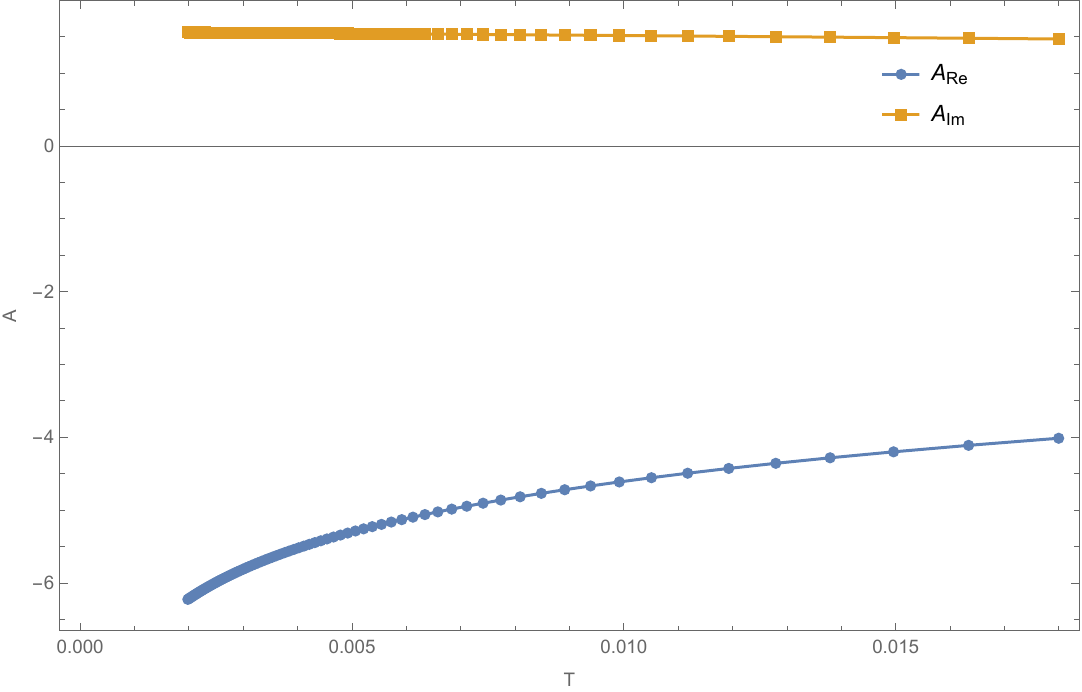}
	\caption{
		Timelike-entanglement area in the large-$r$ regime as a function of the
		boundary interval. The real part shows the logarithmic
		short-boundary-interval behaviour after UV subtraction, while the
		imaginary part approaches a constant. Localized-geometry effects are
		absent in this regime.
	}
	\label{fig:tee-limit-area-T}
\end{figure}

Figure~\ref{fig:tee-limit-area-T} displays the same asymptotic result in the
physical variable $T$. This presentation is useful for comparison with the
fixed-boundary-interval analysis in the exact geometry. In the large-$r$
regime there is no angular competition and no branch selection over
$\theta_0$; the area is determined entirely by the single-valued time map.

We now extract the short-boundary-interval behaviour. In the large-$r$
regime, a short boundary interval is obtained when the turning point lies far
out in the asymptotic region. Thus the limit $T\to0$ is described by taking
$r_0$ large. Keeping $r_\ast$ fixed inside the validity window, the two time integrals
have leading behaviour
\begin{subequations}
	\begin{align}
		I_{\rm tim}
		&=
		\sqrt{Q}\,
		\frac{\sqrt{r_0^2-r_\ast^2}}{r_0r_\ast}
		+
		O(r_0^{-2}),
		\label{eq:Itim-short}
		\\
		I_{\rm sp}
		&=
		\sqrt{Q}\,
		\frac{\sqrt{r_0^2+r_\ast^2}-r_\ast}{r_0r_\ast}
		+
		O(r_0^{-2}).
		\label{eq:Isp-short}
	\end{align}
\end{subequations}
The leading $1/r_\ast$ terms cancel in the difference. Therefore
\begin{equation}
	T(r_0)
	=
	\frac{2\sqrt{Q}}{r_0}
	+
	O(r_0^{-2}),
	\qquad
	r_0(T)
	=
	\frac{2\sqrt{Q}}{T}
	+
	O(1).
	\label{eq:T-short-limit}
\end{equation}
The imaginary part becomes
\begin{equation}
	{\cal A}_{\rm tim}
	=
	i{\cal N}Q
	\left[
	\frac{\pi}{2}
	+
	O(T)
	\right].
	\label{eq:Atim-short-limit}
\end{equation}
For the spacelike part, the regulated area satisfies
\begin{equation}
	{\cal A}_{\rm sp}(R)
	=
	{\cal N}Q
	\left[
	\log R
	+
	\log\left(\frac{T}{\sqrt{Q}}\right)
	+
	O(T^2)
	\right].
	\label{eq:Asp-short-regulated-limit}
\end{equation}
After the subtraction in eq.~\eqref{eq:Asp-ren-limit}, the renormalized area
is
\begin{equation}
	{\cal A}_{\rm tEE,ren}(T)
	=
	{\cal N}Q
	\left[
	\log\left(\frac{T}{\sqrt{Q}}\right)
	+	O(T^2)+
	i(\frac{\pi}{2}
	+
	O(T))
	\right].
	\label{eq:Atee-short-limit}
\end{equation}
Thus
\begin{equation}
	{\cal S}_{\rm tEE}(T)
	=
	\frac{{\cal N}Q}{4G_{10}}
	\left[
	\log\left(\frac{T}{\sqrt{Q}}\right)+	O(T^2)
	+
	i(\frac{\pi}{2}
	+
	O(T))
	\right].
	\label{eq:Stee-short-limit}
\end{equation}

Equation~\eqref{eq:Stee-short-limit} summarizes the large-$r$ region. The
logarithmic real part comes from the boundary-reaching branch after UV
subtraction, while the constant imaginary part comes from the finite
timelike branch. Since the leading angular dependence has dropped out, this
result captures only the universal short-boundary-interval behaviour. The
localized core and the cap/horizon transition region require the exact
black-pole geometry.

\subsection{Timelike entanglement entropy in the exact black-pole geometry}
\label{sec:tee-exact-black-pole}
We now restore the exact black-pole functions $K_y(r,\theta)$ and
$G(r,\theta)$.  No large-$r$ expansion is used in this subsection.  The
purpose is to study how the timelike-entanglement surface changes when the
surface probes the localized core and the cap/horizon transition region of
the internal sphere.

The construction follows the localized lifting prescription of
ref.~\cite{Bena:2024lbh}, adapted to the Lorentzian timelike branches.  For
a fixed angular label $\theta_0$, the effective two-dimensional metric in
eqs.~\eqref{eq:reduced-two-dimensional-metric}--\eqref{eq:H-definition}
defines a Lorentzian branch problem in the $(t,r)$ plane.  The solution
contains a spacelike branch, which reaches the asymptotic boundary, and a
timelike branch, which closes at the turning point $r_0$.  We denote this
branch configuration by
\begin{equation}
	{\cal B}_{\theta_0,r_0}
	=
	\left(
	{\cal B}_{\rm sp},
	{\cal B}_{\rm tim}
	\right)_{\theta_0,r_0}.
	\label{eq:tee-reduced-branch-family}
\end{equation}
Here $\theta_0$ labels the effective branch family, while $r_0$ fixes the
turning point of the timelike branch.  The same branch data determine the
boundary interval,
\begin{equation}
	{\cal B}_{\theta_0,r_0}
	\quad\longrightarrow\quad
	T(r_0,\theta_0).
	\label{eq:tee-branch-to-time}
\end{equation}

The important point is that the ten-dimensional area is not obtained by
setting $\theta=\theta_0$.  The branch profile is first found at the angular
label $\theta_0$, and this profile is then lifted over the physical internal
angle $\theta$.  Let
\begin{equation}
	{\cal N}=(2\pi)^6 V_4 ,
	\label{eq:tee-compact-factor-exact}
\end{equation}
where $V_4$ is the coordinate volume of the internal $T^4$.  The factor
${\cal N}$ collects the volume of the compact directions that are integrated
trivially, while the remaining nontrivial internal dependence is carried by
the $\theta$ integral.  The lifted timelike-entanglement area is then
\begin{equation}
\begin{aligned}
A_{\rm tEE}^{\rm lift}(r_0,\theta_0)
={}&
{\cal N}
\int_0^{\pi/2} d\theta\,
\sin\theta\cos\theta
\Bigg[
\int_{{\cal B}_{\rm sp}} dr\,
\sqrt{
\Delta_{\rm sp}
(r,\theta;\theta_0,r_0)+i0
}
\\
&\qquad\qquad
+
\int_{{\cal B}_{\rm tim}} dr\,
\sqrt{
\Delta_{\rm tim}
(r,\theta;\theta_0,r_0)+i0
}
\Bigg].
\end{aligned}
\label{eq:tee-localized-rt-analogue}
\end{equation}
The lifted area functions are evaluated with the full black-pole geometry,
\begin{subequations}
\begin{align}
	\Delta_{\rm sp}(r,\theta;\theta_0,r_0)
	&=
	H(r,\theta)
	-
	F(r,\theta)K_{\rm sp}(r;\theta_0,r_0)^2,
	\label{eq:exact-tee-Delta-sp}
	\\
	\Delta_{\rm tim}(r,\theta;\theta_0,r_0)
	&=
	H(r,\theta)
	-
	F(r,\theta)K_{\rm tim}(r;\theta_0,r_0)^2,
	\label{eq:exact-tee-Delta-tim}
\end{align}
\end{subequations}
where
\begin{equation}
	F(r,\theta)=Q\,r^2K_y(r,\theta)G(r,\theta),
	\qquad
	H(r,\theta)=\frac{Q^2G(r,\theta)^2}{r^2+\ell^2}.
	\label{eq:exact-tee-H-theta}
\end{equation}
The branch kernels $K_{\rm sp}$ and $K_{\rm tim}$ are determined at
$\theta_0$, whereas the metric functions in
eq.~\eqref{eq:exact-tee-H-theta} are evaluated at the physical internal
angle $\theta$.  Thus $\theta_0$ fixes the Lorentzian path in the effective
$(t,r)$ geometry, while $\theta$ determines the ten-dimensional weighting of
that path.  This distinction is the localized ingredient absent in the BTZ
uplift.

The selected timelike-entanglement surface is obtained only after the
boundary interval has been fixed.  For each target interval, one first solves
\begin{equation}
	T(r_0,\theta_0)=T_{\rm target}
	\label{eq:tee-localized-fixedT}
\end{equation}
for all admissible angular labels and radial branches.  On this
fixed-boundary-interval family, the selected branch is
\begin{equation}
	(\theta_0^\ast,b^\ast)
	=
	\underset{\theta_0,b}{\arg\min}\,
	\operatorname{Re}
	A_{\rm tEE}^{\rm lift}
	\left(
	r_{0,b}(T_{\rm target},\theta_0),
	\theta_0
	\right).
	\label{eq:tee-localized-minimum}
\end{equation}
The imaginary part is then evaluated on the same selected surface.  Thus the
minimisation is over the real part of the lifted area, but only after the
Lorentzian branches have been restricted to a common boundary interval.

We now implement the localized timelike prescription in the exact black-pole
geometry. For each angular label $\theta_0$, the Lorentzian branch equations
are solved in the reduced $(t,r)$ geometry, giving the radial and temporal
profiles of the spacelike and timelike branches. These profiles are then
held fixed while the area is lifted over the physical internal angle
$\theta$. Thus $\theta_0$ selects the reduced branch family, whereas
$\theta$ enters the lifted area through the exact functions
$K_y(r,\theta)$ and $G(r,\theta)$. The final area is therefore not obtained
by setting $\theta=\theta_0$ everywhere: the reduced branch profile fixes the
Lorentzian path, while the lift determines how this path contributes to the
real and imaginary parts of the timelike-entanglement area.

For the exact calculation we use the reduced metric functions
$F(r;\theta_0)$ and $H(r;\theta_0)$ defined in
eqs.~\eqref{eq:reduced-two-dimensional-metric}--\eqref{eq:H-definition},
with the full black-pole functions evaluated at $\theta=\theta_0$. The
turning-point value is
\begin{equation}
	F_0=F(r_0;\theta_0).
	\label{eq:exact-tee-F0}
\end{equation}
We denote the exact branch slopes by
\begin{subequations}
	\begin{align}
		K_{\rm tim}(r;\theta_0,r_0)
		&\equiv
		t_{\rm tim}'(r),
		\label{eq:exact-tee-Ktim}
		\\
		K_{\rm sp}(r;\theta_0,r_0)
		&\equiv
		t_{\rm sp}'(r).
		\label{eq:exact-tee-Ksp}
	\end{align}
\end{subequations}

These are the general slopes in eqs.~\eqref{eq:ttim-prime} and
\eqref{eq:tsp-prime}, evaluated with the exact black-pole functions. The
timelike branch ends at the radial turning point $r=r_0$, while the
spacelike branch reaches the asymptotic boundary.

For each pair $(r_0,\theta_0)$, the boundary interval is
\begin{equation}
	T(r_0,\theta_0)
	=
	2\left[
	I_{\rm tim}(r_0,\theta_0)
	-
	I_{\rm sp}(r_0,\theta_0)
	\right],
	\label{eq:exact-tee-boundary-interval}
\end{equation}
where
\begin{subequations}
	\begin{align}
		I_{\rm tim}(r_0,\theta_0)
		&=
		\int_{\epsilon}^{r_0}
		dr\,K_{\rm tim}(r;\theta_0,r_0),
		\label{eq:exact-tee-Itim}
		\\
		I_{\rm sp}(r_0,\theta_0)
		&=
		\int_{\epsilon}^{R_{\rm max}}
		dr\,K_{\rm sp}(r;\theta_0,r_0).
		\label{eq:exact-tee-Isp}
	\end{align}
\end{subequations}
The parameter $\epsilon$ regulates the small-$r$ endpoint in the localized
core region. The cutoff $R_{\rm max}$ regulates the boundary-reaching
spacelike branch. The logarithmic dependence on $R_{\rm max}$ is removed by
the ten-dimensional UV subtraction described below.

For the numerical examples in this subsection we use
\begin{equation}
	x_E=0.2,
	\qquad
	\sigma=-1,
	\qquad
	Q_1=Q_5=R_y=1.
	\label{eq:exact-tee-parameters}
\end{equation}
The transition angle is determined from eq.~\eqref{eq:theta-star}. For
this choice, $\theta_\star\simeq0.759$. We use angular labels on both sides
of $\theta_\star$ so that the cap-side sector, the horizon-side sector and
the cap/horizon transition region are all included.

The first effect of the exact geometry appears in the time map. In the
large-$r$ regime, the boundary interval is single-valued as a function of
the turning point in the short-boundary-interval domain. In the exact
black-pole geometry, the functions $K_y(r,\theta_0)$ and $G(r,\theta_0)$
modify $F(r;\theta_0)$ and $H(r;\theta_0)$ throughout the radial direction.
At fixed $\theta_0$, the curve $T(r_0,\theta_0)$ can increase, reach a
maximum and then decrease. The same boundary interval can then be realized
by more than one radial branch.

For each angular label we define
\begin{equation}
	T_{\rm max}(\theta_0)
	=
	\max_{r_0}T(r_0,\theta_0).
	\label{eq:exact-tee-Tmax}
\end{equation}
A branch at fixed target interval exists only when
\begin{equation}
	T_{\rm target}\leq T_{\rm max}(\theta_0).
	\label{eq:exact-tee-admissible-domain}
\end{equation}
Thus the fixed-boundary-interval problem is already constrained by the time
map before any area comparison is made.

\begin{figure}[H]
	\centering
	\includegraphics[width=0.48\textwidth]{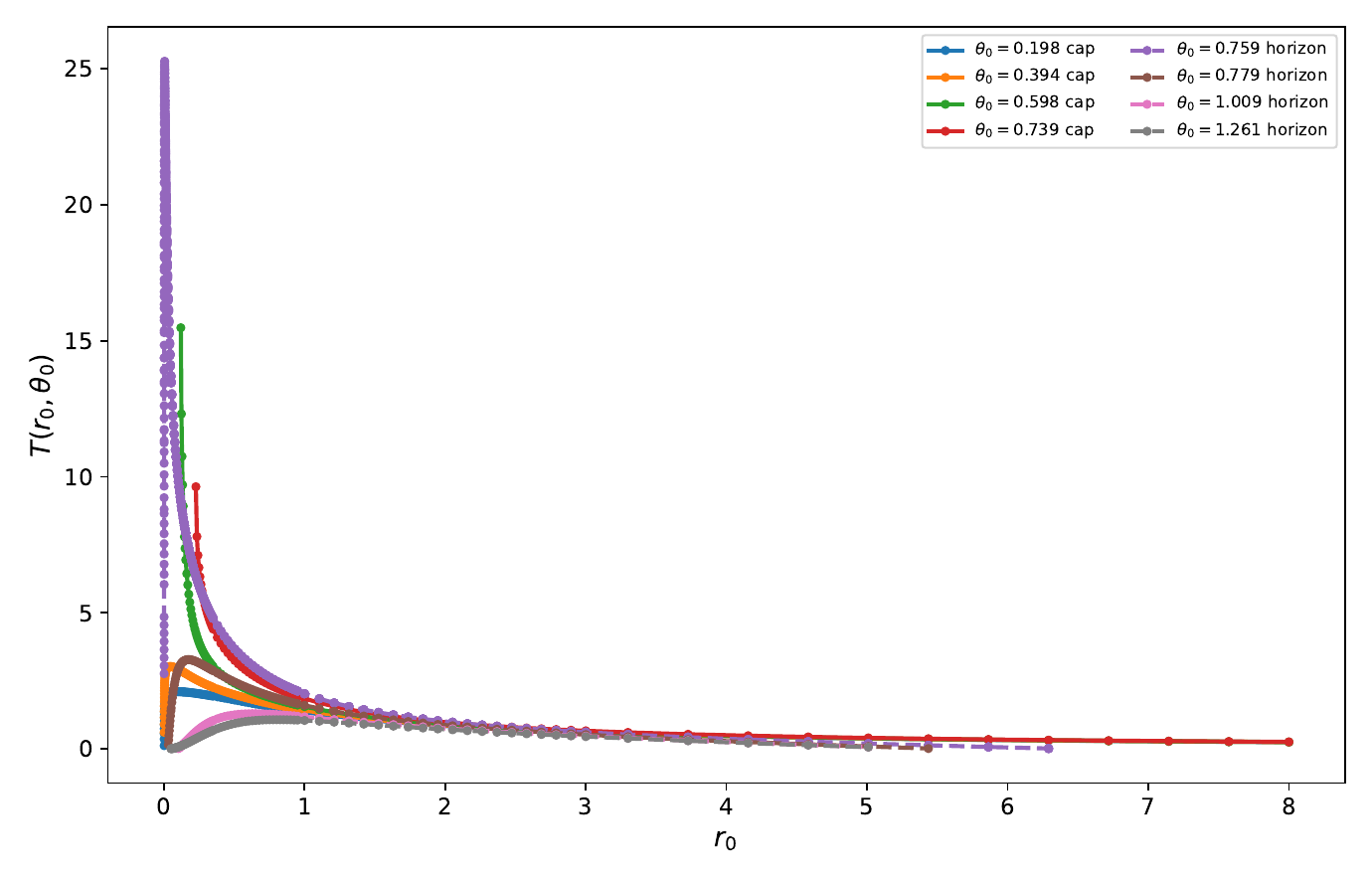}
	\hfill
	\includegraphics[width=0.48\textwidth]{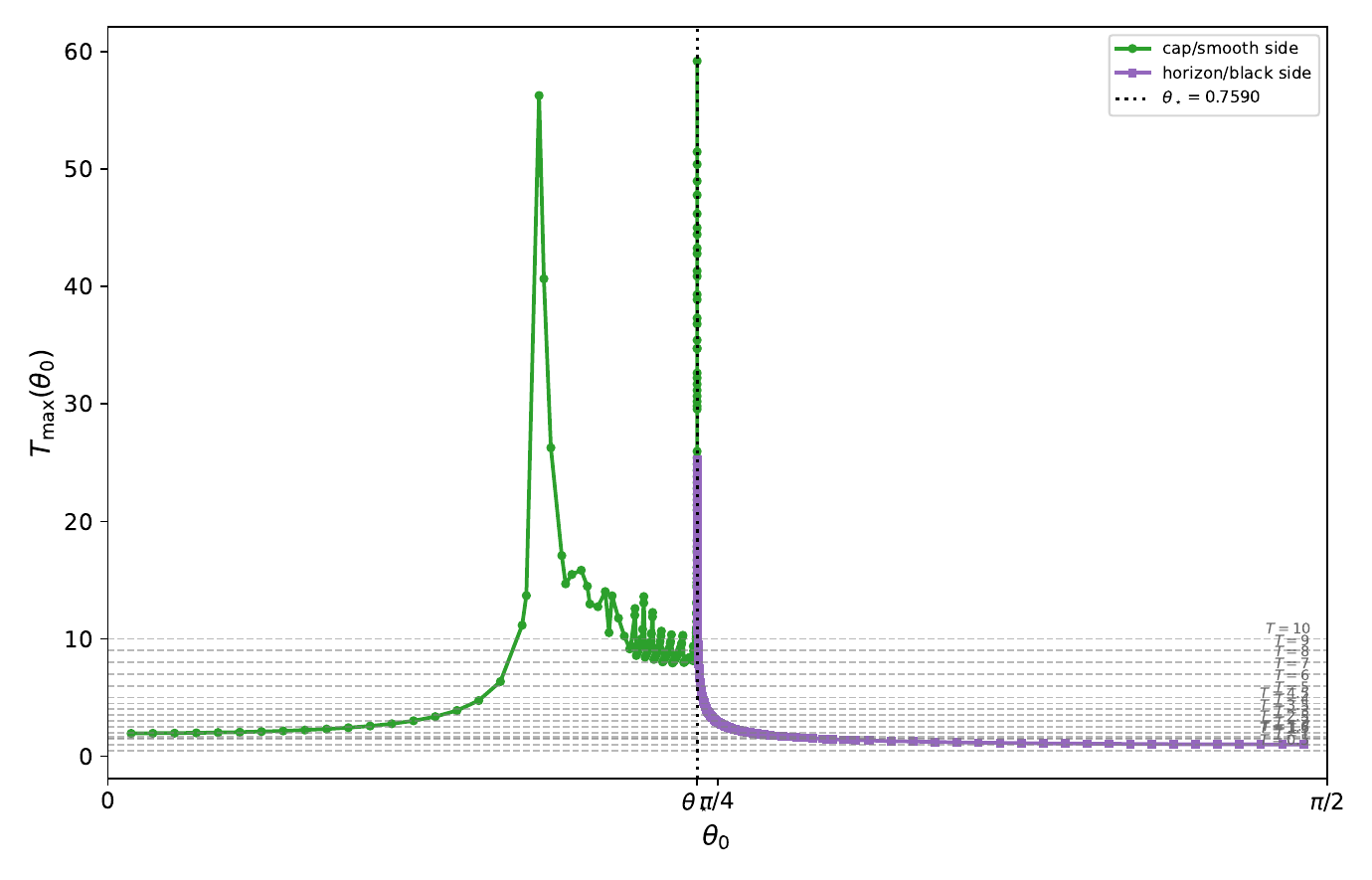}
	\caption{
		Exact black-pole time map for timelike entanglement. Left: representative
		families $T(r_0,\theta_0)$ at fixed angular label. When the curve is
		non-monotonic, the same $T_{\rm target}$ can be reached by two radial
		branches. Right: maximum accessible interval $T_{\rm max}(\theta_0)$.
		The vertical dashed line marks $\theta_\star$, and the horizontal dashed
		lines denote representative target intervals. Only angular labels with
		$T_{\rm max}(\theta_0)\geq T_{\rm target}$ can contribute at that target
		interval.
	}
	\label{fig:tee-exact-temporal-structure}
\end{figure}

Figure~\ref{fig:tee-exact-temporal-structure} shows that the physical
boundary quantity is $T$, not the bulk parameter $r_0$. When the time map is
non-monotonic, two different turning points can describe the same boundary
interval. These two radial branches are not equivalent: they enter
different radial parts of the exact black-pole geometry and therefore can
have different lifted areas. The right panel shows that long target
intervals are supported mainly by angular labels near the cap/horizon
transition region.

\begin{figure}[H]
	\centering
	\includegraphics[width=0.48\textwidth]{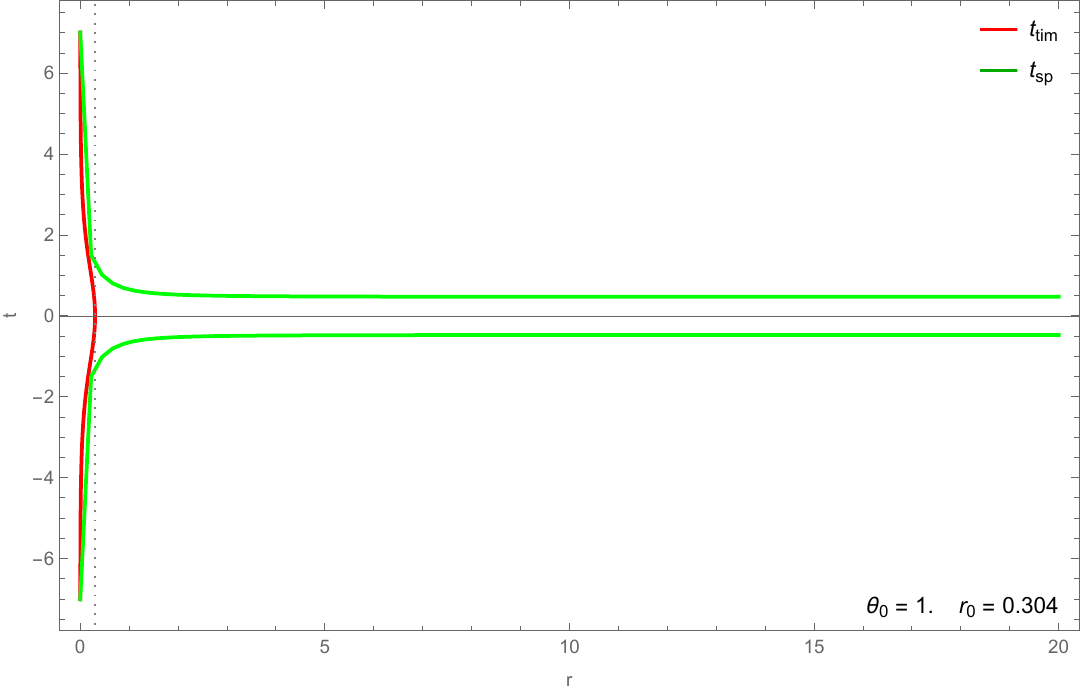}
	\hfill
	\includegraphics[width=0.48\textwidth]{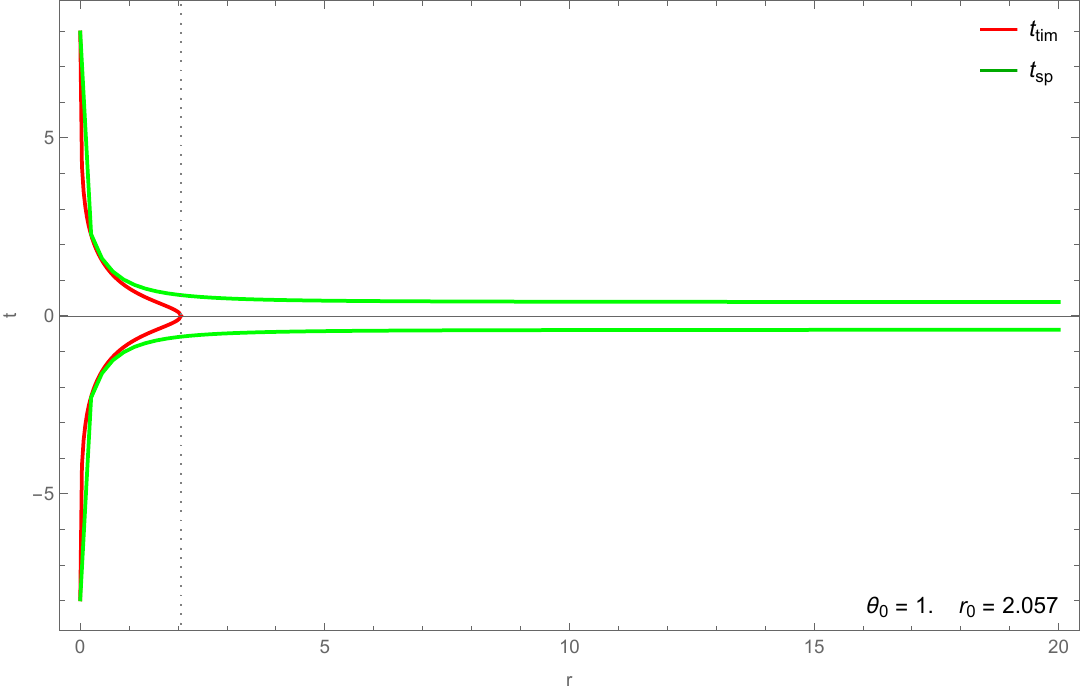}
	\caption{Exact $t$-$r$ branch profiles at fixed angular label $\theta_0$ for two different turning points $r_0$. The profiles are shown before imposing the fixed-boundary-interval condition.}
	\label{fig:tee-exact-tr-profiles}
\end{figure}

The exact branch profiles also reflect the different endpoint behaviour of
the two angular sectors as we explain in Appendix \ref{app:transition-region}. On the cap side, $\theta_0<\theta_\star$, the
time-direction function has a finite small-$r$ limit,
\begin{equation}
	F(r;\theta_0)
	=
	F_{\rm cap}(\theta_0)
	+
	O(r^2),
	\qquad
	F_{\rm cap}(\theta_0)>0.
	\label{eq:tee-cap-F}
\end{equation}
Consequently, the branch kernels are finite at the lower endpoint,
\begin{subequations}
	\begin{align}
		K_{\rm tim}(r;\theta_0,r_0)
		&=
		K_{\rm tim}^{(0)}(\theta_0,r_0)
		+
		O(r^2),
		\label{eq:tee-cap-Ktim}
		\\
		K_{\rm sp}(r;\theta_0,r_0)
		&=
		K_{\rm sp}^{(0)}(\theta_0,r_0)
		+
		O(r^2).
		\label{eq:tee-cap-Ksp}
	\end{align}
\end{subequations}
The cap-side branch profiles therefore have mild small-$r$ behaviour. On the horizon side, $\theta_0>\theta_\star$, the local form is instead
\begin{equation}
	F(r;\theta_0)
	=
	f_2(\theta_0)\,r^2
	+
	O(r^4),
	\qquad
	f_2(\theta_0)>0.
	\label{eq:tee-horizon-F}
\end{equation}
The branch kernels then behave as
\begin{subequations}
	\begin{align}
		K_{\rm tim}(r;\theta_0,r_0)
		&=
		\frac{\alpha_{\rm tim}(\theta_0,r_0)}{r}
		+
		O(r),
		\label{eq:tee-horizon-Ktim}
		\\
		K_{\rm sp}(r;\theta_0,r_0)
		&=
		\frac{\alpha_{\rm sp}(\theta_0,r_0)}{r}
		+
		O(r).
		\label{eq:tee-horizon-Ksp}
	\end{align}
\end{subequations}
The individual upper-branch profiles therefore contain logarithmic terms,
\begin{subequations}
	\begin{align}
		t_{\rm tim}^{+}(r)
		&\sim
		\alpha_{\rm tim}(\theta_0,r_0)
		\log\frac{1}{r},
		\label{eq:tee-horizon-ttim-log}
		\\
		t_{\rm sp}^{+}(r)
		&\sim
		\alpha_{\rm sp}(\theta_0,r_0)
		\log\frac{1}{r},
		\qquad
		r\rightarrow0.
		\label{eq:tee-horizon-tsp-log}
	\end{align}
\end{subequations}
These logarithms explain the large vertical extent of horizon-side
$t$-$r$ profiles shown in Figure \ref{fig:tee-exact-tr-profiles}. They do not imply an infinite boundary interval. The
boundary interval is the difference of the timelike and spacelike integrals
in eq.~\eqref{eq:exact-tee-boundary-interval}; the common leading endpoint
terms cancel in this difference. Near the turning point, the behaviour is independent of this small-$r$
sector distinction. The timelike branch is real only where
$F(r;\theta_0)\leq F_0$. Close to $r=r_0$,
\begin{equation}
	F_0-F(r;\theta_0)
	=
	F'(r_0;\theta_0)(r_0-r)
	+
	\cdots.
	\label{eq:tee-turning-F}
\end{equation}
Therefore
\begin{equation}
	K_{\rm tim}(r;\theta_0,r_0)
	\sim
	\frac{1}{\sqrt{r_0-r}}.
	\label{eq:tee-turning-Ktim}
\end{equation}
This square-root behaviour is integrable and gives a smooth closing of the
timelike branch at $r=r_0$.

The upper half of the exact branch profile is written as
\begin{equation}
	t_{\rm tim}^{+}(r)
	=
	\int_{r}^{r_0}
	d\tilde r\,K_{\rm tim}(\tilde r;\theta_0,r_0),
	\qquad
	t_{\rm tim}^{+}(r_0)=0,
	\label{eq:exact-tee-ttim-profile}
\end{equation}
and
\begin{equation}
	t_{\rm sp}^{+}(r)
	=
	\frac{T}{2}
	+
	\int_{r}^{R_{\rm max}}
	d\tilde r\,K_{\rm sp}(\tilde r;\theta_0,r_0),
	\qquad
	t_{\rm sp}^{+}(R_{\rm max})=\frac{T}{2}.
	\label{eq:exact-tee-tsp-profile}
\end{equation}
At the lower regulator,
\begin{equation}
	t_{\rm tim}^{+}(\epsilon)=I_{\rm tim},
	\qquad
	t_{\rm sp}^{+}(\epsilon)=\frac{T}{2}+I_{\rm sp}.
	\label{eq:endpoint-profile-values}
\end{equation}
Using eq.~\eqref{eq:exact-tee-boundary-interval}, these values agree:
\begin{equation}
	t_{\rm tim}^{+}(\epsilon)=t_{\rm sp}^{+}(\epsilon).
	\label{eq:endpoint-profile-matching}
\end{equation}
The lower half is obtained by time reflection. The complete Lorentzian
surface is therefore anchored at $t=\pm T/2$ on the asymptotic boundary.

Figure~\ref{fig:tee-exact-tr-profiles} illustrates how the radial turning
point changes the Lorentzian branch geometry. Moving $r_0$ inward makes the
surface enter deeper into the localized region, where the exact functions
$K_y(r,\theta_0)$ and $G(r,\theta_0)$ control the profile. Since the two
curves generally correspond to different boundary intervals, they should not
be compared as competing surfaces until the common condition
$T(r_0,\theta_0)=T_{\rm target}$ is imposed.

Before performing the ten-dimensional lift, we introduce the reduced branch
area functions at fixed $\theta_0$. These quantities describe the sign
structure of the reduced branch geometry before the internal angular
integral is taken:
\begin{subequations}
	\begin{align}
		\Delta_{\rm sp}^{\rm red}(r;\theta_0,r_0)
		&=
		H(r;\theta_0)
		-
		F(r;\theta_0)K_{\rm sp}(r;\theta_0,r_0)^2,
		\label{eq:exact-tee-Delta-red-sp}
		\\
		\Delta_{\rm tim}^{\rm red}(r;\theta_0,r_0)
		&=
		H(r;\theta_0)
		-
		F(r;\theta_0)K_{\rm tim}(r;\theta_0,r_0)^2.
		\label{eq:exact-tee-Delta-red-tim}
	\end{align}
\end{subequations}
Using eqs.~\eqref{eq:exact-tee-Ktim} and \eqref{eq:exact-tee-Ksp}, one finds
\begin{subequations}
	\begin{align}
		\Delta_{\rm sp}^{\rm red}
		&=
		H(r;\theta_0)\,
		\frac{F(r;\theta_0)}{F(r;\theta_0)+F_0},
		\label{eq:exact-tee-Delta-red-sp-simplified}
		\\
		\Delta_{\rm tim}^{\rm red}
		&=
		-
		H(r;\theta_0)\,
		\frac{F(r;\theta_0)}{F_0-F(r;\theta_0)}.
		\label{eq:exact-tee-Delta-red-tim-simplified}
	\end{align}
\end{subequations}
In the allowed domain of the timelike branch, $F(r;\theta_0)<F_0$, and hence
\begin{equation}
	\Delta_{\rm sp}^{\rm red}>0,
	\qquad
	\Delta_{\rm tim}^{\rm red}<0.
	\label{eq:exact-tee-red-signs}
\end{equation}
The corresponding reduced density magnitudes are
\begin{align}
	\rho_{\rm sp}^{\rm red}(r;\theta_0,r_0)
	&=
	\sqrt{\Delta_{\rm sp}^{\rm red}(r;\theta_0,r_0)},
	\label{eq:exact-tee-rho-red-sp}
	\\
	\rho_{\rm tim}^{\rm red}(r;\theta_0,r_0)
	&=
	\sqrt{-\Delta_{\rm tim}^{\rm red}(r;\theta_0,r_0)}.
	\label{eq:exact-tee-rho-red-tim}
\end{align}
At the reduced-branch level, the spacelike branch is real and the timelike
branch is imaginary. This simple separation is lost after the lift, because
the area is then evaluated with the physical internal angle $\theta$ rather
than only with the angular label $\theta_0$.

The reduced sign structure in eq.~\eqref{eq:exact-tee-red-signs} is not
preserved pointwise after the ten-dimensional lift.  In the lifted area, the
branch kernels $K_{\rm tim}(r;\theta_0,r_0)$ and
$K_{\rm sp}(r;\theta_0,r_0)$ remain those determined by the reduced problem
at the angular label $\theta_0$, but the metric functions are evaluated at
the physical internal angle $\theta$, as in
eq.~\eqref{eq:exact-tee-H-theta}.  Equivalently, the relevant lifted area
functions are the quantities defined in
eqs.~\eqref{eq:exact-tee-Delta-tim}--\eqref{eq:exact-tee-Delta-sp}.
At $\theta=\theta_0$ they reproduce the reduced-branch signs, but away from
$\theta_0$ the exact functions $K_y(r,\theta)$ and $G(r,\theta)$ can change
their sign.  Therefore the timelike and spacelike branches can both
contribute to the real and imaginary parts of the lifted area.  The branch
label specifies how the curve is constructed; the sign of the lifted area
function decides whether a local contribution is real or imaginary.

The area is evaluated with the principal square-root continuation,
\begin{equation}
	\sqrt{x+i0}
	=
	\begin{cases}
		\sqrt{x}, & x>0,\\
		i\sqrt{|x|}, & x<0.
	\end{cases}
	\label{eq:exact-tee-principal-root}
\end{equation}
Thus positive regions of $\Delta_{\rm tim}$ or $\Delta_{\rm sp}$ contribute
to the real part, while negative regions contribute to the imaginary part.
This is the precise sense in which the exact ten-dimensional lift mixes the
reduced timelike and spacelike branch contributions. The lifted timelike-branch area is
\begin{equation}
	A_{\rm tim}^{\rm full}(r_0,\theta_0)
	=
	{\cal N}
	\int_0^{\pi/2}
	d\theta\,\sin\theta\cos\theta
	\int_{\epsilon}^{r_0}
	dr\,
	\sqrt{\Delta_{\rm tim}(r,\theta;\theta_0,r_0)+i0}.
	\label{eq:exact-tee-Atim-full}
\end{equation}
This contribution is finite because the timelike branch ends at $r=r_0$ and
does not reach the asymptotic boundary. The lifted spacelike-branch area is
\begin{equation}
	A_{\rm sp}^{\rm full}(R;r_0,\theta_0)
	=
	{\cal N}
	\int_0^{\pi/2}
	d\theta\,\sin\theta\cos\theta
	\int_{\epsilon}^{R}
	dr\,
	\sqrt{\Delta_{\rm sp}(r,\theta;\theta_0,r_0)+i0}.
	\label{eq:exact-tee-Asp-full}
\end{equation}
This branch reaches the asymptotic boundary and contains the logarithmic UV
divergence. At large $r$,
\begin{equation}
	\sqrt{\Delta_{\rm sp}(r,\theta;\theta_0,r_0)+i0}
	=
	\frac{Q}{r}
	+
	O(r^{-3}).
	\label{eq:exact-tee-sp-large-r}
\end{equation}
The divergent contribution is therefore universal:
\begin{equation}
	A_{\rm div}(R)
	=
	{\cal N}Q
	\left(
	\int_0^{\pi/2}
	d\theta\,\sin\theta\cos\theta
	\right)
	\log R
	=
	\frac{{\cal N}Q}{2}\log R.
	\label{eq:exact-tee-Adiv}
\end{equation}
The renormalized spacelike-branch area is
\begin{equation}
	A_{\rm sp,ren}^{\rm full}(r_0,\theta_0)
	=
	\lim_{R\to\infty}
	\left[
	A_{\rm sp}^{\rm full}(R;r_0,\theta_0)
	-
	\frac{{\cal N}Q}{2}\log R
	\right].
	\label{eq:exact-tee-Asp-ren-full}
\end{equation}
Equivalently, for numerical evaluation one may subtract the asymptotic tail
inside the integral:
\begin{equation}
	A_{\rm sp,ren}^{\rm full}(r_0,\theta_0)
	=
	{\cal N}
	\int_0^{\pi/2}
	d\theta\,\sin\theta\cos\theta
	\int_{\epsilon}^{\infty}
	dr\,
	\left[
	\sqrt{\Delta_{\rm sp}(r,\theta;\theta_0,r_0)+i0}
	-
	\sqrt{H(r,\theta)}
	\right].
	\label{eq:exact-tee-Asp-ren-local}
\end{equation}
Here
\begin{equation}
	\sqrt{H(r,\theta)}
	=
	\frac{Q}{r}
	+
	O(r^{-3})
	\label{eq:exact-tee-local-tail}
\end{equation}
at large $r$, so the subtraction removes the logarithmic UV divergence. The
finite contribution from the localized region is kept.

The timelike branch does not require a UV subtraction:
\begin{equation}
	A_{\rm tim,ren}^{\rm full}(r_0,\theta_0)
	=
	A_{\rm tim}^{\rm full}(r_0,\theta_0).
	\label{eq:exact-tee-Atim-ren-full}
\end{equation}
The full renormalized timelike-entanglement area is
\begin{equation}
	A_{\rm tEE}^{\rm full}(r_0,\theta_0)
	=
	A_{\rm sp,ren}^{\rm full}(r_0,\theta_0)
	+
	A_{\rm tim}^{\rm full}(r_0,\theta_0).
	\label{eq:exact-tee-Afull}
\end{equation}
In general,
\begin{equation}
	A_{\rm tEE}^{\rm full}
	=
	\Re A
	+
	i\,\Im A.
	\label{eq:exact-tee-Afull-complex}
\end{equation}
In the numerical figures, the plotted areas are divided by the common
compact-space factor ${\cal N}$. This plotting convention does not affect
the branch comparison or the fixed-boundary-interval selection. The
analytic expressions above keep the full area.

The corresponding timelike entanglement entropy is
\begin{equation}
	S_{\rm tEE}^{\rm full}(r_0,\theta_0)
	=
	\frac{A_{\rm tEE}^{\rm full}(r_0,\theta_0)}{4G_{10}}.
	\label{eq:exact-tee-entropy}
\end{equation}

The exact geometry changes the area as a function of the boundary interval
in two ways. First, the non-monotonic time map allows more than one radial
branch at the same $T$. Second, the ten-dimensional lift restores the full
internal angular dependence, so both lifted branches can carry real and
imaginary contributions. The resulting complex area can therefore become
folded as a function of $T$, even before the fixed-boundary-interval
selection is imposed.

\begin{figure}[H]
	\centering
	\includegraphics[width=0.48\textwidth]{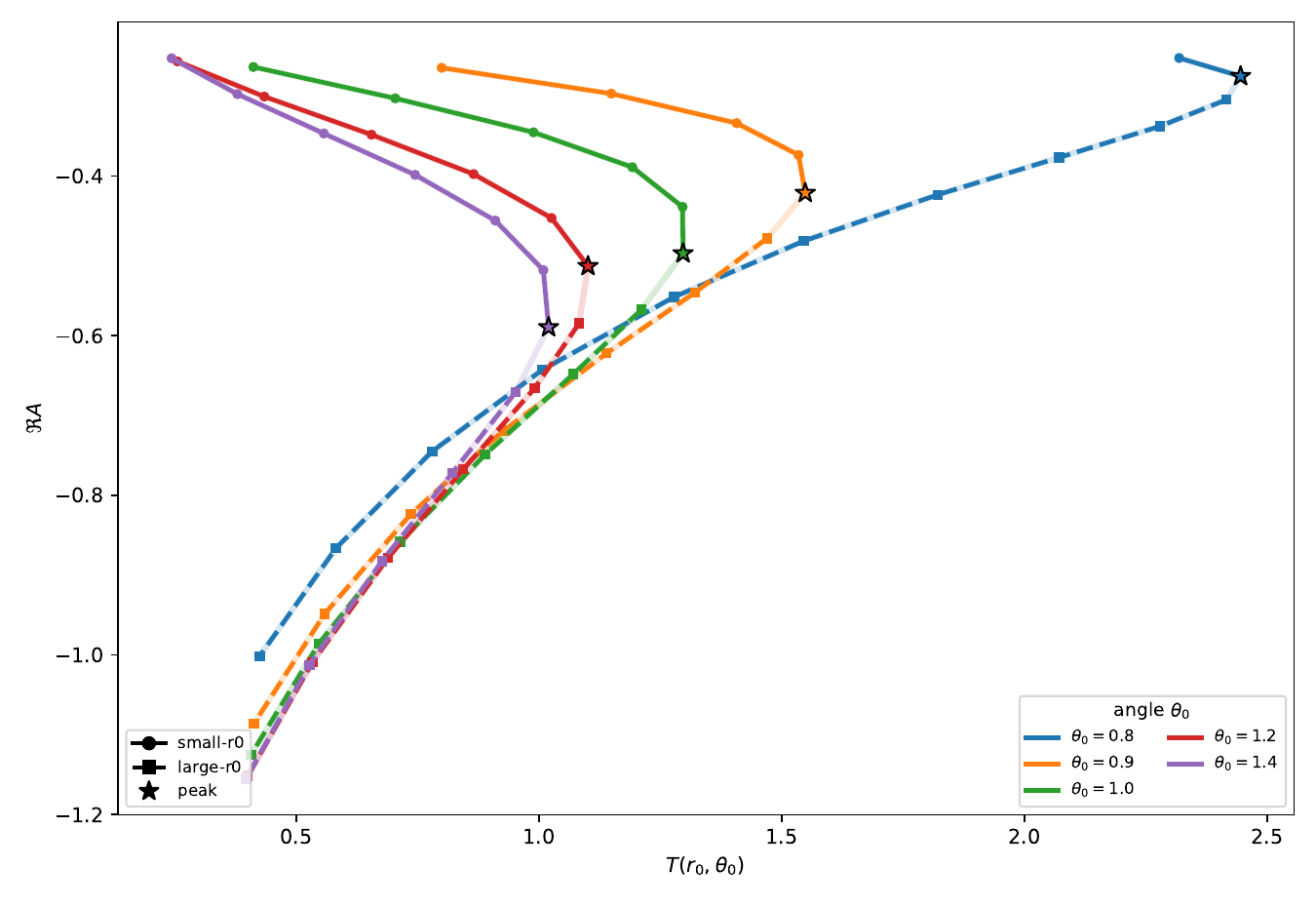}
	\hfill
	\includegraphics[width=0.48\textwidth]{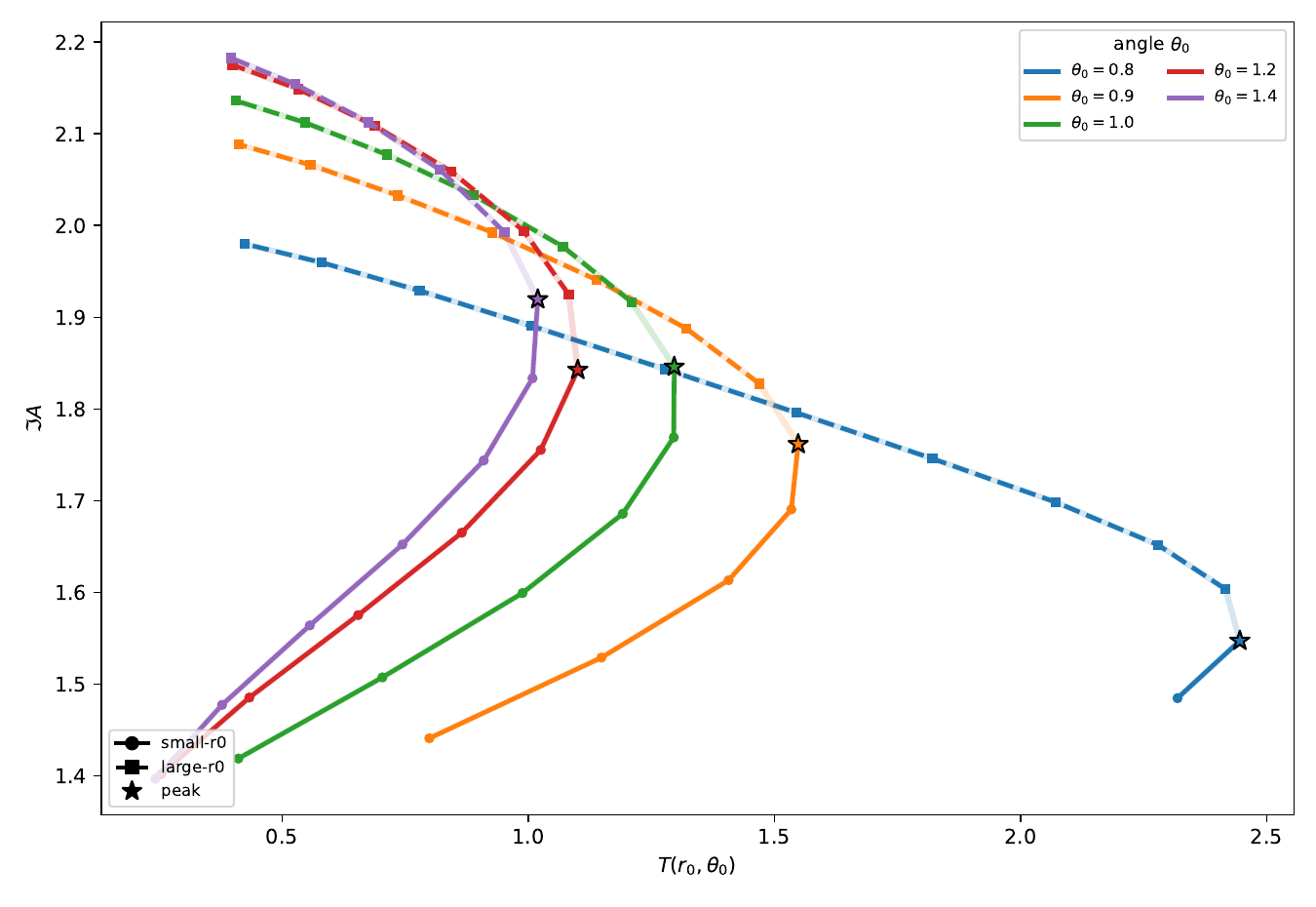}
	\caption{
		Exact lifted timelike-entanglement area before fixed-boundary-interval
		selection. The left panel shows the real part as a function of the
		boundary interval along representative angular families. The right panel
		shows the imaginary part along the same families. For a fixed angular
		label, the solid curve denotes the smaller-$r_0$ branch and the dashed
		curve denotes the larger-$r_0$ branch. Different branches can give
		different complex areas for the same boundary interval.
	}
	\label{fig:tee-exact-folded-area}
\end{figure}

Figure~\ref{fig:tee-exact-folded-area} completes the exact-area construction
before selection. The real part measures the UV-renormalized lifted area.
The imaginary part is controlled by the signs of
$\Delta_{\rm tim}$ and $\Delta_{\rm sp}$ after the internal angular
dependence is restored. The folding of the area curves follows from the
exact localized geometry. A fixed-boundary-interval comparison is needed
before choosing the relevant surface.

The final step is the fixed-boundary-interval selection. For each
$T_{\rm target}$, the branch family is first restricted to solutions of
\begin{equation}
T(r_0,\theta_0)=T_{\rm target}.
	\label{eq:tee-fixed-boundary-equation}
\end{equation}
Only these branches describe the same boundary interval. The selected
timelike-entanglement surface is then obtained by minimizing the real part
of $A_{\rm tEE}^{\rm full}$ within this fixed-boundary-interval family.

\subsection{Fixed-boundary-interval minimisation for TEE}
\label{sec:tee-fixed-boundary-interval}

We now select the timelike-entanglement surface at fixed boundary interval.
In the exact black-pole geometry, one should not minimize freely over
$r_0$ and $\theta_0$. The pair $(r_0,\theta_0)$ first determines the
boundary interval through
eq.~\eqref{eq:exact-tee-boundary-interval}. Therefore two surfaces can be
compared only when they correspond to the same value of $T$.

This point is important because the exact time map can be non-monotonic. At
fixed angular label $\theta_0$, the equation
$T(r_0,\theta_0)=T_{\rm target}$ can have more than one radial solution.
These solutions are different radial branches of the same boundary problem.
The area comparison must therefore be performed only after all admissible
roots of the time equation have been found.

For each angular label, let $r_{0,\rm peak}(\theta_0)$ denote the point where
the temporal family reaches its maximum:
\begin{equation}
	\left.
	\frac{\partial T(r_0,\theta_0)}{\partial r_0}
	\right|_{r_0=r_{0,\rm peak}}
	=
	0.
	\label{eq:tee-r0-peak}
\end{equation}
The two sides of this point define the two radial branches,
\begin{equation}
	r_0<r_{0,\rm peak}(\theta_0)
	\quad
	\hbox{smaller-}r_0\hbox{ branch},
	\qquad
	r_0>r_{0,\rm peak}(\theta_0)
	\quad
	\hbox{larger-}r_0\hbox{ branch}.
	\label{eq:tee-small-large-branches}
\end{equation}
These names refer only to the two sides of $r_{0,\rm peak}(\theta_0)$. A
point on the larger-$r_0$ branch can still move inward as
$T_{\rm target}$ is increased.

For a prescribed target interval, the fixed-boundary-interval condition is
\begin{equation}
	T(r_0,\theta_0)
	=
	T_{\rm target}.
	\label{eq:tee-fixed-boundary-equation1}
\end{equation}
An angular label contributes only if its temporal family can reach this
target interval. Using $T_{\rm max}(\theta_0)$ from
eq.~\eqref{eq:exact-tee-Tmax}, the admissible angular domain is
\begin{equation}
	{\cal D}(T_{\rm target})
	=
	\left\{
	\theta_0\in\left[0,\frac{\pi}{2}\right]\;:\;
	T_{\rm max}(\theta_0)\geq T_{\rm target}
	\right\}.
	\label{eq:tee-admissible-angular-domain}
\end{equation}
For every $\theta_0\in{\cal D}(T_{\rm target})$, we solve
eq.~\eqref{eq:tee-fixed-boundary-equation1}. If two roots are present, both
are kept:
\begin{equation}
	r_0
	=
	r_0^{(b)}(T_{\rm target},\theta_0),
	\qquad
	b\in\{\mathrm{smaller},\mathrm{larger}\}.
	\label{eq:tee-fixed-boundary-roots}
\end{equation}
On these fixed-boundary-interval branches we evaluate
\begin{equation}
	A_b(T_{\rm target},\theta_0)
	=
	A_{\rm tEE}^{\rm full}
	\left(
	r_0^{(b)}(T_{\rm target},\theta_0),
	\theta_0
	\right).
	\label{eq:tee-fixed-boundary-branch-area}
\end{equation}
The selected surface is obtained by minimizing the real part of the lifted
area among the admissible branches:
\begin{equation}
	(\theta_0^\ast,b^\ast)
	=
	\underset{
		\theta_0\in{\cal D}(T_{\rm target}),\;b
	}{\operatorname{arg\,min}}\,
	\operatorname{Re}
	A_b(T_{\rm target},\theta_0).
	\label{eq:tee-fixed-boundary-minimum}
\end{equation}
The selected turning point is
\begin{equation}
	r_0^\ast(T_{\rm target})
	=
	r_0^{(b^\ast)}
	\left(
	T_{\rm target},\theta_0^\ast
	\right).
	\label{eq:tee-selected-r0}
\end{equation}
The imaginary part is not minimized independently. It is evaluated on the
same selected surface:
\begin{equation}
	A_{\rm selected}(T_{\rm target})
	=
	A_{\rm tEE}^{\rm full}
	\left(
	r_0^\ast(T_{\rm target}),
	\theta_0^\ast(T_{\rm target})
	\right).
	\label{eq:tee-selected-area}
\end{equation}
For the plots we divide out the compact-space factor and define
\begin{equation}
	A_s(T_{\rm target})
	\equiv
	\frac{A_{\rm selected}(T_{\rm target})}{\mathcal N}.
	\label{eq:tee-selected-normalized-area-def}
\end{equation}
Thus
\begin{equation}
	A_s(T_{\rm target})
	=
	\Re A_s(T_{\rm target})
	+
	i\,\Im A_s(T_{\rm target}).
	\label{eq:tee-selected-area-complex}
\end{equation}
The corresponding selected timelike entanglement entropy is
\begin{equation}
	S_{\rm tEE}^{\rm selected}(T_{\rm target})
	=
	\frac{A_{\rm selected}(T_{\rm target})}{4G_{10}}.
	\label{eq:tee-selected-entropy}
\end{equation}

This is the fixed-boundary-interval form of the minimal-area rule. The real
part of the renormalized lifted area is minimized only after the boundary
interval has been fixed. For the numerical plots in this subsection we use \eqref{eq:exact-tee-parameters}.

\begin{figure}[H]
	\centering
	
	\begin{subfigure}[t]{0.25\textwidth}
		\centering
		\includegraphics[width=\linewidth]{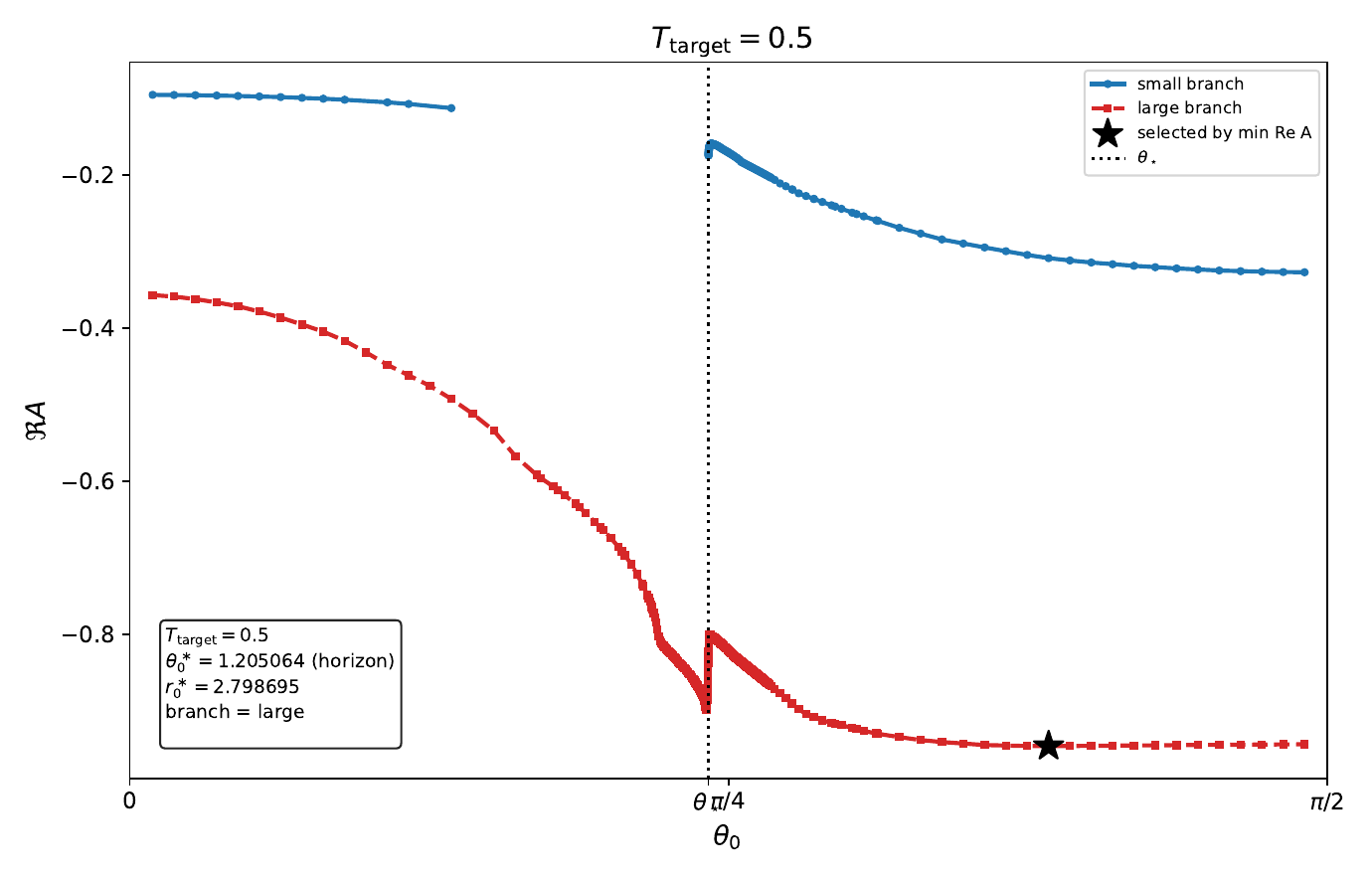}
	\end{subfigure}
	\hfill
	\begin{subfigure}[t]{0.25\textwidth}
		\centering
		\includegraphics[width=\linewidth]{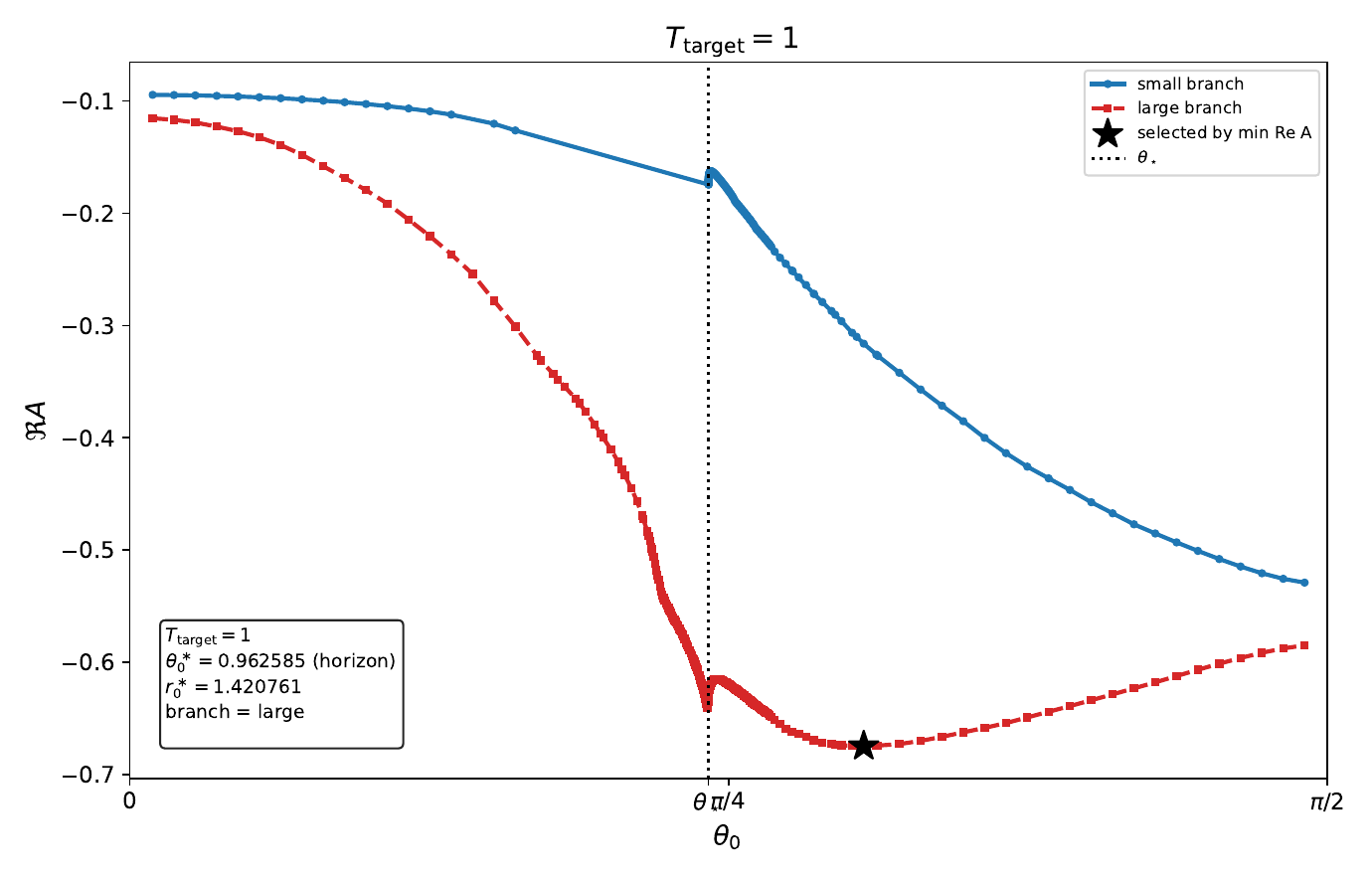}
	\end{subfigure}
	\hfill
	\begin{subfigure}[t]{0.25\textwidth}
		\centering
		\includegraphics[width=\linewidth]{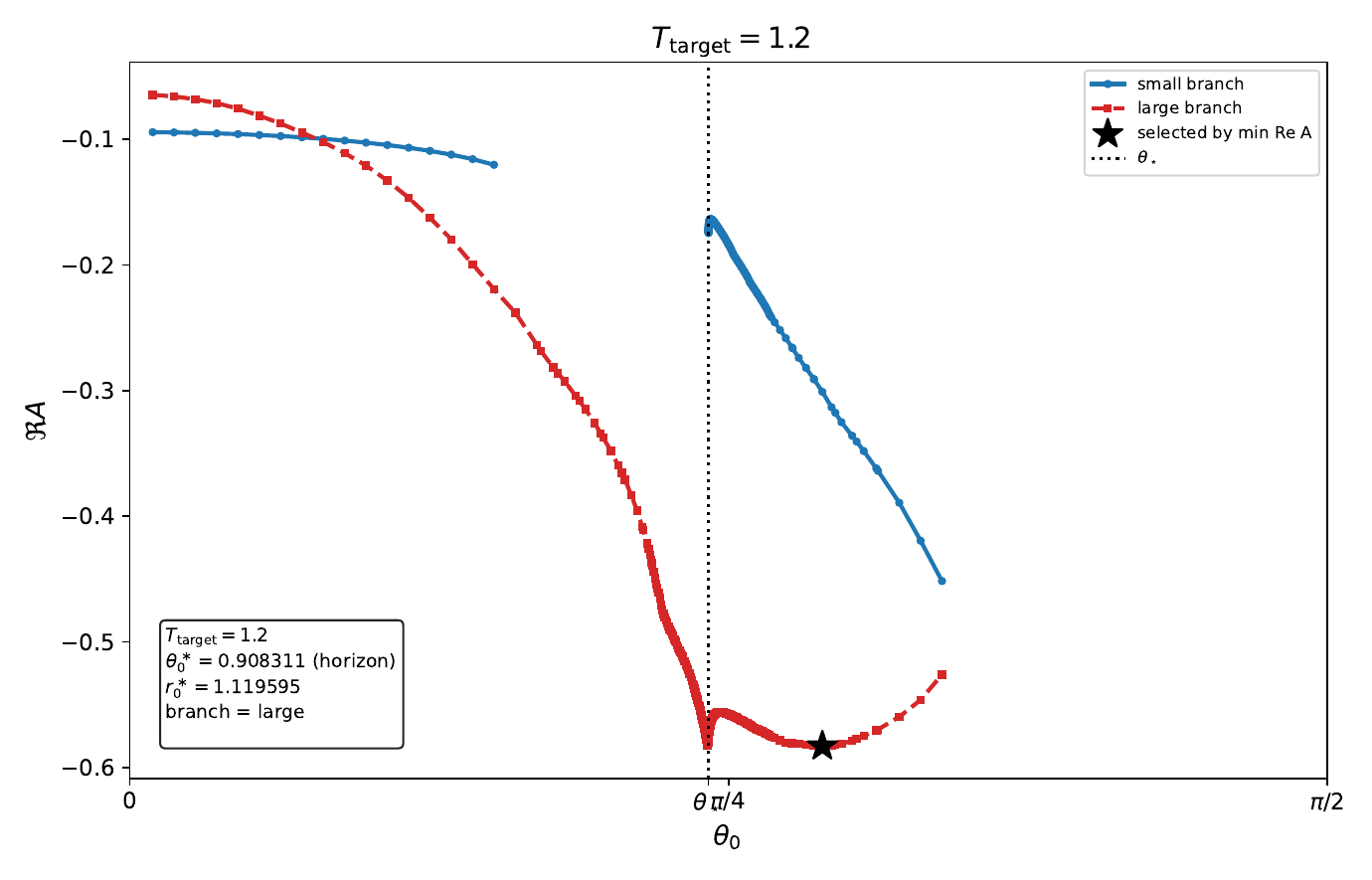}
	\end{subfigure}
	
	\medskip
	
	\begin{subfigure}[t]{0.25\textwidth}
		\centering
		\includegraphics[width=\linewidth]{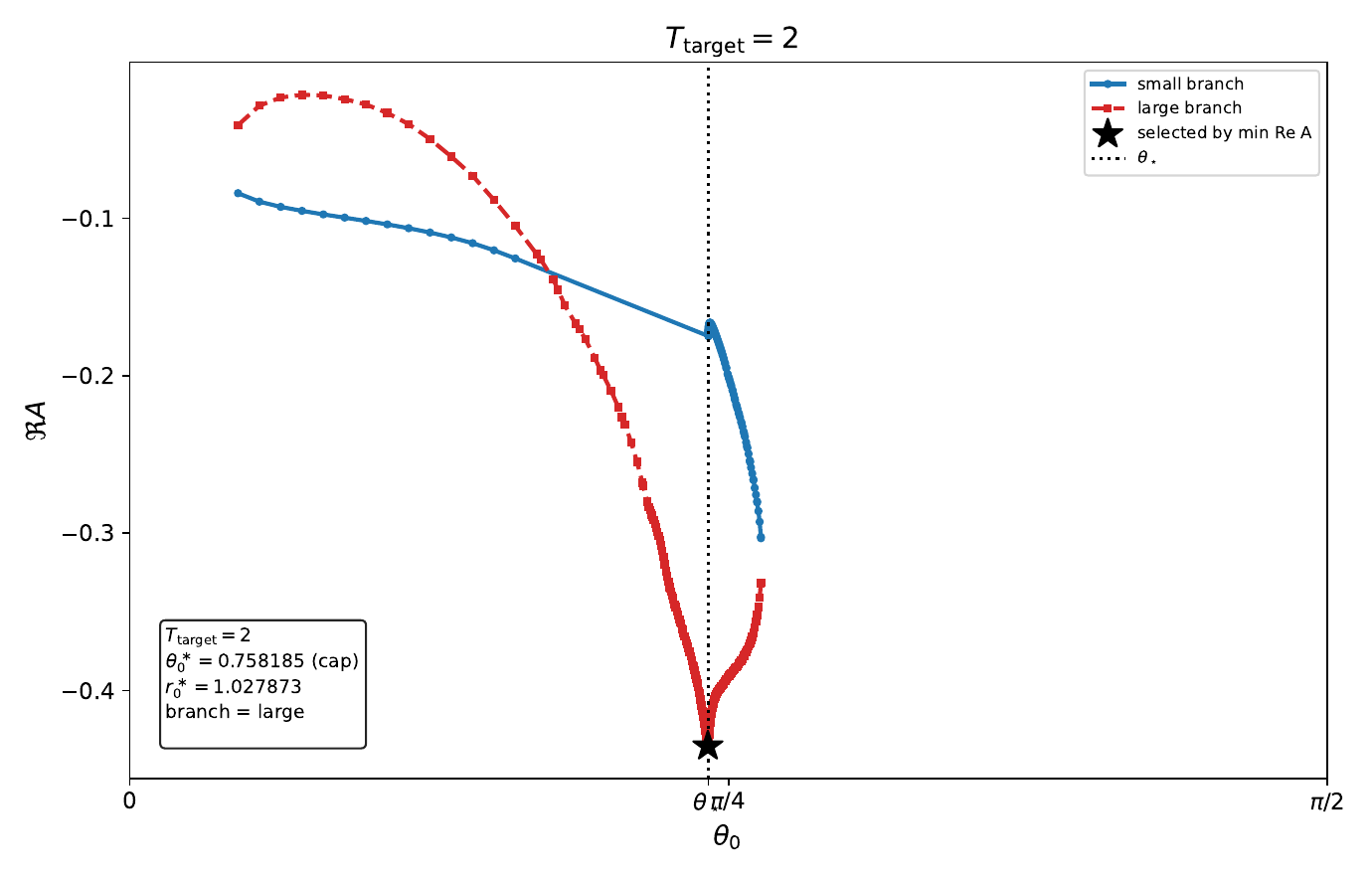}
	\end{subfigure}
	\hfill
	\begin{subfigure}[t]{0.25\textwidth}
		\centering
		\includegraphics[width=\linewidth]{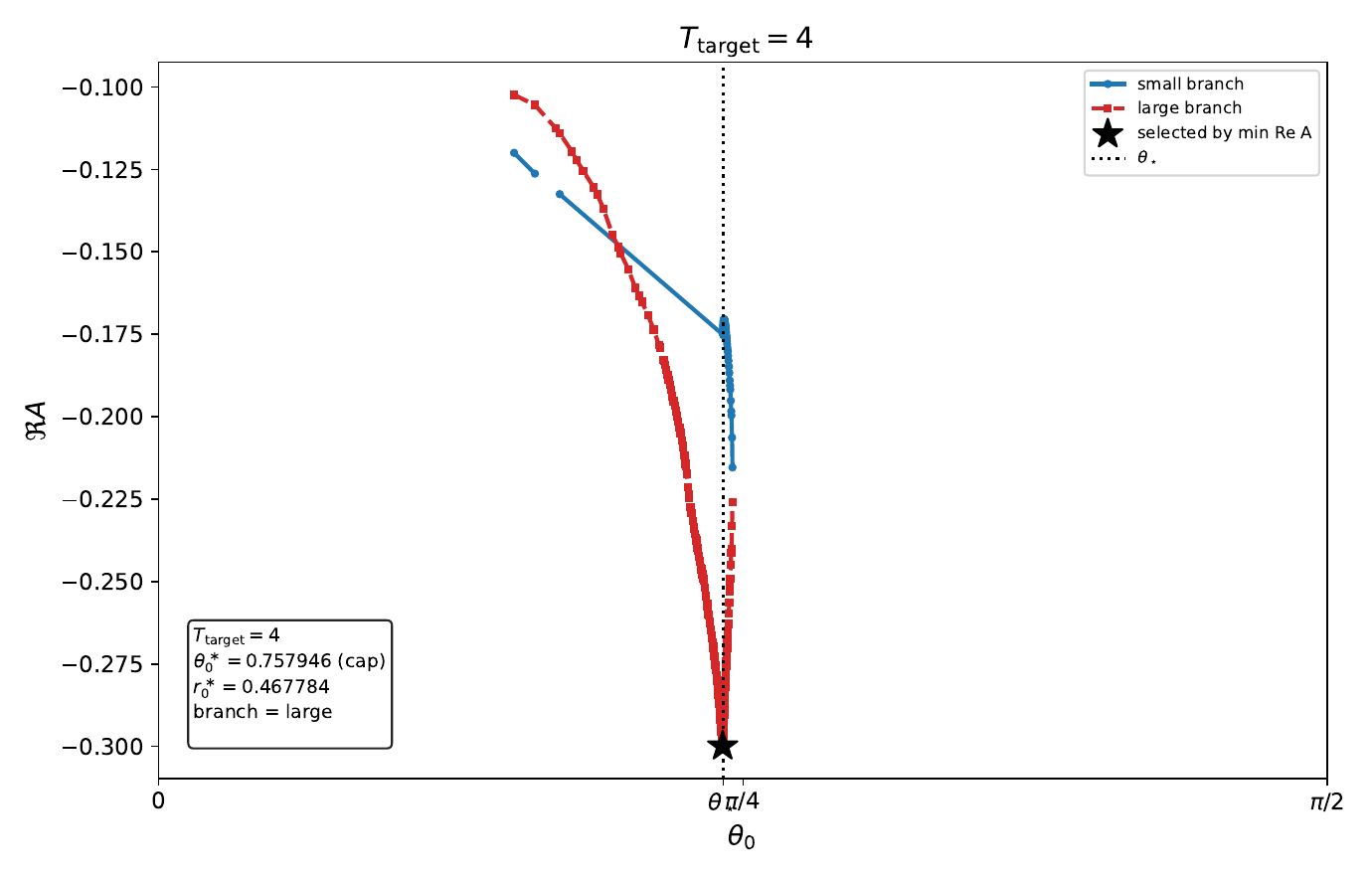}
	\end{subfigure}
	\hfill
	\begin{subfigure}[t]{0.25\textwidth}
		\centering
		\includegraphics[width=\linewidth]{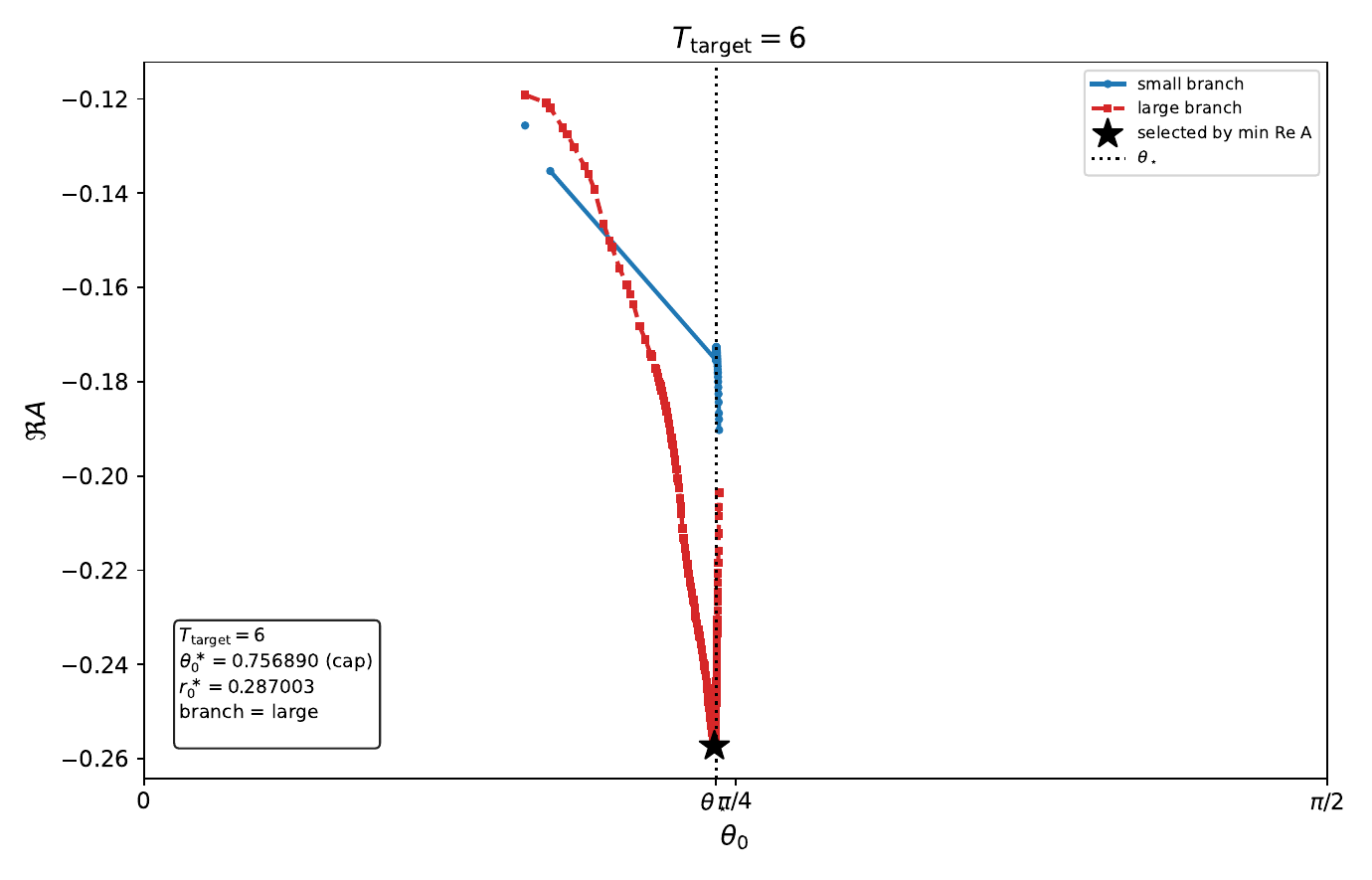}
	\end{subfigure}
	
	\medskip
	
	\begin{subfigure}[t]{0.25\textwidth}
		\centering
		\includegraphics[width=\linewidth]{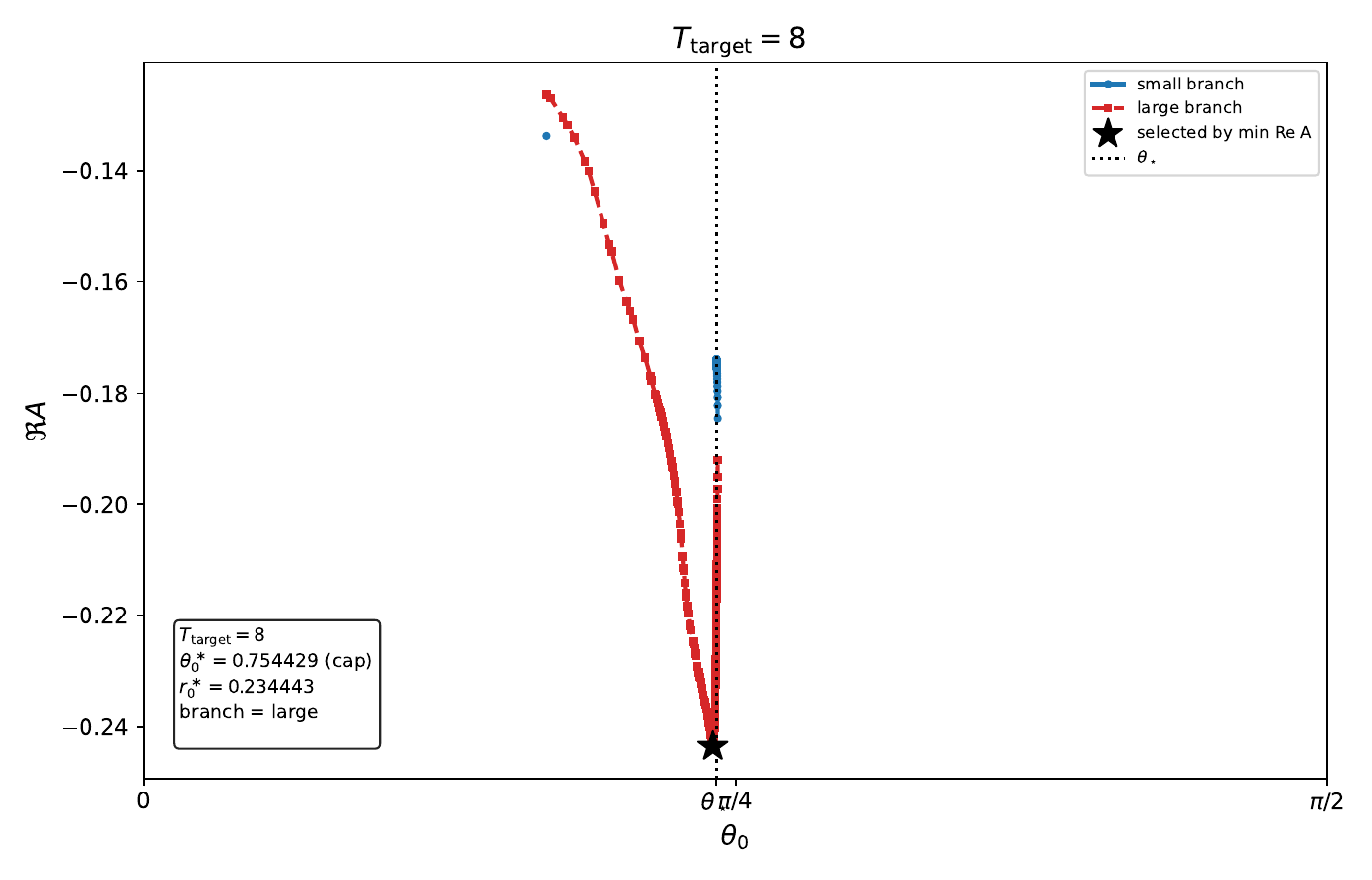}
	\end{subfigure}
	\hfill
	\begin{subfigure}[t]{0.25\textwidth}
		\centering
		\includegraphics[width=\linewidth]{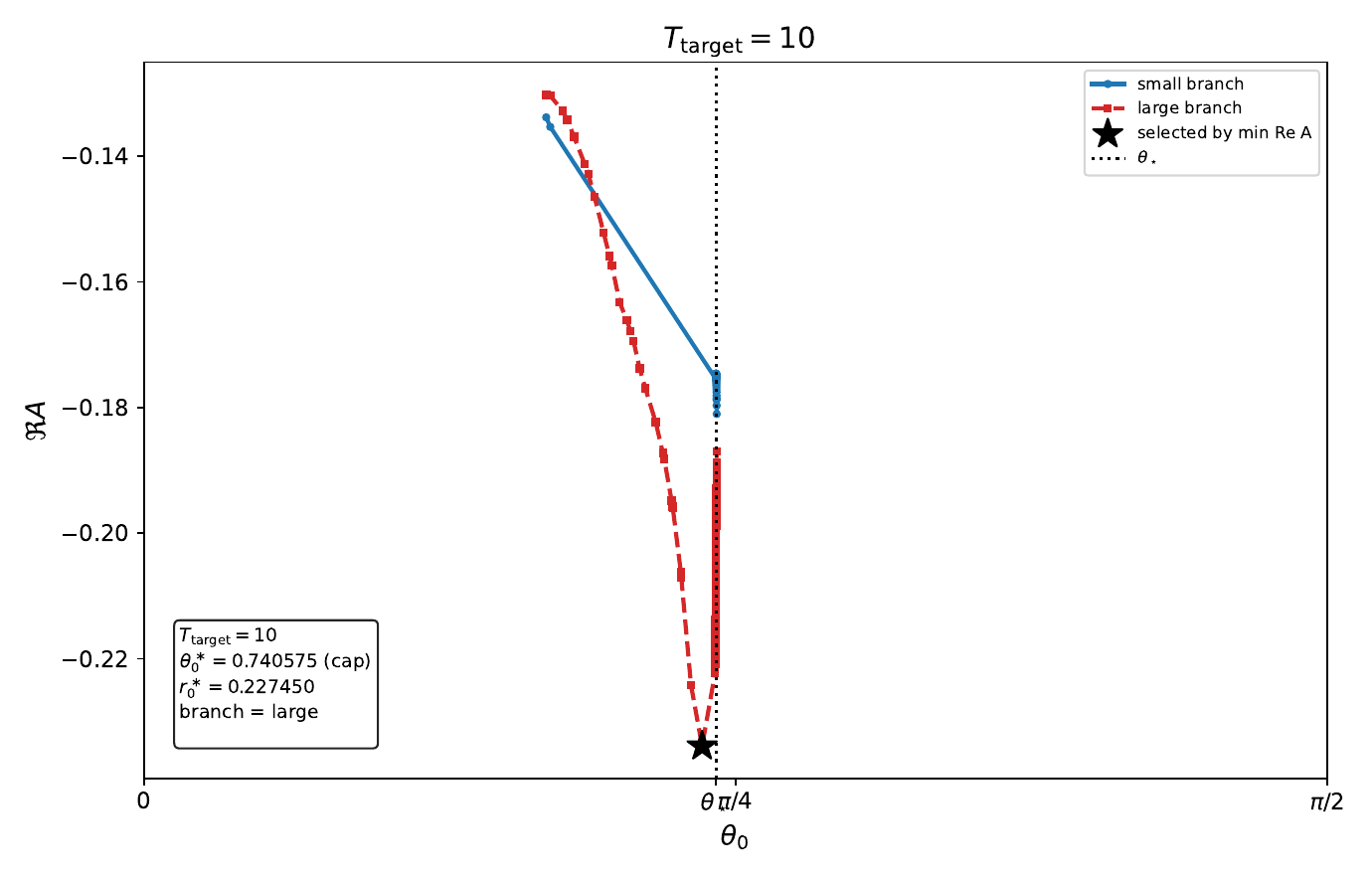}
	\end{subfigure}
	\hfill
	\begin{subfigure}[t]{0.25\textwidth}
		\centering
		\includegraphics[width=\linewidth]{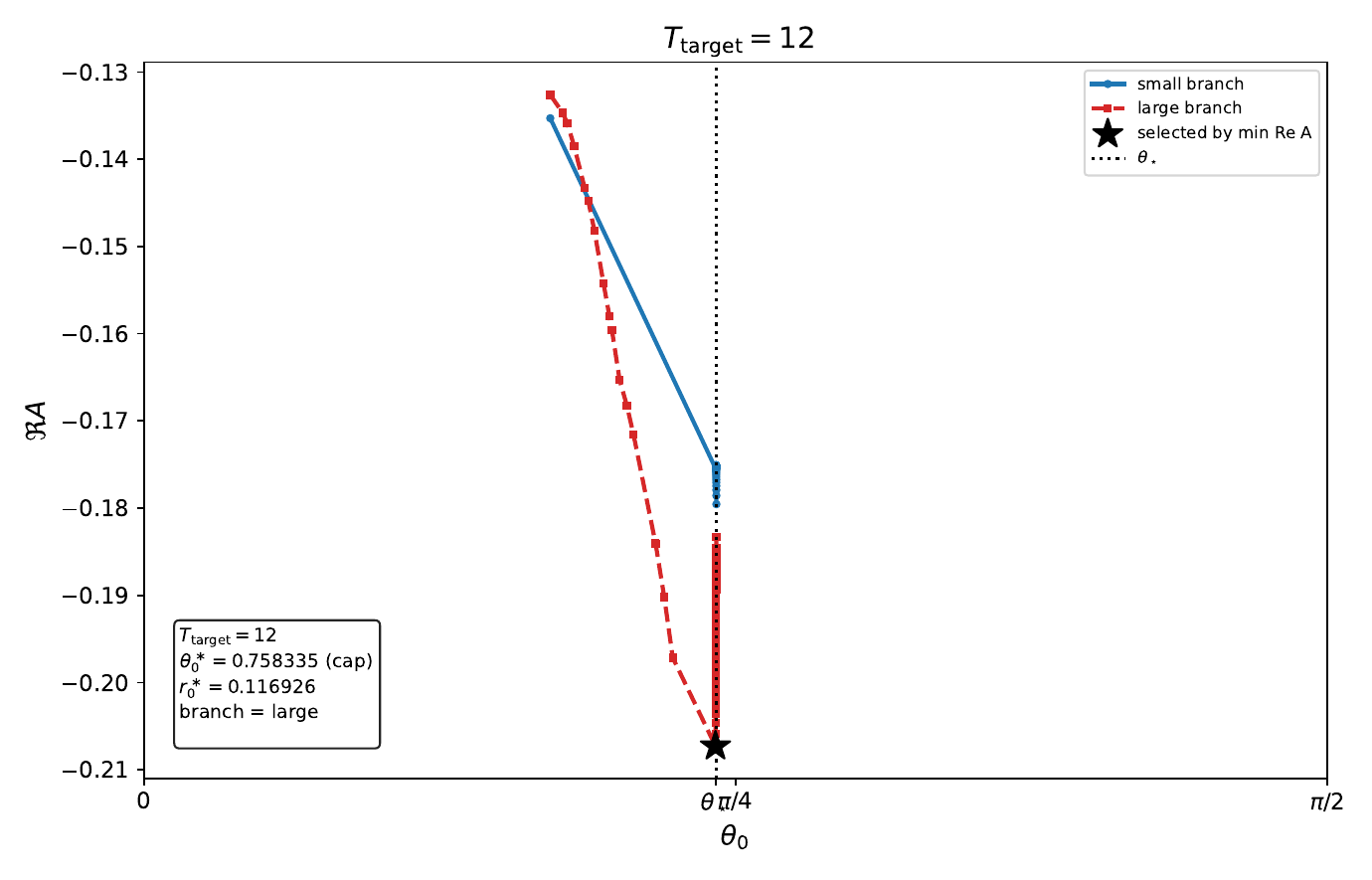}
	\end{subfigure}
	
	\caption{
		Fixed-boundary-interval minimisation of the real part of the exact lifted
		area. Each point satisfies $T(r_0,\theta_0)=T_{\rm target}$. The blue
		curves denote the smaller-$r_0$ branch and the red dashed curves denote
		the larger-$r_0$ branch. The black star marks the selected surface.
		Blank angular intervals indicate that no root of the time equation exists.
	}
	\label{fig:tee-fixed-boundary-minimisation}
\end{figure}

Figure~\ref{fig:tee-fixed-boundary-minimisation} shows that increasing
$T_{\rm target}$ restricts the admissible angular domain. At small target
intervals, many angular labels can solve the time equation. At larger
target intervals, the condition
$T_{\rm max}(\theta_0)\geq T_{\rm target}$ keeps only labels close to the
cap/horizon transition region. Within this restricted domain, the selected
surface is the one with the smallest real part of the lifted area. In the
range displayed, the selected branch is the larger-$r_0$ branch, although
the selected value $r_0^\ast$ decreases as $T_{\rm target}$ grows.

The movement toward $\theta_\star$ has a local geometric origin. As shown
in appendix~\ref{app:transition-region}, near the transition angle there is
a crossover scale
\begin{equation}
	r_c(\theta_0)
	\simeq
	\ell\,|\theta_0-\theta_\star|.
	\label{eq:tee-transition-crossover}
\end{equation}
This scale separates the deep cap-side or horizon-side region from the
transition radial interval. In that interval,
\begin{equation}
	K_{\rm tim}\simeq\frac{C_{\rm tim}}{r},
	\qquad
	K_{\rm sp}\simeq\frac{C_{\rm sp}}{r},
	\qquad
	C_{\rm tim}>C_{\rm sp}.
	\label{eq:tee-transition-kernels}
\end{equation}
Therefore $I_{\rm tim}-I_{\rm sp}$ receives a positive logarithmic
contribution. As $\theta_0$ approaches $\theta_\star$, $r_c(\theta_0)$
decreases and the transition interval becomes longer. This is why larger
boundary intervals are supported near the cap/horizon transition region.

After the selection, the complex area is evaluated on the selected surface.
The selected data are
\begin{equation}
	\theta_0^\ast(T_{\rm target}),
	\qquad
	r_0^\ast(T_{\rm target}),
	\qquad
	b^\ast(T_{\rm target}).
	\label{eq:tee-selected-data}
\end{equation}
The real part is the quantity used in the minimisation. The imaginary part
is then evaluated on the same surface and carries independent information
about the lifted sign structure.

\begin{figure}[H]
	\centering
	\includegraphics[width=0.48\textwidth]{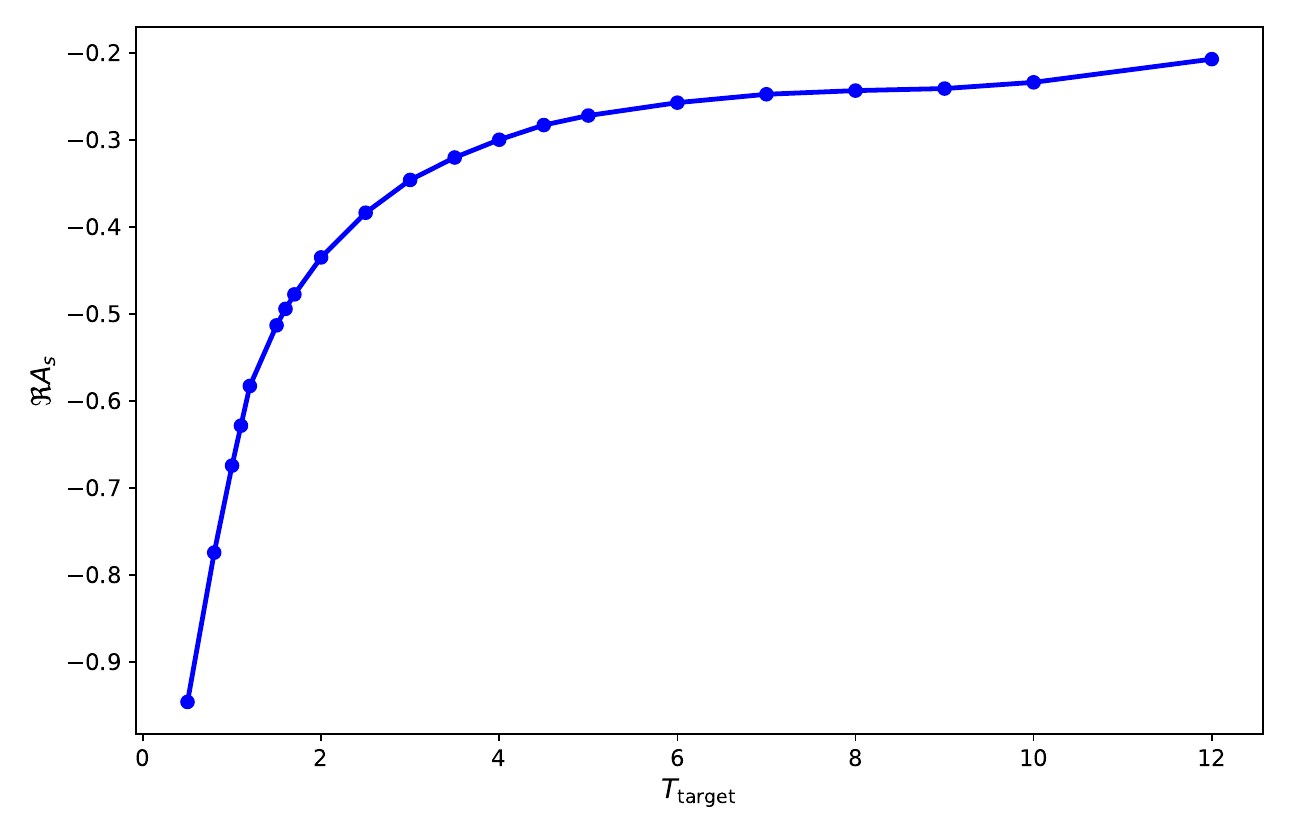}
	\includegraphics[width=0.48\textwidth]{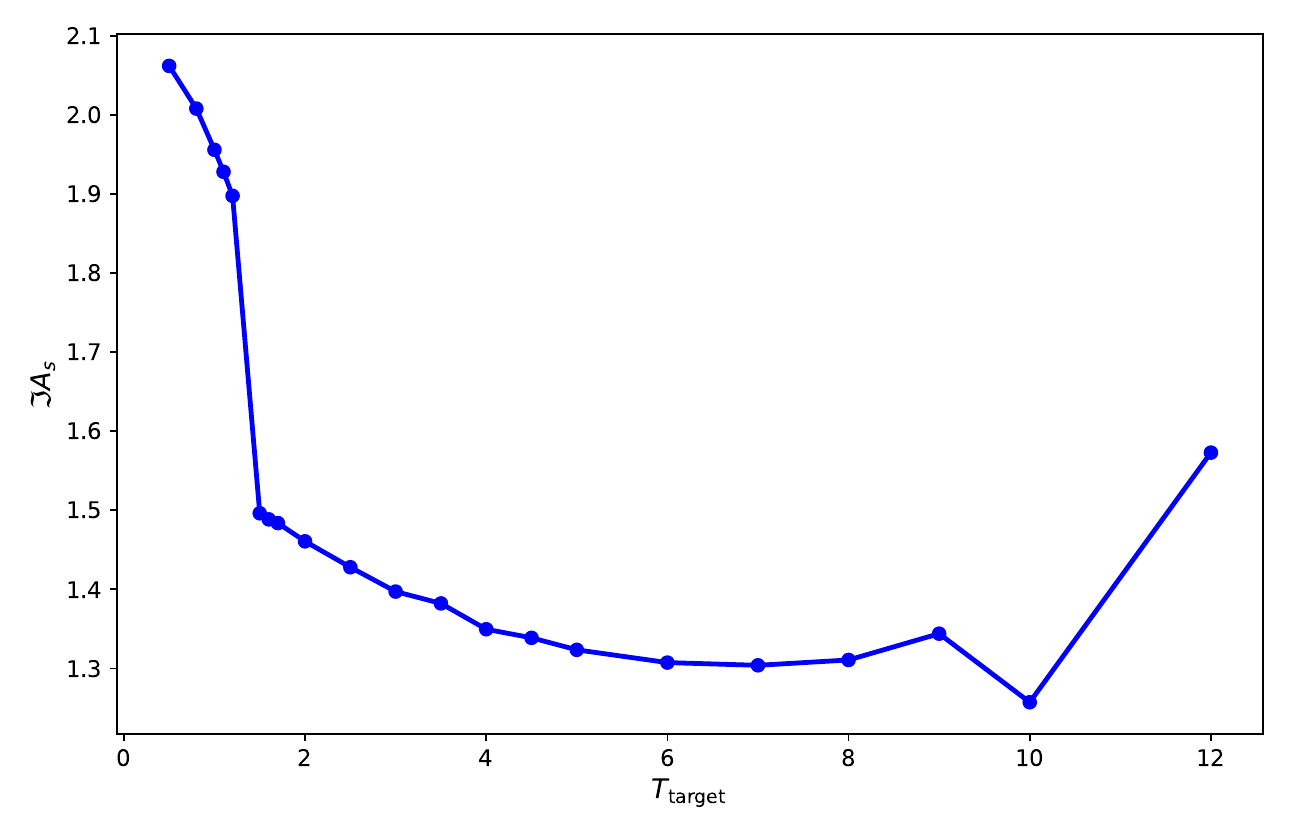}
	\caption{
		Selected real and imaginary parts of the normalized lifted area. The real
		part is minimized at fixed $T_{\rm target}$. The imaginary part is
		evaluated on the same selected surface.
	}
	\label{fig:tee-selected-real-imag}
\end{figure}

Figure~\ref{fig:tee-selected-real-imag} shows that the selected real part
changes slowly at larger target intervals. This behaviour reflects the
fixed-boundary-interval selection: once the admissible angular labels are
pushed close to the transition region, increasing $T_{\rm target}$ mainly
moves the selected surface within a narrow part of the exact geometry. The
imaginary part has a different origin. It is determined by the signs of the
lifted area functions
$\Delta_{\rm tim}$ and $\Delta_{\rm sp}$ in
eqs.~\eqref{eq:exact-tee-Delta-tim} and
\eqref{eq:exact-tee-Delta-sp}. Thus it can vary differently from the real
part.

The selected imaginary part receives contributions from both lifted
branches. Using $A_s=A_{\rm selected}/\mathcal N$, we write
\begin{equation}
	\Im A_s
	=
	\Im A_{{\rm tim},s}
	+
	\Im A_{{\rm sp},s}.
	\label{eq:tee-selected-imag-branch-decomposition}
\end{equation}
Here $\Im A_{{\rm tim},s}$ and $\Im A_{{\rm sp},s}$ are the contributions
from the lifted timelike and spacelike branches on the selected surface.

\begin{figure}[H]
	\centering
	\includegraphics[width=0.65\textwidth]{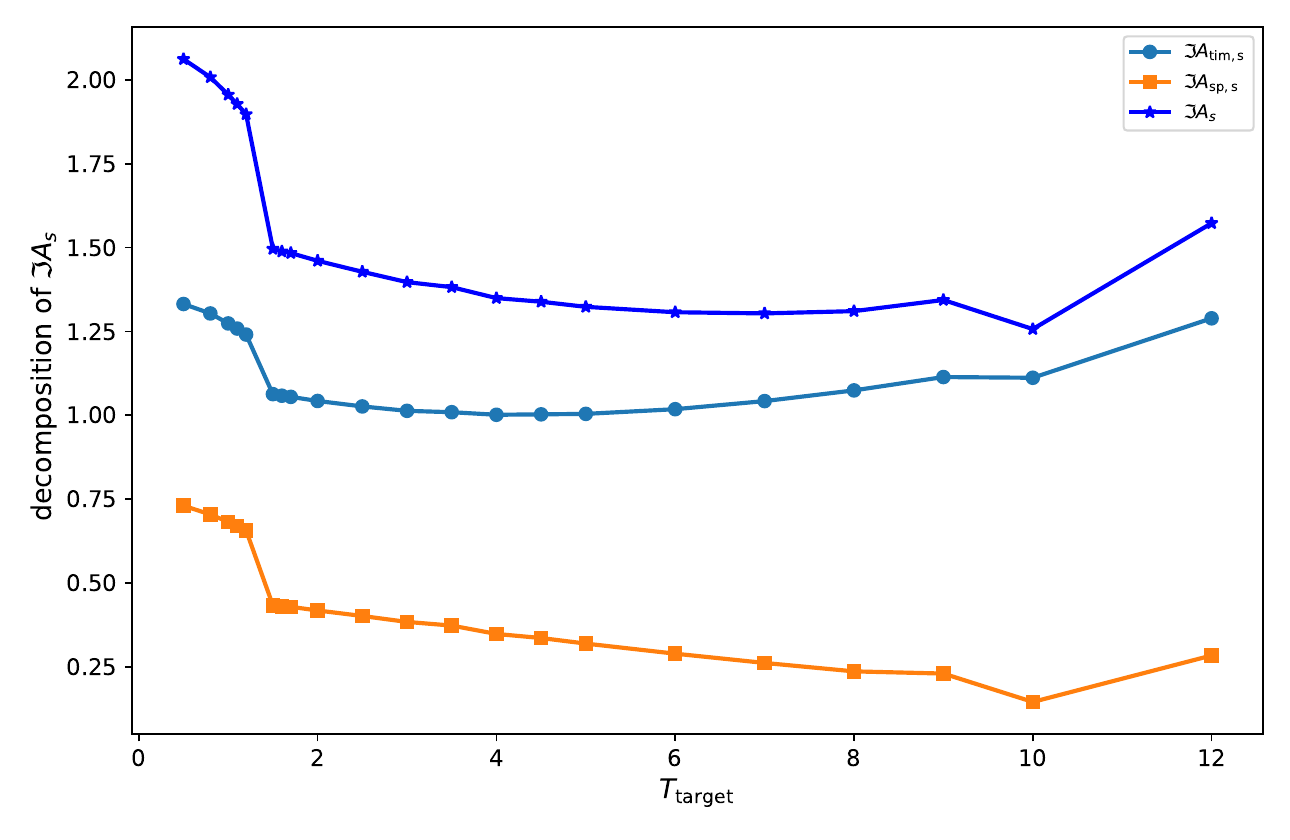}
	\caption{
		Branch decomposition of the selected imaginary part. Both lifted branches
		can contribute because the sign is determined after the full angular lift.
	}
	\label{fig:tee-selected-imag-decomposition}
\end{figure}

Figure~\ref{fig:tee-selected-imag-decomposition} shows that the timelike
branch gives the dominant imaginary contribution in the displayed range,
while the spacelike branch is smaller but non-zero. This is a
localized-geometry effect. At fixed $\theta_0$, the reduced spacelike
branch has a real density, but after the lift the same branch is evaluated
over the full internal angle $\theta$, where parts of the lifted area
function can become negative.

We next examine where the selected surface lies in angular and radial
directions.

\begin{figure}[H]
	\centering
	\includegraphics[width=0.48\textwidth]{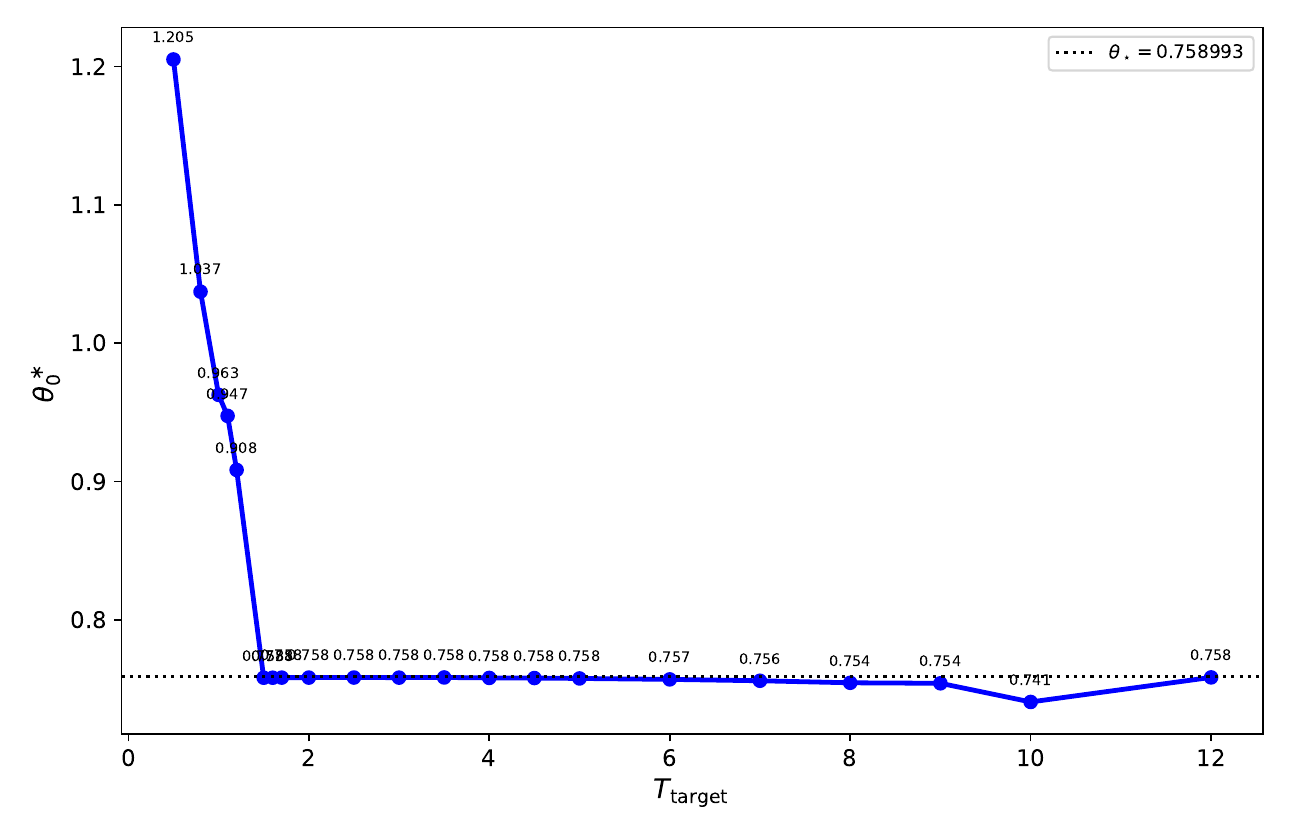}
	\hfill
	\includegraphics[width=0.48\textwidth]{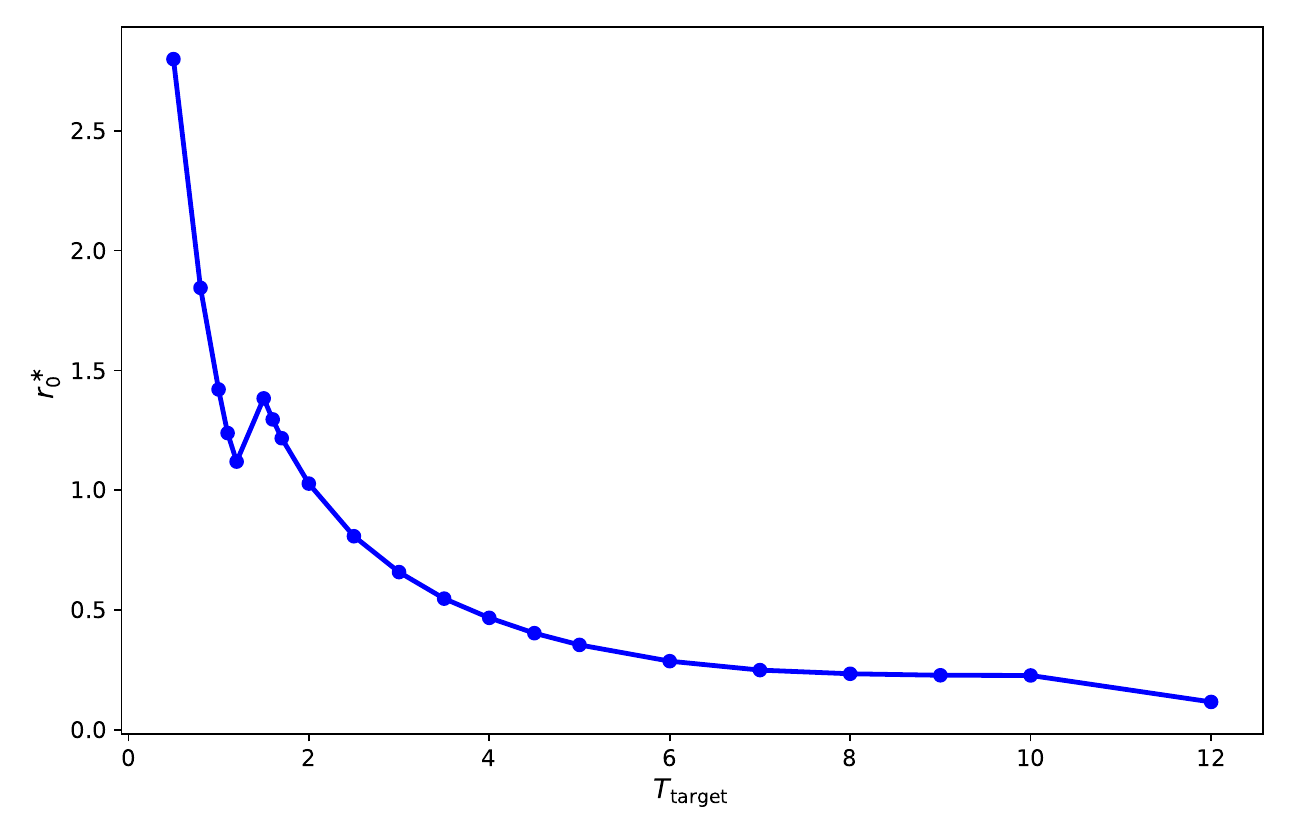}
	\caption{
		Selected angular label and selected turning point. The left panel
		compares $\theta_0^\ast(T_{\rm target})$ with $\theta_\star$. The right
		panel shows $r_0^\ast(T_{\rm target})$.
	}
	\label{fig:tee-selected-angle-r0}
\end{figure}

Figure~\ref{fig:tee-selected-angle-r0} shows two correlated effects. As the
target interval grows, the selected angular label moves toward
$\theta_\star$, while the selected turning point moves inward. The selected
surface remains on the larger-$r_0$ branch in the displayed range, because
it lies on the larger-$r_0$ side of $r_{0,\rm peak}(\theta_0)$. At the same
time, the numerical value of $r_0^\ast$ decreases, so the selected surface
becomes more sensitive to the exact localized geometry.

Finally, we reconstruct the effective $t-r$ profiles after the selection.
For each target interval, the selected values
$\theta_0^\ast$, $r_0^\ast$ and $b^\ast$ determine the branch profile:
\begin{equation}
	t_{\rm tim}^{+}(r)
	=
	\int_{r}^{r_0^\ast}
	d\tilde r\,
	K_{\rm tim}
	\left(
	\tilde r;\theta_0^\ast,r_0^\ast
	\right),
	\qquad
	t_{\rm tim}^{+}(r_0^\ast)=0,
	\label{eq:tee-selected-ttim-profile}
\end{equation}
and
\begin{equation}
	t_{\rm sp}^{+}(r)
	=
	\frac{T_{\rm target}}{2}
	+
	\int_{r}^{R_{\rm max}}
	d\tilde r\,
	K_{\rm sp}
	\left(
	\tilde r;\theta_0^\ast,r_0^\ast
	\right),
	\qquad
	t_{\rm sp}^{+}(R_{\rm max})
	=
	\frac{T_{\rm target}}{2}.
	\label{eq:tee-selected-tsp-profile}
\end{equation}
The lower branches are obtained by time reflection:
\begin{equation}
	t_{\rm tim}^{-}(r)=-t_{\rm tim}^{+}(r),
	\qquad
	t_{\rm sp}^{-}(r)=-t_{\rm sp}^{+}(r).
	\label{eq:tee-selected-reflection}
\end{equation}

\begin{figure}[H]
	\centering
	\begin{subfigure}[t]{0.25\textwidth}
		\centering
		\includegraphics[width=\linewidth]{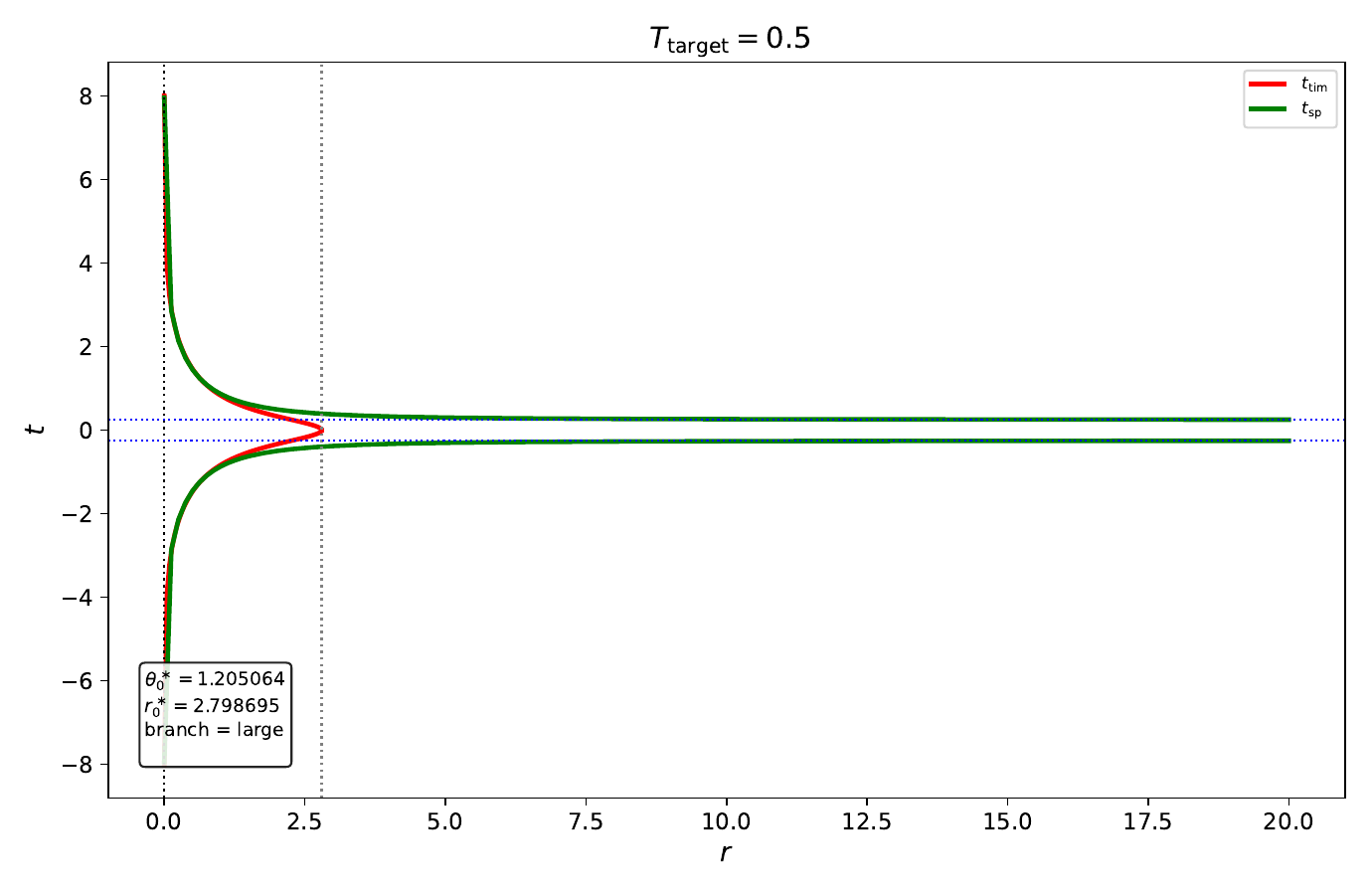}
	\end{subfigure}
	\hfill
	\begin{subfigure}[t]{0.25\textwidth}
		\centering
		\includegraphics[width=\linewidth]{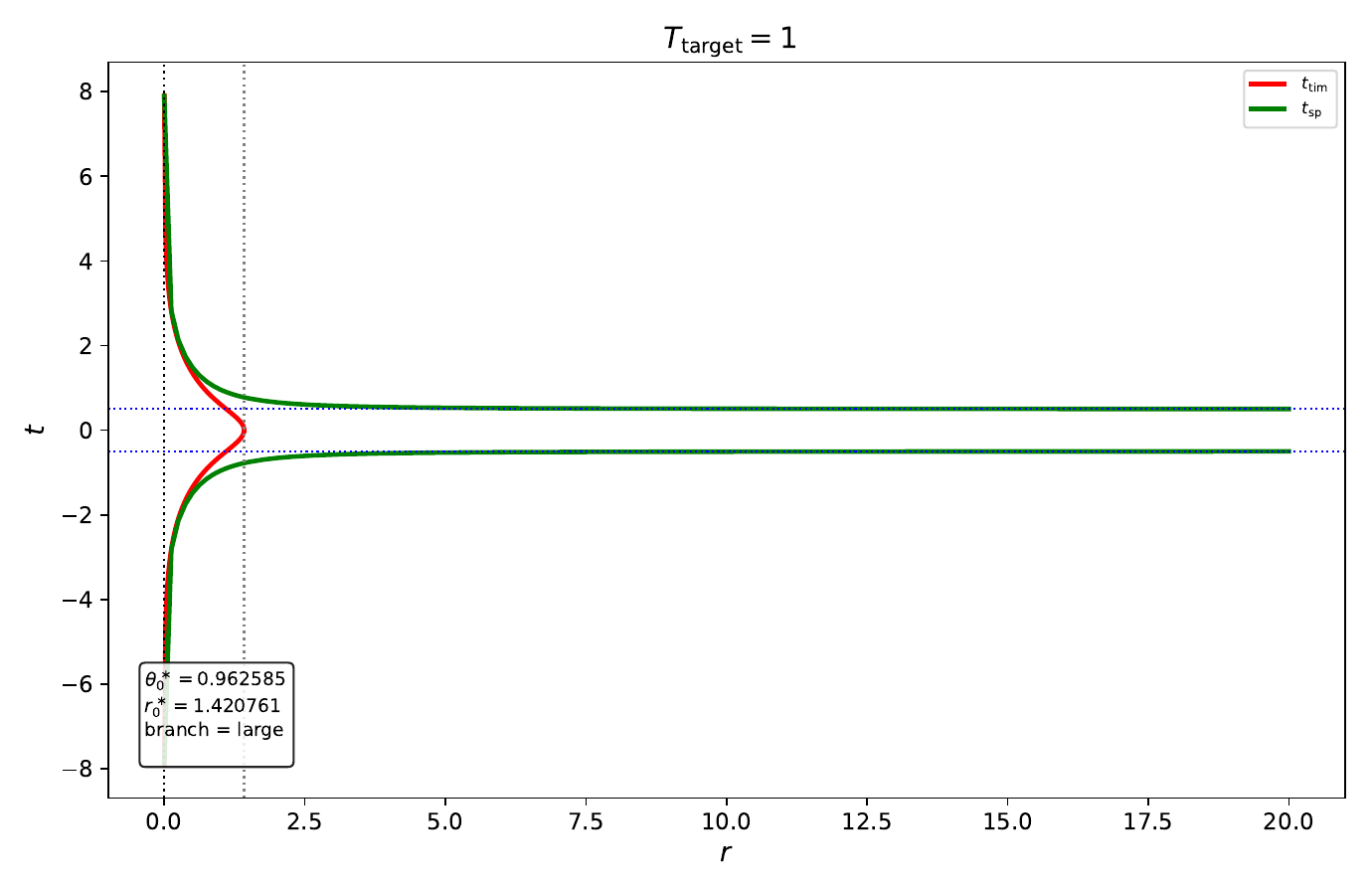}
	\end{subfigure}
	\hfill
	\begin{subfigure}[t]{0.25\textwidth}
		\centering
		\includegraphics[width=\linewidth]{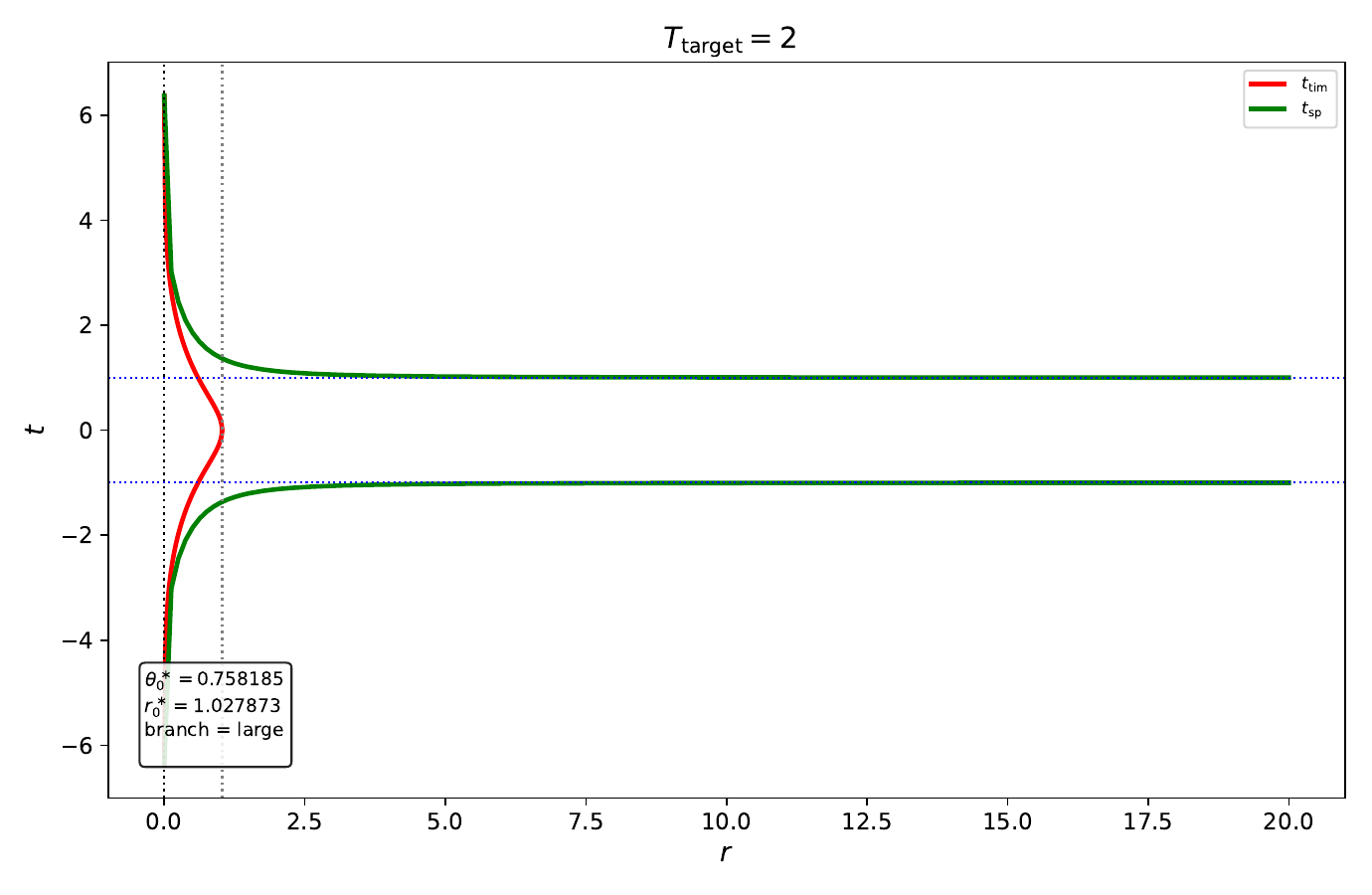}
	\end{subfigure}
	
	\medskip
	
	\begin{subfigure}[t]{0.25\textwidth}
		\centering
		\includegraphics[width=\linewidth]{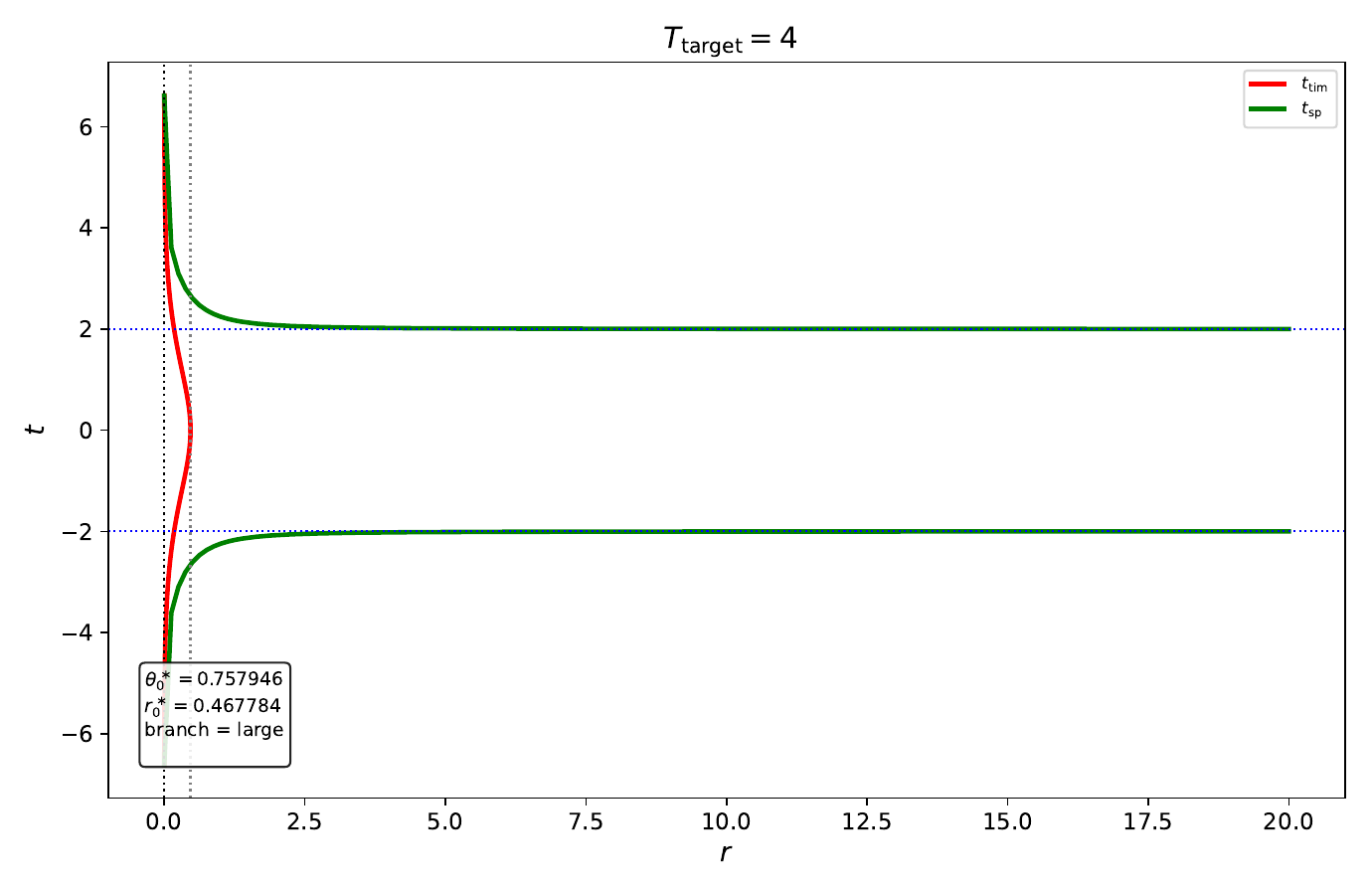}
	\end{subfigure}
	\hfill
	\begin{subfigure}[t]{0.25\textwidth}
		\centering
		\includegraphics[width=\linewidth]{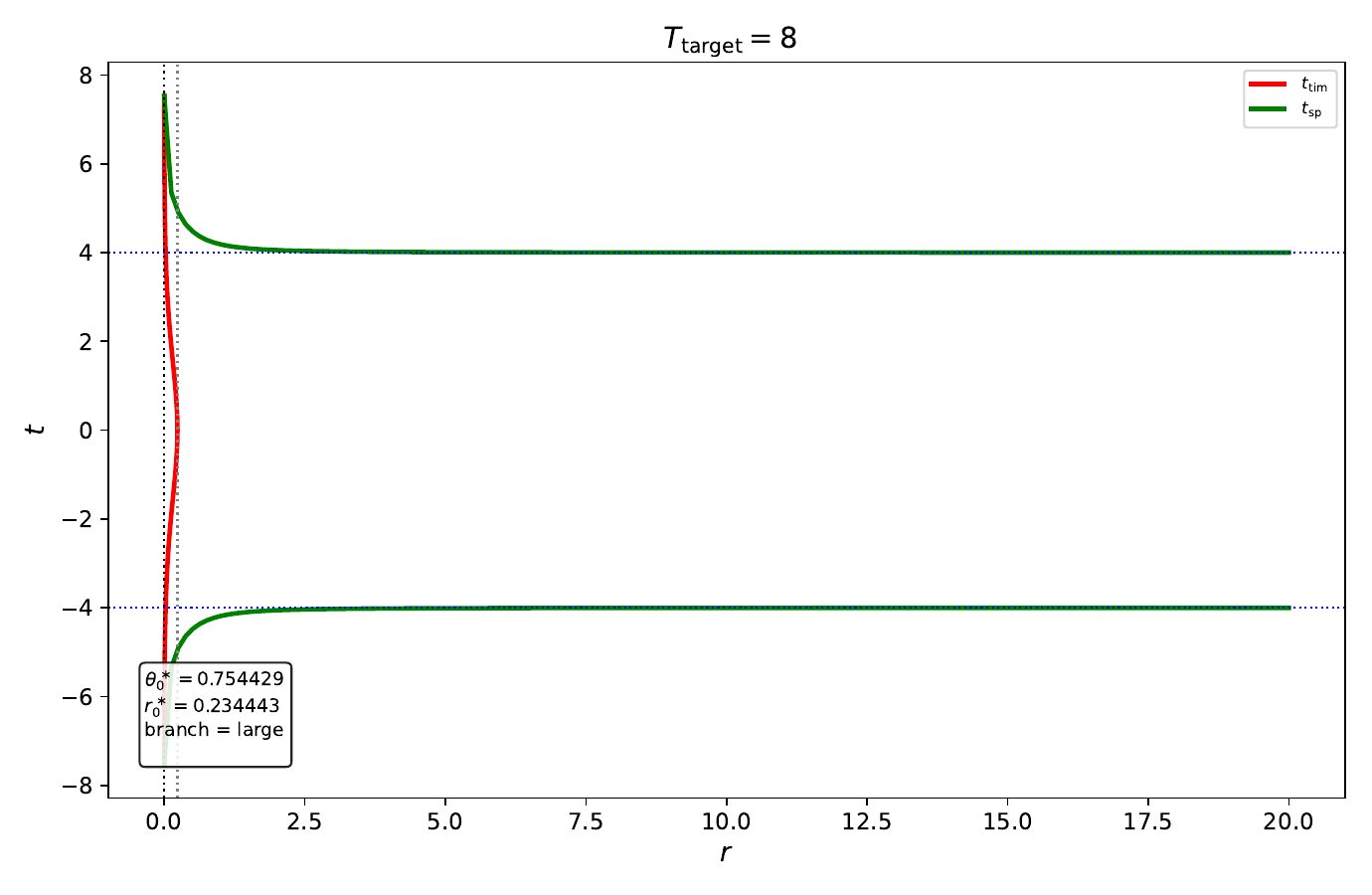}
	\end{subfigure}
	\hfill
	\begin{subfigure}[t]{0.25\textwidth}
		\centering
		\includegraphics[width=\linewidth]{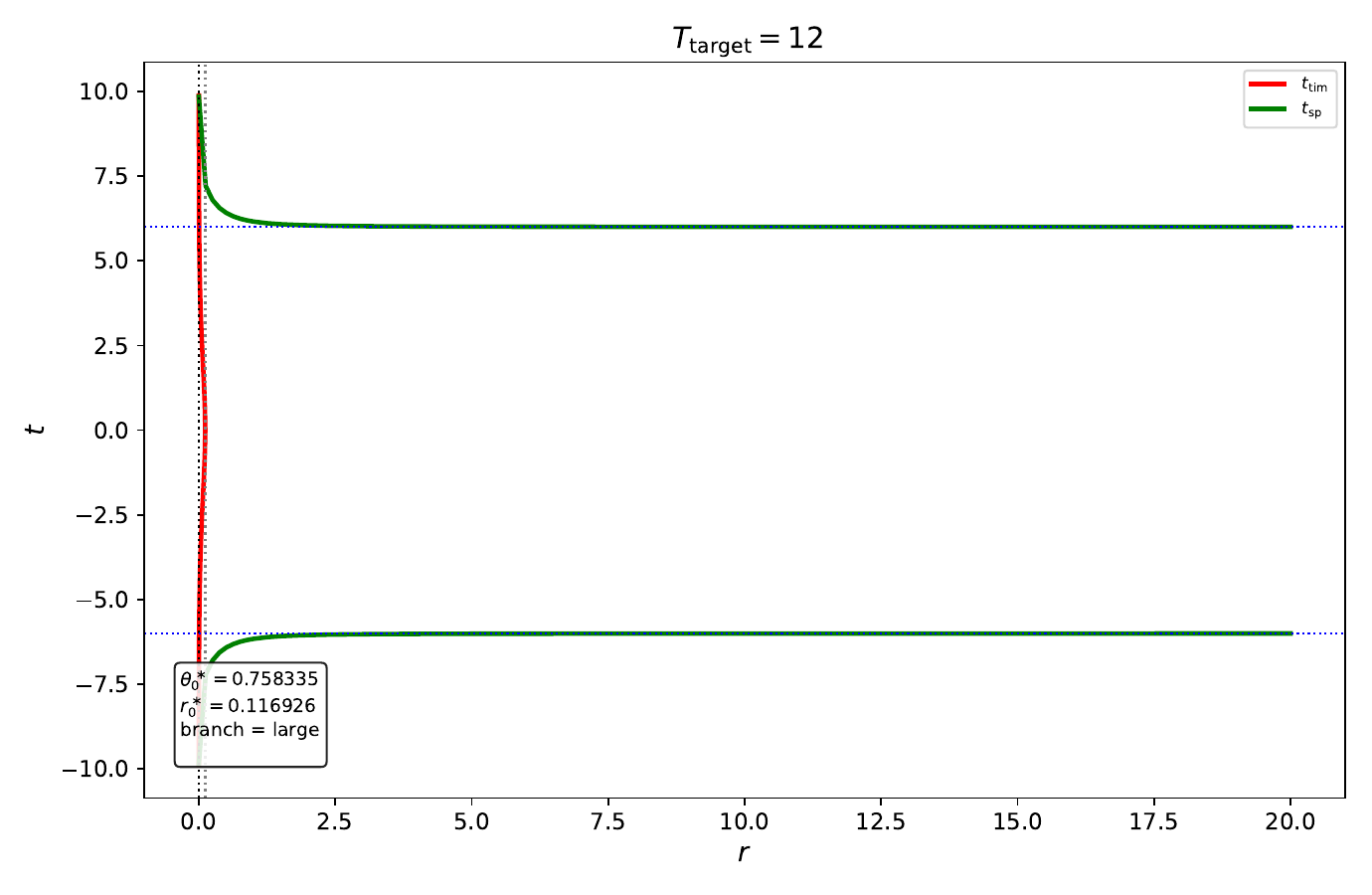}
	\end{subfigure}
	\caption{
		Post-selection $t-r$ profiles. Each panel is reconstructed using the
		selected values $\theta_0^\ast$, $r_0^\ast$ and $b^\ast$ at the indicated
		target interval. The timelike branch ends at the selected turning point,
		and the spacelike branch reaches the asymptotic boundary at
		$t=\pm T_{\rm target}/2$.
	}
	\label{fig:tee-selected-tr-profiles}
\end{figure}

Figure~\ref{fig:tee-selected-tr-profiles} shows the radial movement of the
selected surface. At larger target intervals, the selected turning point
moves inward and the profile becomes more sensitive to the exact functions
$K_y(r,\theta)$ and $G(r,\theta)$. This is the geometric reason why the
cap/horizon transition region controls the large-interval selected
surfaces.

The fixed-boundary-interval analysis gives the following picture. The
large-$r$ calculation describes the short-boundary-interval regime, where
the time map is single-valued and the leading angular dependence is absent.
In the exact black-pole geometry, the time map becomes non-monotonic and the
admissible angular domain moves toward $\theta_\star$ at larger
$T_{\rm target}$. The real part selects the lowest-area surface within this
domain. The imaginary part is evaluated on that same surface and is
controlled by the sign structure of the lifted area over the internal
sphere.

\subsection{Summary of the timelike-entanglement analysis}
\label{sec:tee-summary-results}

We now summarize the timelike-entanglement result before turning to
timelike subregion complexity. The main conclusion is that the exact
black-pole geometry changes the Lorentzian branch structure in a way that
is absent in the large-$r$ regime. The large-$r$ calculation gives the
short-boundary-interval limit, while the exact calculation shows how the
selected surface becomes sensitive to the localized core and to the
cap/horizon transition region.

In the large-$r$ regime, the leading angular dependence of
$K_y(r,\theta)$ and $G(r,\theta)$ drops out. The branch equations become
independent of the angular label at leading order, and the time map is
single-valued in the short-boundary-interval domain. In this limit,
\begin{equation}
	T\rightarrow0
	\qquad\Longleftrightarrow\qquad
	r_0\rightarrow\infty.
	\label{eq:tee-summary-short-time}
\end{equation}
The boundary-reaching spacelike branch gives the logarithmic
UV-renormalized real part, while the finite timelike branch gives the
leading imaginary contribution. This result gives the universal
near-boundary behaviour of the timelike surface, but it does not contain
the localized cap/horizon structure.

The exact black-pole calculation is different because the metric functions
depend on the internal sphere. The angular label $\theta_0$ defines the
reduced Lorentzian branch profile in the $(t,r)$ geometry. The physical
internal angle $\theta$ is then integrated over in the ten-dimensional
area. Therefore the full complex area is not obtained by replacing
$\theta$ by $\theta_0$. The reduced branch profile and the lifted
ten-dimensional area play different roles.

This distinction has three main consequences. First, the exact time map
can become non-monotonic. At fixed $\theta_0$, the same target interval can
be reached by two radial branches, one on each side of
$r_{0,\rm peak}(\theta_0)$. These branches are distinct candidates only
after the common condition \eqref{eq:tee-localized-fixedT} has been imposed.

Second, the maximum accessible boundary interval depends on the angular
label. An angular label contributes at a given $T_{\rm target}$ only if
\begin{equation}
	T_{\rm max}(\theta_0)\geq T_{\rm target}.
\end{equation}
As $T_{\rm target}$ is increased, this condition restricts the allowed
angular labels toward the cap/horizon transition region. The local origin
of this behaviour is explained in appendix~\ref{app:transition-region}.
Near $\theta_\star$, the crossover scale
\begin{equation}
	r_c(\theta_0)\simeq \ell|\theta_0-\theta_\star|
\end{equation}
becomes small. The transition radial interval then gives a positive
logarithmic contribution to $I_{\rm tim}-I_{\rm sp}$, allowing larger
boundary intervals to be supported near $\theta_\star$.

Third, the ten-dimensional lift changes the separation between real and
imaginary contributions. At the reduced-branch level, the spacelike branch
is real and the timelike branch is imaginary. After the lift, both
branches can contribute to both parts of the area. The sign of the lifted
area functions $\Delta_{\rm tim}$ and $\Delta_{\rm sp}$ over the
$(r,\theta)$ integration region decides whether a local contribution enters
the real or imaginary part. This is why the imaginary part contains
information that is not contained in the reduced branch label alone.

The fixed-boundary-interval selection follows these steps. For each
$T_{\rm target}$, one first solves the time equation for all allowed
angular labels and radial branches. One then minimizes the real part of
the renormalized lifted area among those branches. The imaginary part is
evaluated on the same selected surface and is not minimized separately.

In the numerical range studied here, the selected surface lies on the
larger-$r_0$ branch, where ``larger'' means the side
$r_0>r_{0,\rm peak}(\theta_0)$. This does not imply that the selected
surface remains in the asymptotic region. As $T_{\rm target}$ increases, the
selected value $r_0^\ast$ decreases, while the selected angular label
$\theta_0^\ast$ moves toward $\theta_\star$. The selected surface therefore
moves inward and becomes increasingly sensitive to the exact localized
geometry.

The selected real and imaginary parts have different meanings. The real
part is the quantity used to select the surface. Its slow variation at
larger target intervals reflects the fact that the allowed angular labels
are already confined close to the transition region. The imaginary part is
controlled by the sign structure of the lifted area functions over the
internal sphere. It can therefore vary differently from the real part and
need not be monotonic.

The timelike-entanglement analysis therefore gives a clear picture. The
large-$r$ regime reproduces the universal short-boundary-interval behaviour.
The exact black-pole geometry introduces angular dependence, a
non-monotonic time map, an angular restriction through
$T_{\rm max}(\theta_0)$, and a lifted area in which both branches can carry
real and imaginary contributions. The selected surface moves toward the
cap/horizon transition region as the target interval is increased. This is
the localized-geometry effect that will be compared with timelike
subregion complexity in the next section.

\section{Timelike subregion complexity}
\label{sec:timelike-complexity-prescription}

We now associate a volume to the same Lorentzian branch geometry used in the
timelike-entanglement calculation. For timelike entanglement, the spacelike
and timelike branches enter the lifted area. For timelike subregion
complexity, the same branches bound a Lorentzian region whose
ten-dimensional lift defines a volume. The construction follows the
localized lifting prescription of ref.~\cite{Bena:2024lbh}, with the area
functional replaced by a volume functional.

The reduced branch data are inherited from
section~\ref{sec:tee-branch-equations}. We use the metric functions
$F(r;\theta_0)$ and $H(r;\theta_0)$ defined in
eqs.~\eqref{eq:reduced-two-dimensional-metric}--\eqref{eq:H-definition},
the turning-point value $F_0=F(r_0;\theta_0)$, and the branch kernels
$K_{\rm tim}$ and $K_{\rm sp}$ defined in
eqs.~\eqref{eq:exact-tee-Ktim} and \eqref{eq:exact-tee-Ksp}. For each pair
$(r_0,\theta_0)$, the time map
eq.~\eqref{eq:exact-tee-boundary-interval} assigns a boundary interval
$T(r_0,\theta_0)$. Thus $(r_0,\theta_0)$ labels a Lorentzian branch
configuration before any fixed-boundary-interval comparison is made.

Let $\Sigma_T(r_0,\theta_0)$ denote the region in the effective $(t,r)$
plane bounded by the upper and lower Lorentzian branches. This region is
determined by the reduced geometry at the angular label $\theta_0$. The
ten-dimensional lift is then performed over the physical internal angle
$\theta$. The lifted volume is
\begin{equation}
	V_{\rm lift}
	\left[
	\Sigma_T(r_0,\theta_0)\times S^3\times T^4
	\right]
	=
	{\cal N}
	\int_{\Sigma_T(r_0,\theta_0)}
	dt\,dr
	\int_0^{\pi/2}
	d\theta\,
	\sin\theta\cos\theta\,
	\mu(r,\theta).
	\label{eq:complexity-lifted-region-volume}
\end{equation}
Here
\begin{equation}
	{\cal N}=(2\pi)^6V_4.
	\label{eq:complexity-N}
\end{equation}
This is the volume counterpart of the localized lifted-area prescription:
$\theta_0$ fixes the reduced branch region, while $\theta$ enters the lifted
density through the full black-pole functions $K_y(r,\theta)$ and
$G(r,\theta)$.

Equivalently, the local lifted volume element is
\begin{equation}
	dV_{\rm lift}
	=
	{\cal N}\,
	dt\,dr\,d\theta\,
	\sin\theta\cos\theta\,
	\mu(r,\theta).
	\label{eq:complexity-local-lifted-volume-element}
\end{equation}
The density $\mu(r,\theta)$ is the effective volume weight of a radial shell
after the compact directions are included. It is evaluated using the
physical internal angle $\theta$, because the ten-dimensional lift keeps the
full angular dependence of the localized black-pole metric. Using
$F(r,\theta)=Qr^2K_y(r,\theta)G(r,\theta)$ and
$H(r,\theta)=Q^2G(r,\theta)^2/(r^2+\ell^2)$, one finds
\begin{equation}
	\mu(r,\theta)
	=
	\sqrt{F(r,\theta)H(r,\theta)}
	=
	Q^{3/2}
	\frac{
		r\,\sqrt{K_y(r,\theta)}\,G(r,\theta)^{3/2}
	}{
		\sqrt{r^2+\ell^2}
	}.
	\label{eq:complexity-local-density}
\end{equation}
After the internal angular integration, the volume density becomes a radial
weight,
\begin{equation}
	\bar\mu(r)
	=
	{\cal N}Q^{3/2}
	\int_0^{\pi/2}
	d\theta\,\sin\theta\cos\theta\,
	\frac{
		r\,\sqrt{K_y(r,\theta)}\,G(r,\theta)^{3/2}
	}{
		\sqrt{r^2+\ell^2}
	}.
	\label{eq:complexity-lifted-density}
\end{equation}
The radial weight $\bar\mu(r)$ is independent of the branch label
$\theta_0$. The $\theta_0$ dependence enters through the branch kernels and
through the profiles $t_{\rm tim}^{+}$ and $t_{\rm sp}^{+}$. Thus the
internal angular integral gives the ten-dimensional volume weight of each
radial shell, while the Lorentzian branches determine how much of that shell
belongs to $\Sigma_T(r_0,\theta_0)$.

The density $\bar\mu(r)$ is positive in the region covered by the branch
construction. Therefore timelike subregion complexity is a real volume
observable. This differs from timelike entanglement entropy, where the
lifted area can be complex. In the volume prescription, the sign enters
only after the UV subtraction through the finite branch-volume combination:
the spacelike weighted volume contributes with a plus sign, while the
timelike weighted volume is subtracted.

The UV-subtracted finite volume derived below has the form
\begin{equation}
	C_T^{\rm lift}(r_0,\theta_0)
	=
	\frac{1}{G_{10}L_{\rm AdS}}
	\left[
	T(r_0,\theta_0)M_\infty
	+
	2I_{\rm sp}^{V}(r_0,\theta_0)
	-
	2I_{\rm tim}^{V}(r_0,\theta_0)
	\right].
	\label{eq:complexity-localized-finite-volume}
\end{equation}
The two weighted branch volumes are positive separately, but they enter the
renormalized observable with opposite signs. This signed finite difference
is the basic quantity minimized in the timelike-complexity calculation.

As in the timelike-entanglement analysis, the boundary interval must be fixed
before the branch volumes are compared. For a prescribed target interval,
one first solves
\begin{equation}
T(r_0,\theta_0)=T_{\rm target}
	\label{eq:complexity-localized-fixedT}
\end{equation}
for all admissible angular labels and radial branches. If the time map is
non-monotonic, the same $\theta_0$ can give more than one radial root; we
denote these roots by $r_{0,b}(T_{\rm target},\theta_0)$. The selected
timelike complexity is then
\begin{equation}
	C_T^{\rm selected}(T_{\rm target})
	=
	\min_{\theta_0,b}
	C_T^{\rm lift}
	\left(
	r_{0,b}(T_{\rm target},\theta_0),
	\theta_0
	\right).
	\label{eq:complexity-localized-minimum}
\end{equation}
Thus the complexity calculation is the volume analogue of the localized
lifted-area minimisation, with the additional Lorentzian requirement that
the boundary interval is generated by the branch data and must be fixed
before the volume comparison is made.

For a time-reflection symmetric configuration, the lifted volume associated
with one branch is
\begin{equation}
	V_b^{\rm lift}
	=
	2
	\int_{{\cal D}_b}
	dr\,
	\bar\mu(r)\,
	t_b^+(r),
	\label{eq:complexity-branch-volume}
\end{equation}
where $b=\mathrm{tim},\mathrm{sp}$. The radial domains are
\begin{equation}
	{\cal D}_{\rm tim}=[\epsilon,r_0],
	\qquad
	{\cal D}_{\rm sp}=[\epsilon,R].
	\label{eq:complexity-branch-domains}
\end{equation}
For the timelike branch,
\begin{equation}
	t_b^+(r)
	=
	t_{\rm tim}^{+}(r;r_0,\theta_0),
	\qquad
	b=\mathrm{tim},
	\label{eq:complexity-tb-tim}
\end{equation}
where the profile ends at $r=r_0$. For the spacelike branch at cutoff $R$,
\begin{equation}
	t_b^+(r)
	=
	t_{\rm sp}^{+}(r;R,r_0,\theta_0),
	\qquad
	b=\mathrm{sp}.
	\label{eq:complexity-tb-sp}
\end{equation}
The cutoff is shown explicitly because the spacelike branch reaches the
asymptotic boundary.

It is useful to introduce the radial primitive
\begin{equation}
	{\cal M}(r)
	=
	\int_{\epsilon}^{r}
	ds\,\bar\mu(s),
	\qquad
	{\cal M}'(r)=\bar\mu(r).
	\label{eq:complexity-M-primitive}
\end{equation}
This primitive is the lifted radial volume accumulated between the lower
cutoff and the point $r$. It allows the branch volumes to be written in
terms of the branch kernels rather than the branch profiles.

For the timelike branch,
\begin{align}
	V_{\rm tim}^{\rm lift}(r_0,\theta_0)
	&=
	2
	\int_{\epsilon}^{r_0}
	dr\,
	\bar\mu(r)
	\int_{r}^{r_0}
	d\tilde r\,
	K_{\rm tim}(\tilde r;\theta_0,r_0)
	\nonumber\\
	&=
	2
	\int_{\epsilon}^{r_0}
	d\tilde r\,
	{\cal M}(\tilde r)
	K_{\rm tim}(\tilde r;\theta_0,r_0).
\end{align}
Thus
\begin{equation}
	V_{\rm tim}^{\rm lift}(r_0,\theta_0)
	=
	2I_{\rm tim}^{V}(r_0,\theta_0),
	\label{eq:complexity-Vtim-Itim}
\end{equation}
with
\begin{equation}
	I_{\rm tim}^{V}(r_0,\theta_0)
	=
	\int_{\epsilon}^{r_0}
	dr\,
	{\cal M}(r)\,
	K_{\rm tim}(r;\theta_0,r_0).
	\label{eq:complexity-ItimV}
\end{equation}
This contribution is finite because the timelike branch ends at the turning
point.

For the spacelike branch,
\begin{align}
	V_{\rm sp}^{\rm lift}(R;r_0,\theta_0)
	&=
	2
	\int_{\epsilon}^{R}
	dr\,
	\bar\mu(r)
	\left[
	\frac{T(r_0,\theta_0)}{2}
	+
	\int_{r}^{R}
	d\tilde r\,
	K_{\rm sp}(\tilde r;\theta_0,r_0)
	\right]
	\nonumber\\
	&=
	T(r_0,\theta_0){\cal M}(R)
	+
	2I_{\rm sp}^{V}(R;r_0,\theta_0),
	\label{eq:complexity-Vsp-Isp}
\end{align}
where
\begin{equation}
	I_{\rm sp}^{V}(R;r_0,\theta_0)
	=
	\int_{\epsilon}^{R}
	dr\,
	{\cal M}(r)\,
	K_{\rm sp}(r;\theta_0,r_0).
	\label{eq:complexity-IspV-cutoff}
\end{equation}
The term proportional to $T{\cal M}(R)$ is the volume of the boundary-time
strip carried by the boundary-reaching branch. It contains the leading
large-$R$ divergence.

The subtraction is fixed by the large-$r$ expansion of the lifted density,
\begin{equation}
	\bar\mu(r)
	=
	A_\mu
	+
	\frac{B_\mu}{r^2}
	+
	\frac{C_\mu}{r^4}
	+
	O(r^{-6}).
	\label{eq:complexity-density-large-r}
\end{equation}
The leading coefficient is
\begin{equation}
	A_\mu
	=
	\frac{{\cal N}Q^{3/2}}{2}.
	\label{eq:complexity-density-A}
\end{equation}
The coefficients $B_\mu$ and $C_\mu$ are determined by the subleading
large-$r$ expansion of $K_y(r,\theta)$ and $G(r,\theta)$ after the angular
integral. In the large-$r$ limit $K_y(r,\theta)$ and $G(r,\theta)$ are given in \eqref{eq:G-asymptotic}, one obtains
\begin{align}
	B_\mu
	=
	\frac{{\cal N}Q^{3/2}}{4}
	\left(
	\ell_2^2-\ell^2
	\right),\qquad
	C_\mu
	=
	\frac{{\cal N}Q^{3/2}}{16}
	\left(
	-\ell_2^4
	-
	2\ell_2^2\ell^2
	+
	3\ell^4
	\right).
	\label{eq:complexity-density-C}
\end{align}

These coefficients determine the finite large-$r$ tail of the spacelike
volume. They are needed for cutoff-stable numerical results.

The primitive has the asymptotic form
\begin{equation}
	{\cal M}(R)
	=
	A_\mu R
	+
	{\cal M}_\infty
	-
	\frac{B_\mu}{R}
	-
	\frac{C_\mu}{3R^3}
	+
	O(R^{-5}).
	\label{eq:complexity-M-large-r}
\end{equation}
At a finite large cutoff $R_{\rm max}$, the finite constant can be extracted
as
\begin{equation}
	{\cal M}_\infty
	=
	{\cal M}(R_{\rm max})
	-
	A_\mu R_{\rm max}
	+
	\frac{B_\mu}{R_{\rm max}}
	+
	\frac{C_\mu}{3R_{\rm max}^{3}}
	+
	O(R_{\rm max}^{-5}).
	\label{eq:complexity-Minf-extraction}
\end{equation}
The term $A_\mu R$ is the universal asymptotic growth. The constant
${\cal M}_\infty$ is the finite part of the radial primitive after this
growth is removed.

The leading divergent contribution to the spacelike lifted volume is
\begin{equation}
	V_{\rm ct}^{\rm lift}(R;r_0,\theta_0)
	=
	A_\mu\,T(r_0,\theta_0)\,R.
	\label{eq:complexity-volume-counterterm}
\end{equation}
The finite lifted volume is therefore defined by
\begin{equation}
	\Delta V_{\rm fin}^{\rm lift}(r_0,\theta_0)
	=
	\lim_{R\to\infty}
	\left[
	V_{\rm sp}^{\rm lift}(R;r_0,\theta_0)
	-
	V_{\rm tim}^{\rm lift}(r_0,\theta_0)
	-
	A_\mu\,T(r_0,\theta_0)R
	\right].
	\label{eq:complexity-ren-volume-definition}
\end{equation}
Using the branch-volume expressions above, this becomes
\begin{equation}
	\Delta V_{\rm fin}^{\rm lift}(r_0,\theta_0)
	=
	T(r_0,\theta_0){\cal M}_\infty
	+
	2I_{\rm sp}^{V}(r_0,\theta_0)
	-
	2I_{\rm tim}^{V}(r_0,\theta_0),
	\label{eq:complexity-finite-volume}
\end{equation}
where
\begin{equation}
	I_{\rm sp}^{V}(r_0,\theta_0)
	=
	\lim_{R\to\infty}
	I_{\rm sp}^{V}(R;r_0,\theta_0).
	\label{eq:complexity-IspV-finite}
\end{equation}

For numerical evaluation, the finite cutoff integral must be supplemented
by the analytic large-$r$ tail. The cutoff-stable form is
\begin{align}
	\Delta V_{\rm fin}^{\rm lift}
	&=
	T{\cal M}_\infty
	+
	2
	\int_{\epsilon}^{R_{\rm max}}
	dr\,
	{\cal M}(r)K_{\rm sp}(r;\theta_0,r_0)
	+
	2I_{\rm sp,tail}^{V}
	\nonumber\\
	&\hspace{1.0cm}
	-
	2
	\int_{\epsilon}^{r_0}
	dr\,
	{\cal M}(r)K_{\rm tim}(r;\theta_0,r_0).
	\label{eq:complexity-tail-improved-volume}
\end{align}
Here $I_{\rm sp,tail}^{V}$ is the finite contribution from
$R_{\rm max}<r<\infty$, computed from the large-$r$ expansion of
$\bar\mu(r)$ and the spacelike kernel. This term removes the residual
cutoff dependence that would remain in a purely finite-cutoff computation.

The timelike subregion complexity is the finite volume in units of
$G_{10}L_{\rm AdS}$:
\begin{equation}
	C_T^{\rm lift}(r_0,\theta_0)
	=
	\frac{
		\Delta V_{\rm fin}^{\rm lift}(r_0,\theta_0)
	}{
		G_{10}L_{\rm AdS}
	}.
	\label{eq:complexity-Clift-definition}
\end{equation}
Equivalently,
\begin{equation}
	C_T^{\rm lift}(r_0,\theta_0)
	=
	\frac{1}{G_{10}L_{\rm AdS}}
	\left[
	T(r_0,\theta_0){\cal M}_\infty
	+
	2I_{\rm sp}^{V}(r_0,\theta_0)
	-
	2I_{\rm tim}^{V}(r_0,\theta_0)
	\right].
	\label{eq:complexity-Clift-final}
\end{equation}
In numerical plots we suppress the common positive prefactor. This does
not affect the fixed-boundary-interval minimisation or the selected branch.

Several features of this expression are important. The local density
$\bar\mu(r)$ is positive, so the individual spacelike and timelike branch
volumes are positive. The finite observable, however, is not the sum of
these two volumes. It is the renormalized spacelike-minus-timelike
combination in eq.~\eqref{eq:complexity-Clift-final}, after the universal
boundary contribution has been subtracted. Its value therefore depends on
which Lorentzian branch family is selected at the chosen boundary interval.

The fixed-boundary-interval rule given in
eqs.~\eqref{eq:complexity-localized-fixedT}--\eqref{eq:complexity-localized-minimum}.
It will be implemented explicitly in
section~\ref{sec:complexity-fixed-boundary-minimisation}. There, for each
target interval, the time equation is solved for all admissible angular
labels and radial branches, and only then are the corresponding finite
renormalized volumes compared.

This completes the prescription. The same Lorentzian branches used for
timelike entanglement now define a lifted volume. The observable is real,
but it remains sensitive to the localized black-pole geometry through two
sources: the branch kernels, which depend on the angular label $\theta_0$,
and the lifted radial density, which contains the integral over the physical
internal angle $\theta$. Thus timelike subregion complexity is a finite
Lorentzian branch-volume observable at fixed boundary interval, rather than
a global interior complexity of the complete black-hole spacetime.

\subsection{BTZ benchmark for timelike subregion complexity}
\label{sec:complexity-btz-benchmark}

Before applying the volume prescription to the localized black pole, we
review the BTZ limit. This is the simplest case of the same Lorentzian
branch construction. The internal space contributes only an overall factor,
there is no angular label $\theta_0$, and the branch density is constant.
Thus BTZ fixes the sign convention, the spacelike-minus-timelike volume
combination, and the UV subtraction that will later be used in the localized
geometry.

We consider the non-rotating BTZ metric at fixed boundary spatial position,
\begin{equation}
	ds_2^2
	=
	L^2
	\left[
	-(r^2-r_h^2)\,dt^2
	+
	\frac{dr^2}{r^2-r_h^2}
	\right].
	\label{eq:btz-metric}
\end{equation}
The boundary interval is
\begin{equation}
	-\frac{T}{2}\leq t\leq \frac{T}{2}.
	\label{eq:btz-boundary-interval}
\end{equation}
The Lorentzian surface contains a timelike branch, a spacelike branch, and
their time-reflected lower branches. These branches bound the region whose
finite volume defines the BTZ timelike subregion complexity.

For the branch family used below,
\begin{equation}
	r_0\geq \sqrt{2}\,r_h.
	\label{eq:btz-r0-domain}
\end{equation}
The first-order branch equations are
\begin{subequations}
	\begin{align}
		\left(t_{\rm tim}'(r)\right)^2
		&=
		\frac{
			r_h^2-r_0^2
		}{
			(r^2-r_0^2)(r^2-r_h^2)^2
		},
		\label{eq:btz-ttim-prime}
		\\
		\left(t_{\rm sp}'(r)\right)^2
		&=
		\frac{
			r_0^2-r_h^2
		}{
			(r^2+r_0^2-2r_h^2)(r^2-r_h^2)^2
		}.
		\label{eq:btz-tsp-prime}
	\end{align}
\end{subequations}
These equations determine the positive branch kernels. The global signs are
fixed by time reflection: the upper and lower branches are mirror images,
the timelike branch closes at $r=r_0$, and the spacelike branch reaches the
asymptotic boundary.

Define
\begin{equation}
	a_{\rm B}
	=
	\sqrt{r_0^2-r_h^2},
	\qquad
	b_{\rm B}
	=
	\sqrt{r_0^2-2r_h^2}.
	\label{eq:btz-ab}
\end{equation}
Outside the horizon,
\begin{subequations}
	\begin{align}
		K_{\rm tim}^{\rm out}(r)
		&=
		\frac{
			a_{\rm B}
		}{
			(r^2-r_h^2)\sqrt{r_0^2-r^2}
		},
		\qquad
		r_h<r\leq r_0,
		\label{eq:btz-Ktim-out}
		\\
		K_{\rm sp}^{\rm out}(r)
		&=
		\frac{
			a_{\rm B}
		}{
			(r^2-r_h^2)\sqrt{r^2+b_{\rm B}^2}
		},
		\qquad
		r_h<r<R.
		\label{eq:btz-Ksp-out}
	\end{align}
\end{subequations}
Inside the horizon,
\begin{subequations}
	\begin{align}
		K_{\rm tim}^{\rm in}(r)
		&=
		\frac{
			a_{\rm B}
		}{
			(r_h^2-r^2)\sqrt{r_0^2-r^2}
		},
		\qquad
		0\leq r<r_h,
		\label{eq:btz-Ktim-in}
		\\
		K_{\rm sp}^{\rm in}(r)
		&=
		\frac{
			a_{\rm B}
		}{
			(r_h^2-r^2)\sqrt{r^2+b_{\rm B}^2}
		},
		\qquad
		0\leq r<r_h.
		\label{eq:btz-Ksp-in}
	\end{align}
\end{subequations}
The upper timelike branch outside the horizon is
\begin{equation}
	t_{\rm tim}^{\rm out,+}(r)
	=
	\int_{r}^{r_0}
	ds\,
	K_{\rm tim}^{\rm out}(s),
	\qquad
	r_h<r\leq r_0.
	\label{eq:btz-ttim-out}
\end{equation}
The upper spacelike branch outside the horizon is
\begin{equation}
	t_{\rm sp}^{\rm out,+}(r;R)
	=
	\frac{T}{2}
	+
	\int_{r}^{R}
	ds\,
	K_{\rm sp}^{\rm out}(s),
	\qquad
	r_h<r<R.
	\label{eq:btz-tsp-out}
\end{equation}
The inside profiles are obtained from $K_{\rm tim}^{\rm in}$ and
$K_{\rm sp}^{\rm in}$, with additive constants chosen so that the profiles
join across the regulated horizon. The lower branches are obtained by
\begin{equation}
	t_b^{-}(r)=-t_b^{+}(r),
	\qquad
	b=\mathrm{tim},\mathrm{sp}.
	\label{eq:btz-time-reflection}
\end{equation}

\begin{figure}[t]
	\centering
	\includegraphics[width=0.70\textwidth]{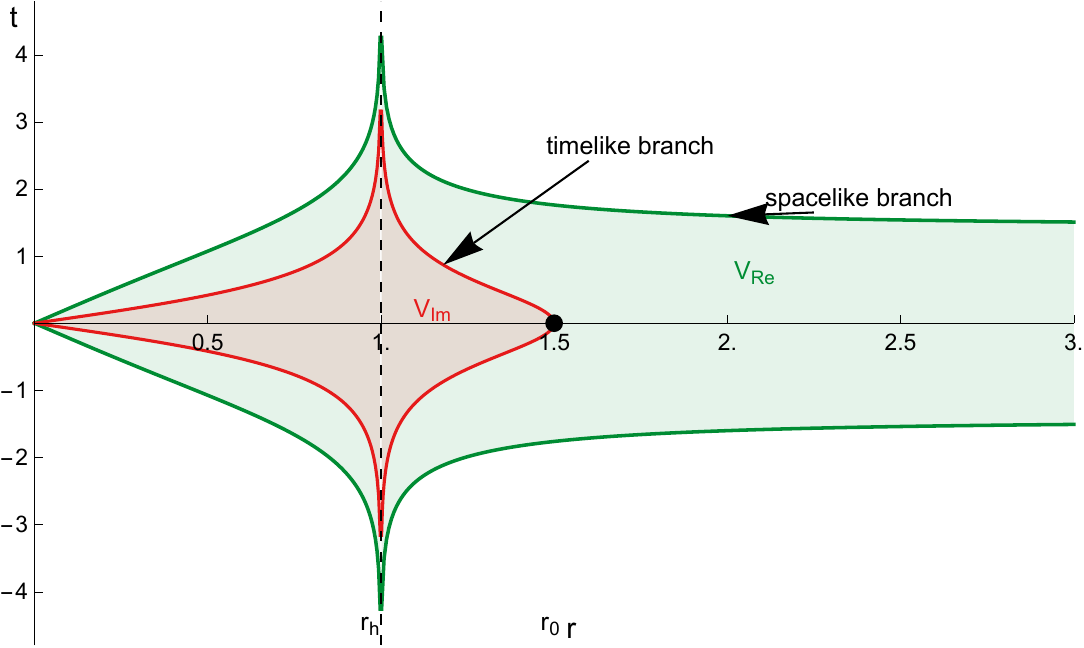}
	\caption{
		BTZ branch convention for timelike subregion complexity. The timelike
		branch closes at $r_0$, while the spacelike branch reaches the
		asymptotic cutoff. The shaded regions denote the regulated branch
		volumes entering the spacelike-minus-timelike prescription.
	}
	\label{fig:complexity-btz-branches}
\end{figure}

Figure~\ref{fig:complexity-btz-branches} illustrates the BTZ version of the
volume prescription. The same prescription will be used for the localized
black pole: the spacelike branch gives the boundary-reaching volume, the
timelike branch gives the finite branch volume, and the UV contribution of
the spacelike branch is subtracted.

For BTZ the volume density is constant,
\begin{equation}
	\sqrt{|g_{tt}g_{rr}|}
	=
	L^2.
	\label{eq:btz-volume-density}
\end{equation}
Therefore a time-reflection symmetric branch has volume
\begin{equation}
	V_b
	=
	2L^2
	\int_{{\cal D}_b}
	dr\,
	t_b^{+}(r).
	\label{eq:btz-branch-volume}
\end{equation}
This is the BTZ limit of the localized formula
\eqref{eq:complexity-branch-volume}. In the localized geometry the constant
$L^2$ is replaced by the radial lifted density $\bar\mu(r)$, and the branch
profiles acquire $\theta_0$ dependence through $K_y(r,\theta_0)$ and
$G(r,\theta_0)$.

We regulate the horizon by cutting the inside and outside regions at
$r_h(1-\varepsilon)$ and $r_h(1+\varepsilon)$. The regulated spacelike
volume is
\begin{equation}
	V_{\rm sp}(R,\varepsilon)
	=
	2L^2
	\left[
	\int_0^{r_h(1-\varepsilon)}
	dr\,
	t_{\rm sp}^{\rm in,+}(r)
	+
	\int_{r_h(1+\varepsilon)}^{R}
	dr\,
	t_{\rm sp}^{\rm out,+}(r;R)
	\right],
	\label{eq:btz-Vsp-reg}
\end{equation}
and the regulated timelike volume is
\begin{equation}
	V_{\rm tim}(\varepsilon)
	=
	2L^2
	\left[
	\int_0^{r_h(1-\varepsilon)}
	dr\,
	t_{\rm tim}^{\rm in,+}(r)
	+
	\int_{r_h(1+\varepsilon)}^{r_0}
	dr\,
	t_{\rm tim}^{\rm out,+}(r)
	\right].
	\label{eq:btz-Vtim-reg}
\end{equation}
The horizon regulator cancels in the difference
$V_{\rm sp}-V_{\rm tim}$. The remaining divergence comes from the
large-$r$ part of the spacelike branch. Since this branch reaches the
cutoff at $t=T/2$, the leading large-$R$ behaviour is
\begin{equation}
	V_{\rm sp}(R,\varepsilon)
	=
	L^2TR+\text{finite}.
	\label{eq:btz-Vsp-large-R}
\end{equation}
Thus the finite BTZ branch volume is
\begin{equation}
	\Delta V_{\rm fin}^{\rm BTZ}
	=
	\lim_{\substack{R\to\infty\\ \varepsilon\to0}}
	\left[
	V_{\rm sp}(R,\varepsilon)
	-
	V_{\rm tim}(\varepsilon)
	-
	L^2TR
	\right].
	\label{eq:btz-finite-volume-definition}
\end{equation}
This is the BTZ analogue of
eq.~\eqref{eq:complexity-ren-volume-definition}. The subtraction
$L^2TR$ is the BTZ form of the localized counterterm
$A_\mu T R$: it removes the universal volume of the boundary-reaching
strip.

The finite result can be written as a function of the turning point:
\begin{equation}
	\Delta V_{\rm fin}^{\rm BTZ}(r_0)
	=
	2L^2
	\left[
	\operatorname{arccoth}
	\left(
	\frac{r_0}{\sqrt{r_0^2-r_h^2}}
	\right)
	-
	\operatorname{artanh}
	\left(
	\sqrt{
		\frac{r_0^2-2r_h^2}{r_0^2-r_h^2}
	}
	\right)
	\right].
	\label{eq:btz-finite-volume-r0}
\end{equation}
The boundary interval and the turning point are related by \cite{Afrasiar:2025timelike}
\begin{equation}
	r_0^2
	=
	r_h^2
	\frac{
		\cosh(r_hT)
	}{
		\sinh^2(r_hT/2)
	}.
	\label{eq:btz-r0-T}
\end{equation}
Introducing
\begin{equation}
	x=\frac{r_hT}{2},
	\label{eq:btz-x}
\end{equation}
we have
\begin{equation}
	\sqrt{r_0^2-r_h^2}
	=
	r_h\coth x,
	\qquad
	\sqrt{r_0^2-2r_h^2}
	=
	r_h\,\operatorname{csch}x.
	\label{eq:btz-root-relations}
\end{equation}
Substituting these relations into eq.~\eqref{eq:btz-finite-volume-r0}
gives \cite{Prihadi:2026scalarhair}
\begin{equation}
	\Delta V_{\rm fin}^{\rm BTZ}(T)
	=
	2L^2
	\left[
	\operatorname{arsinh}
	\left(
	\coth\frac{r_hT}{2}
	\right)
	-
	\operatorname{arsinh}
	\left(
	\operatorname{csch}\frac{r_hT}{2}
	\right)
	\right].
	\label{eq:btz-finite-volume-T}
\end{equation}
The BTZ timelike subregion complexity is
\begin{equation}
	C_T^{\rm BTZ}
	=
	\frac{
		\Delta V_{\rm fin}^{\rm BTZ}
	}{
		G_3L
	}.
	\label{eq:btz-complexity}
\end{equation}

The BTZ calculation therefore gives the clean limit of the localized
prescription. The branch geometry is the same in spirit, but there is no
internal angular structure: the density is constant and there is no
$\theta_0$ selection. In the localized black pole, two changes occur.
First, the reduced branch kernels depend on the angular label through
$K_y(r,\theta_0)$ and $G(r,\theta_0)$. Second, the lifted volume density is
obtained by integrating the exact functions $K_y(r,\theta)$ and
$G(r,\theta)$ over the physical internal angle $\theta$. These two features
make the fixed-boundary-interval problem sensitive to the cap/horizon
structure of the localized geometry.

\subsection{Asymptotic limit for timelike complexity}
\label{sec:complexity-large-r-reference}

We next apply the timelike-complexity prescription to the large-$r$ regime
used in section~\ref{sec:tee-asymptotic-limit}. The Lorentzian branch
geometry is the same as in the timelike-entanglement calculation: a timelike
branch ends at $r_0$, and a spacelike branch reaches the asymptotic cutoff.
The difference is the observable. Here the same branches are inserted into
the lifted volume formula rather than the lifted area formula.

This regime describes short boundary intervals. The turning point lies far
from the localized core, and the branch profiles remain in the radial range
where the leading angular dependence of $K_y(r,\theta)$ and $G(r,\theta)$
drops out. Hence there is no cap-side/horizon-side distinction, no angular
selection, and no non-monotonic time map at this order. These effects enter
only after the exact black-pole functions are restored.

The large-$r$ regime is defined by
\begin{equation}
	r_0>r_\ast\gg \ell,\ell_1,\ell_2.
	\label{eq:complexity-large-r-range}
\end{equation}
Here $r_\ast$ is the lower edge of the radial range in which the large-$r$
expansion is used. The branch profiles are therefore the asymptotic
profiles restricted to $r\geq r_\ast$. The boundary interval is the same
single-valued function $T(r_0)$ obtained in
section~\ref{sec:tee-asymptotic-limit}; increasing $r_0$ corresponds to a
shorter boundary interval.

\begin{figure}[H]
	\centering
	\begin{subfigure}[b]{0.45\textwidth}
		\centering
		\includegraphics[width=\linewidth]{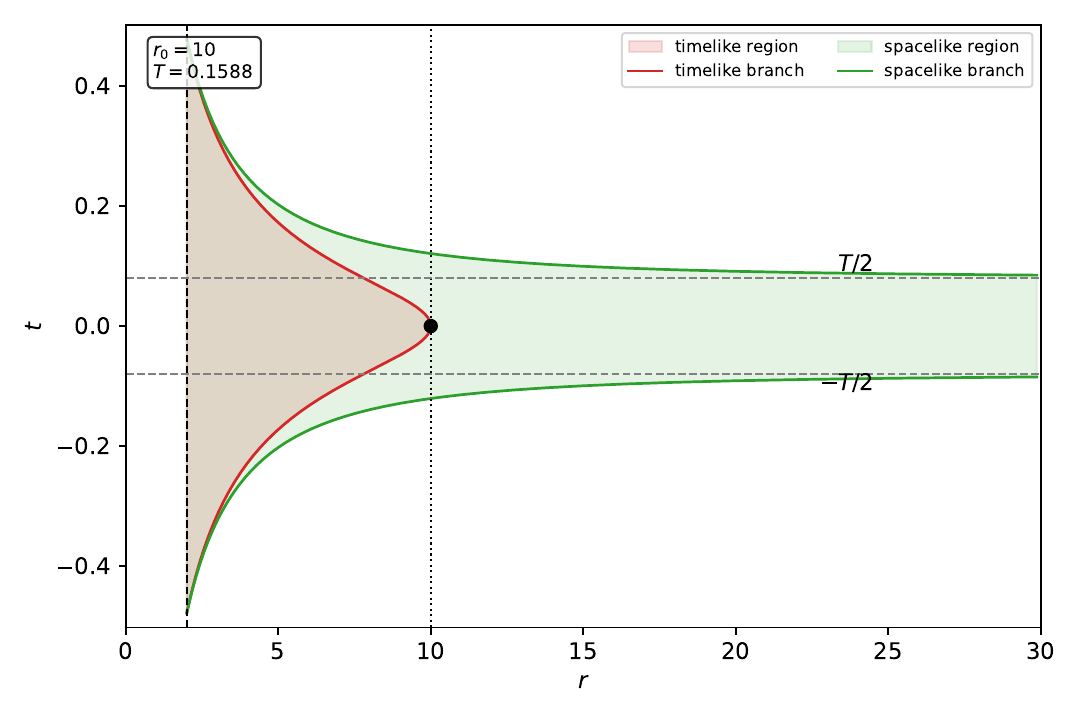}
		\caption{$r_0=10$}
		\label{fig:complexity-large-r-tr-r010}
	\end{subfigure}
	\hfill
	\begin{subfigure}[b]{0.45\textwidth}
		\centering
		\includegraphics[width=\linewidth]{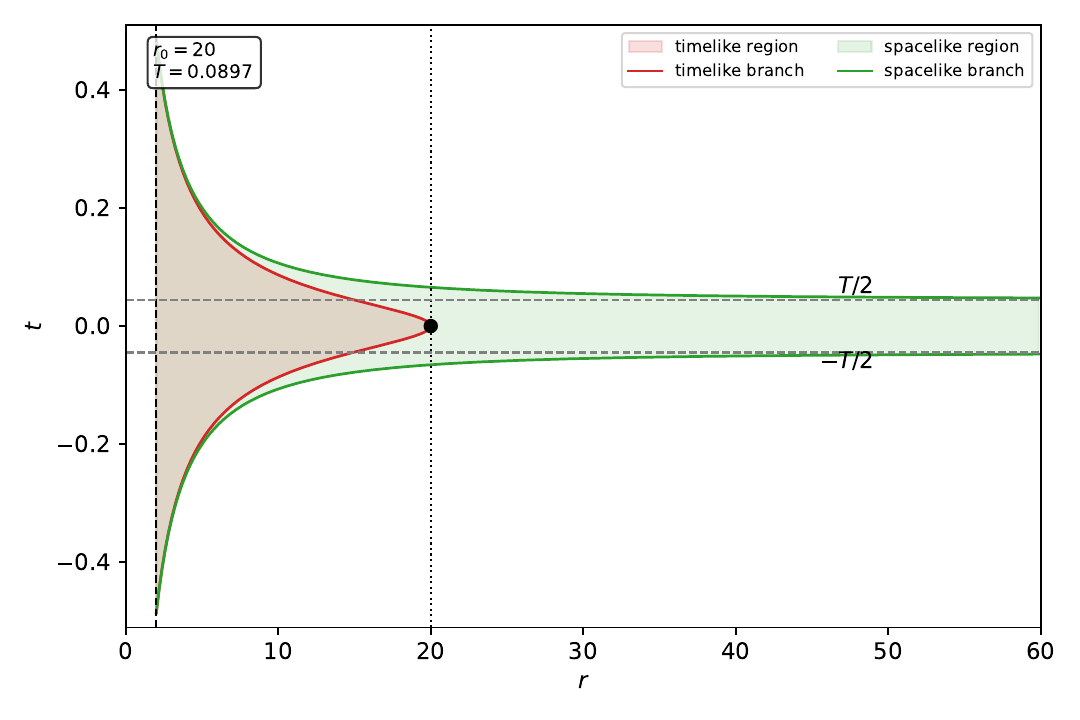}
		\caption{$r_0=20$}
		\label{fig:complexity-large-r-tr-r020}
	\end{subfigure}
	
	\medskip
	
	\begin{subfigure}[b]{0.45\textwidth}
		\centering
		\includegraphics[width=\linewidth]{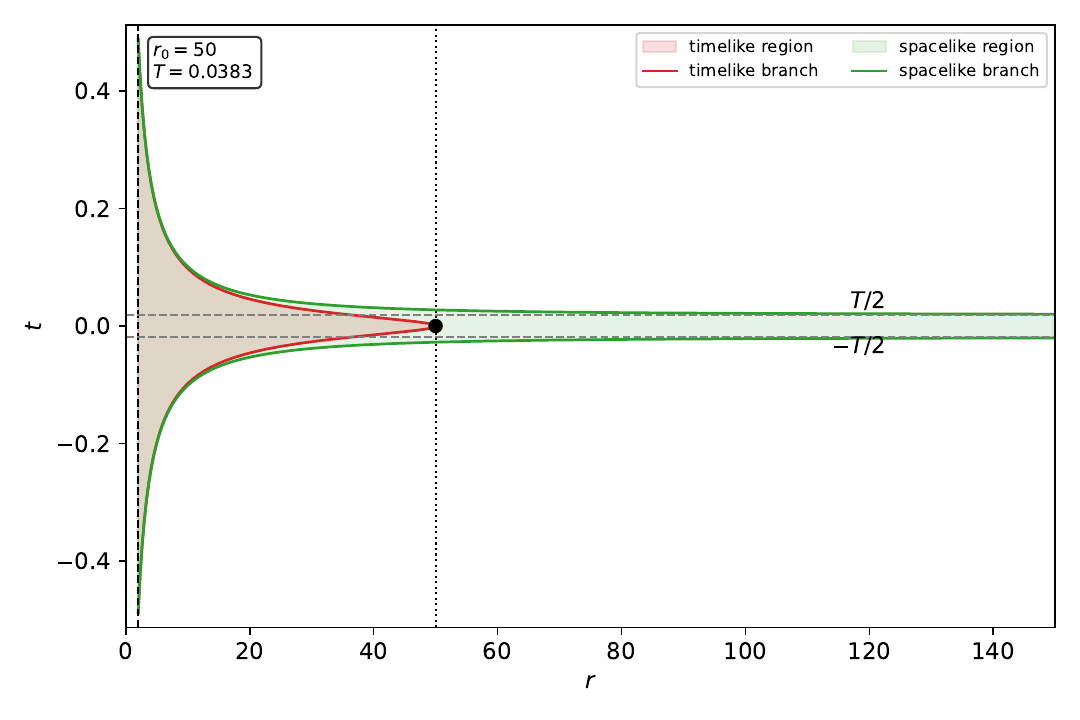}
		\caption{$r_0=50$}
		\label{fig:complexity-large-r-tr-r050}
	\end{subfigure}
	\hfill
	\begin{subfigure}[b]{0.45\textwidth}
		\centering
		\includegraphics[width=\linewidth]{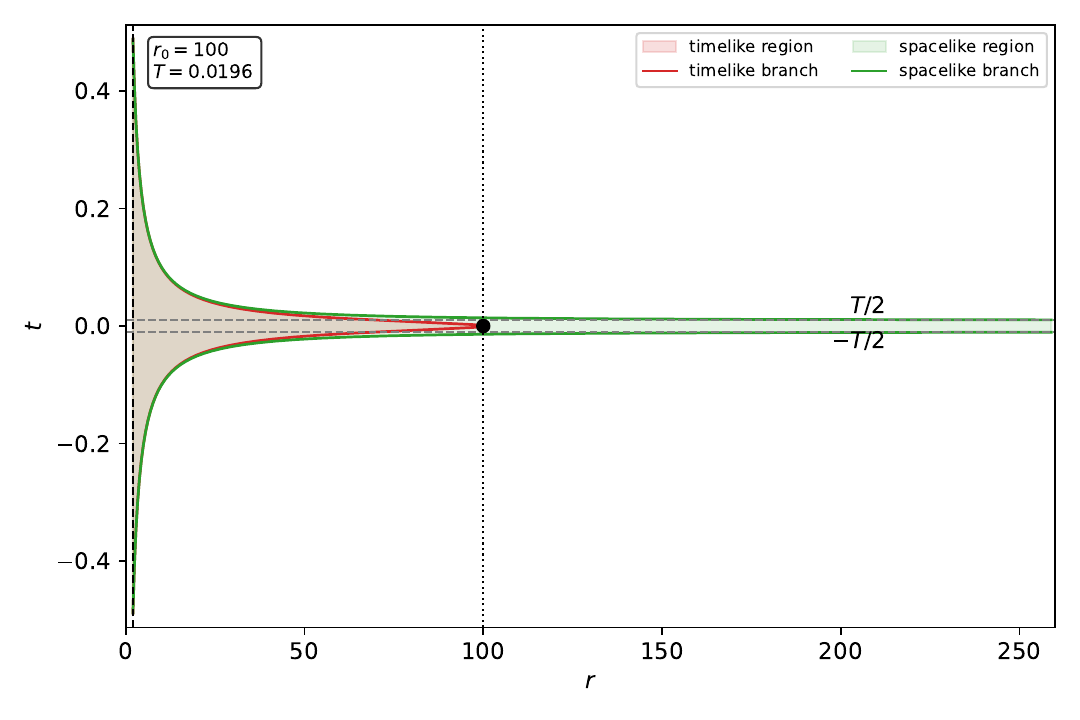}
		\caption{$r_0=100$}
		\label{fig:complexity-large-r-tr-r0100}
	\end{subfigure}
	\caption{
		Large-$r$ $t$-$r$ profiles used for timelike subregion complexity. The
		timelike branch ends at the turning point, while the spacelike branch
		reaches the cutoff. As $r_0$ increases, the corresponding boundary
		interval decreases.
	}
	\label{fig:complexity-large-r-tr-profiles}
\end{figure}

Figure~\ref{fig:complexity-large-r-tr-profiles} shows the Lorentzian region
whose volume is evaluated. The profiles are the same asymptotic branches
used for timelike entanglement, but the assigned quantity is now the
renormalized lifted volume.

The lifted radial density is obtained from
eq.~\eqref{eq:complexity-lifted-density}. Using the large-$r$ forms of
$K_y$ and $G$, one finds
\begin{equation}
	\bar\mu_{\rm large\text{-}r}(r)
	=
	\frac{{\cal N}Q^{3/2}}{2}
	\left(
	\frac{r^2+\ell_2^2}{r^2+\ell^2}
	\right)^{1/2}.
	\label{eq:complexity-large-r-density}
\end{equation}
This density is independent of $\theta_0$. Therefore the large-$r$
complexity cannot distinguish the cap-side and horizon-side sectors. It
only checks the asymptotic volume density and the UV subtraction.

The expansion of eq.~\eqref{eq:complexity-large-r-density} has the same
form as eq.~\eqref{eq:complexity-density-large-r},
\begin{equation}
	\bar\mu_{\rm large\text{-}r}(r)
	=
	A_\mu
	+
	\frac{B_\mu}{r^2}
	+
	\frac{C_\mu}{r^4}
	+
	O(r^{-6}),
	\label{eq:complexity-large-r-density-expansion}
\end{equation}
with the coefficients given in
eqs.~\eqref{eq:complexity-density-A},
\eqref{eq:complexity-density-C}. These coefficients determine the large-$r$
tail of the spacelike volume and are needed for cutoff-stable numerical
evaluation.

We use the primitive
\begin{equation}
	{\cal M}(r)
	=
	\int_{r_\ast}^{r}
	ds\,
	\bar\mu_{\rm large\text{-}r}(s),
	\label{eq:complexity-large-r-M}
\end{equation}
whose large-$r$ expansion has the form
\begin{equation}
	{\cal M}(R)
	=
	A_\mu R
	+
	{\cal M}_{\infty}^{\rm large\text{-}r}
	-
	\frac{B_\mu}{R}
	-
	\frac{C_\mu}{3R^3}
	+
	O(R^{-5}).
	\label{eq:complexity-large-r-M-expansion}
\end{equation}
The constant ${\cal M}_{\infty}^{\rm large\text{-}r}$ depends on the lower
edge $r_\ast$ and gives the finite part of the radial primitive after the
universal large-$R$ growth is removed.

The finite lifted volume follows from the general prescription
eq.~\eqref{eq:complexity-finite-volume}. In the present regime,
\begin{equation}
	\Delta V_{\rm fin}^{\rm large\text{-}r}(r_0)
	=
	T(r_0){\cal M}_{\infty}^{\rm large\text{-}r}
	+
	2I_{\rm sp}^{V,\,{\rm large\text{-}r}}(r_0)
	-
	2I_{\rm tim}^{V,\,{\rm large\text{-}r}}(r_0).
	\label{eq:complexity-large-r-finite-volume}
\end{equation}
Here the weighted branch integrals are evaluated using the asymptotic
kernels of section~\ref{sec:tee-asymptotic-limit} and the density
\eqref{eq:complexity-large-r-density}. The timelike term is finite because
the timelike branch ends at $r_0$. The spacelike term reaches the cutoff
and is completed by the analytic large-$r$ tail, as in
eq.~\eqref{eq:complexity-tail-improved-volume}. This gives a finite result
that is stable under changes of the numerical cutoff.

The corresponding complexity is
\begin{equation}
	C
	=
	\frac{
		\Delta V_{\rm fin}^{\rm large\text{-}r}(r_0)
	}{
		G_{10}L_{\rm AdS}
	}.
	\label{eq:complexity-large-r-C}
\end{equation}
In the plots we suppress the common positive normalization factor, as in the
rest of the complexity analysis.

\begin{figure}[H]
	\centering
	\includegraphics[width=0.48\textwidth]{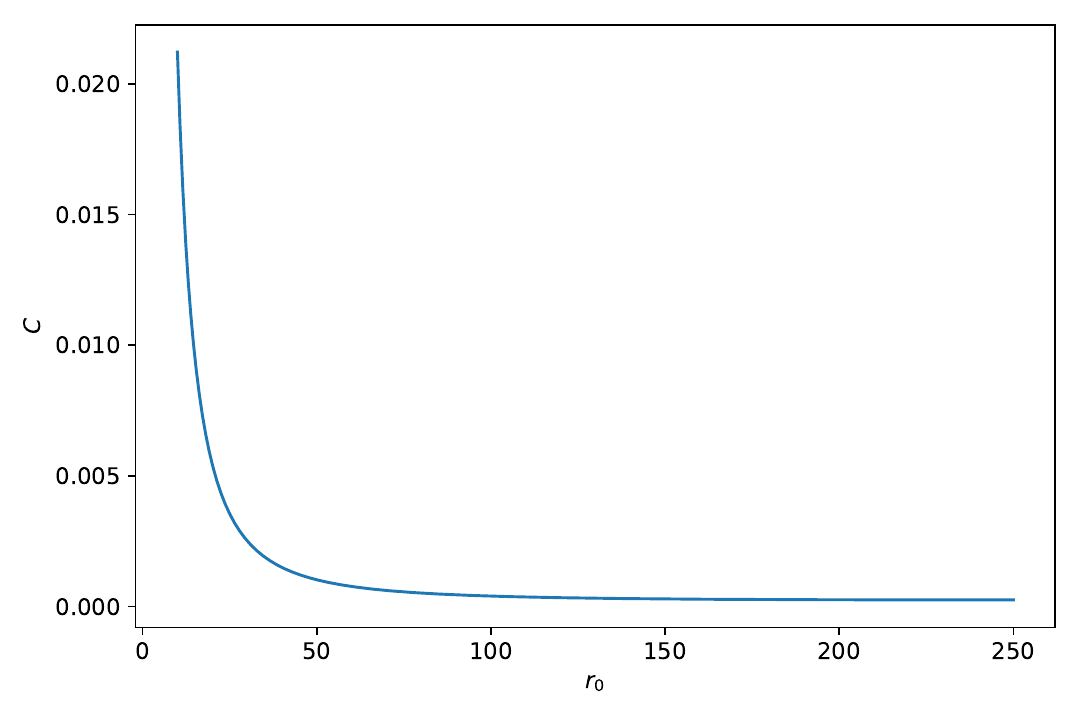}
	\hfill
	\includegraphics[width=0.48\textwidth]{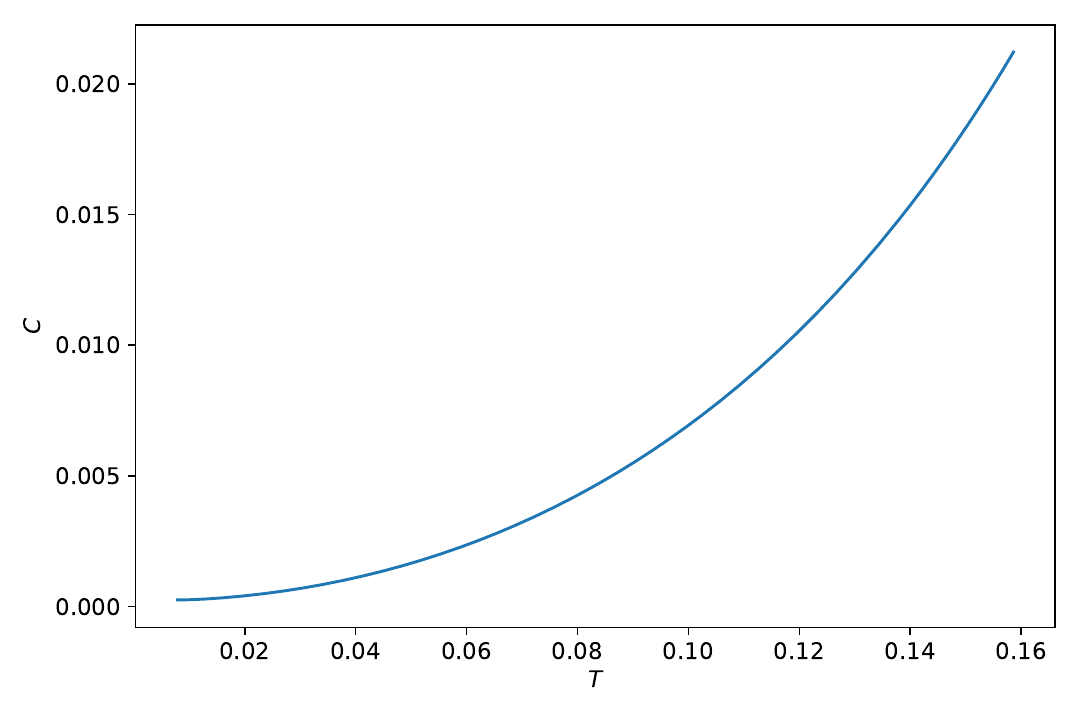}
	\caption{
		Finite timelike-complexity volume in the large-$r$ regime. Left:
		$C$ as a function of the turning point $r_0$.
		Right: the same quantity as a function of the boundary interval. The
		result is single-valued because the large-$r$ time map is single-valued.
	}
	\label{fig:complexity-large-r-complexity}
\end{figure}

Figure~\ref{fig:complexity-large-r-complexity} shows the short-interval
behaviour of the finite volume. As $r_0$ becomes large, the branch remains
near the asymptotic region and the subtracted spacelike and timelike
weighted volumes nearly cancel. The finite part therefore becomes small in
the limit $T\to0$.

This behaviour can be extracted analytically. Combining the large-$r_0$
relation between $T$ and $r_0$ from section~\ref{sec:tee-asymptotic-limit}
with the density in eq.~\eqref{eq:complexity-large-r-density}, the finite
volume behaves as
\begin{equation}
	\Delta V_{\rm fin}^{\rm large\text{-}r}(r_0)
	=
	Q^2(\ell_2^2-\ell^2)
	\frac{\log(r_0/r_\ast)}{r_0^2}
	+
	O(r_0^{-2}).
	\label{eq:complexity-large-r-volume-r0-final}
\end{equation}
Using
\begin{equation}
	r_0(T)
	=
	\frac{2\sqrt Q}{T}
	+
	O(1),
\end{equation}
this becomes
\begin{equation}
	\Delta V_{\rm fin}^{\rm large\text{-}r}(T)
	=
	\frac{Q(\ell_2^2-\ell^2)}{4}
	T^2
	\log
	\left(
	\frac{2\sqrt Q}{r_\ast T}
	\right)
	+
	O(T^2).
	\label{eq:complexity-large-r-volume-T-final}
\end{equation}
Therefore
\begin{equation}
	C_T^{\rm large\text{-}r}(T)
	=
	\frac{Q(\ell_2^2-\ell^2)}
	{4G_{10}L_{\rm AdS}}
	T^2
	\log
	\left(
	\frac{2\sqrt Q}{r_\ast T}
	\right)
	+
	O(T^2).
	\label{eq:complexity-large-r-small-T}
\end{equation}

Thus a short boundary interval is described by a branch whose turning point
lies far out in the large-$r$ region. The finite volume vanishes as
$T^2\log(1/T)$ after the UV subtraction. This confirms that the short-
interval complexity is insensitive to the localized core and to the
cap/horizon transition region.

The large-$r$ calculation therefore gives the asymptotic check of the
timelike-complexity prescription. It verifies the branch signs, the density
tail, and the finite-volume subtraction. The effects specific to the
localized black pole require the exact functions $K_y(r,\theta)$ and
$G(r,\theta)$: once they are restored, the branch kernels depend on
$\theta_0$, the time map can become non-monotonic, and the
fixed-boundary-interval selection becomes a genuine two-parameter problem.

\subsection{Timelike complexity in the exact black-pole geometry}
\label{sec:complexity-exact-branch-structure}

We now restore the exact black-pole functions in the timelike-complexity
calculation. The large-$r$ regime of
section~\ref{sec:complexity-large-r-reference} gives the short-boundary-
interval limit and fixes the UV subtraction. In that regime the leading
angular dependence of $K_y(r,\theta)$ and $G(r,\theta)$ drops out. Here we
keep the exact functions. The branch kernels therefore depend on the
angular label $\theta_0$, while the lifted density $\bar\mu(r)$ contains the
integral over the physical internal angle $\theta$.

The Lorentzian branch construction is the same as in the exact
timelike-entanglement calculation of section~\ref{sec:tee-exact-black-pole}.
We use the reduced metric functions in
eqs.~\eqref{eq:reduced-two-dimensional-metric}--\eqref{eq:H-definition},
the turning-point value $F_0=F(r_0;\theta_0)$, and the exact branch kernels
$K_{\rm tim}$ and $K_{\rm sp}$ in
eqs.~\eqref{eq:exact-tee-Ktim} and \eqref{eq:exact-tee-Ksp}. The boundary
interval is again determined by the exact time map
eq.~\eqref{eq:exact-tee-boundary-interval}. The new quantity is the finite
renormalized lifted volume $C_T^{\rm lift}$ defined in
eq.~\eqref{eq:complexity-Clift-final}. Thus each pair $(r_0,\theta_0)$
defines
\begin{equation}
	(r_0,\theta_0)
	\longrightarrow
	\left(
	T(r_0,\theta_0),
	C_T^{\rm lift}(r_0,\theta_0)
	\right).
	\label{eq:complexity-exact-family-map}
\end{equation}
This is the branch-family map before imposing a common boundary interval.

For the complexity calculation we use
\begin{equation}
	x_E=0.6,
	\qquad
	\sigma=-1,
	\qquad
	Q_1=Q_5=R_y=1.
	\label{eq:complexity-exact-parameters}
\end{equation}
This numerical choice is different from the one used in the
timelike-entanglement plots. The parameter $x_E$ fixes the black-pole
length scales through $\tau$ and changes the transition angle
$\theta_\star$. The cap-side sector is $\theta_0<\theta_\star$, and the
horizon-side sector is $\theta_0>\theta_\star$.

The endpoint behaviour follows from the exact timelike-entanglement
analysis. On the cap side, the branch profiles have regular lower-endpoint
behaviour. On the horizon side, individual profiles can contain logarithmic
terms near the lower endpoint, but the leading endpoint terms cancel in the
boundary interval $T(r_0,\theta_0)$. Hence a large vertical extent in a
$t$-$r$ plot does not by itself imply an infinite boundary interval.

The turning point $r_0$ fixes where the timelike branch closes. Changing
$r_0$ changes the radial part of the geometry followed by the branch.
Changing $\theta_0$ changes the effective branch kernels through
$K_y(r,\theta_0)$ and $G(r,\theta_0)$. These two labels therefore affect
the finite volume in different ways.

\begin{figure}[H]
	\centering
	\begin{subfigure}{0.48\textwidth}
		\centering
		\includegraphics[width=\textwidth]{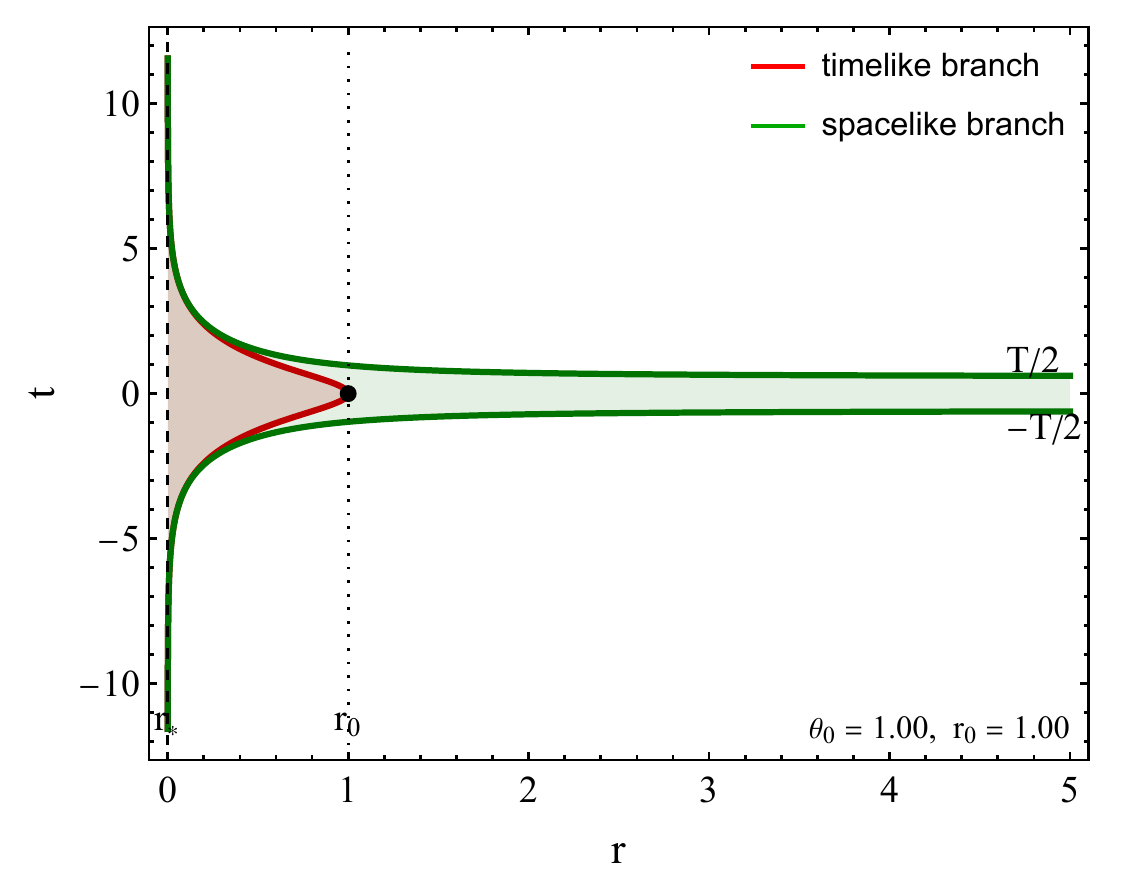}
		\caption{}
		\label{fig:complexity-exact-tr-loc1}
	\end{subfigure}
	\hfill
	\begin{subfigure}{0.48\textwidth}
		\centering
		\includegraphics[width=\textwidth]{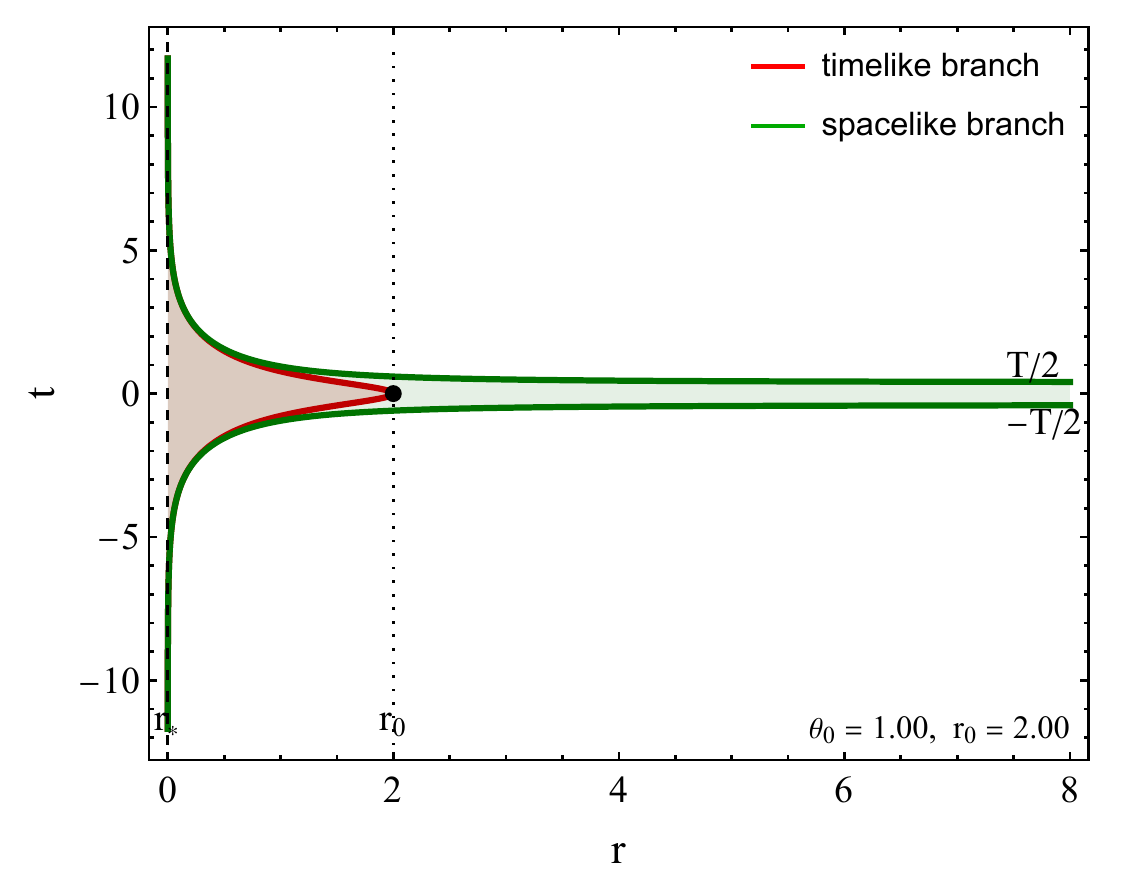}
		\caption{}
		\label{fig:complexity-exact-tr-loc2}
	\end{subfigure}
	
	\medskip
	
	\begin{subfigure}{0.48\textwidth}
		\centering
		\includegraphics[width=\textwidth]{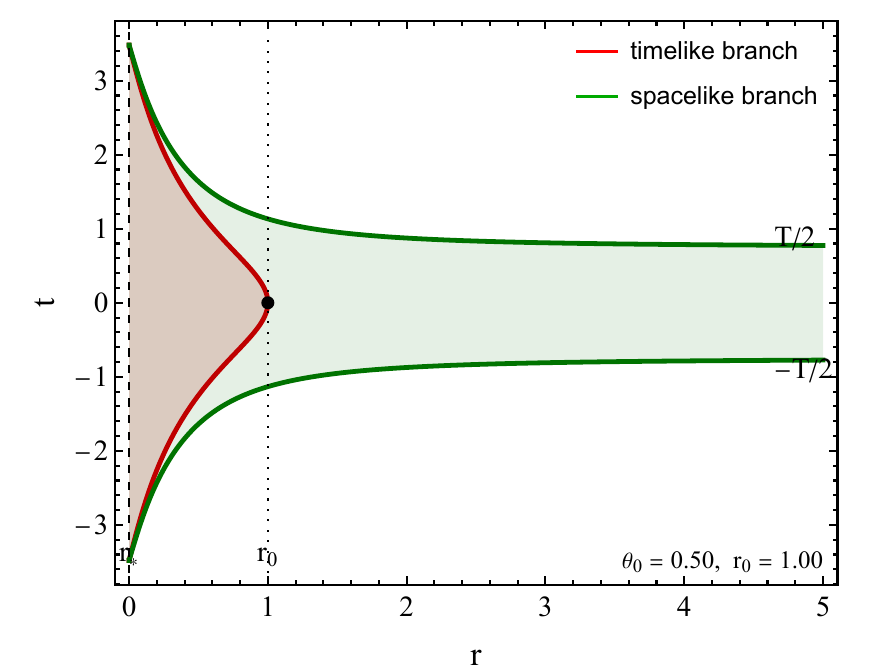}
		\caption{}
		\label{fig:complexity-exact-tr-loc3}
	\end{subfigure}
	\hfill
	\begin{subfigure}{0.48\textwidth}
		\centering
		\includegraphics[width=\textwidth]{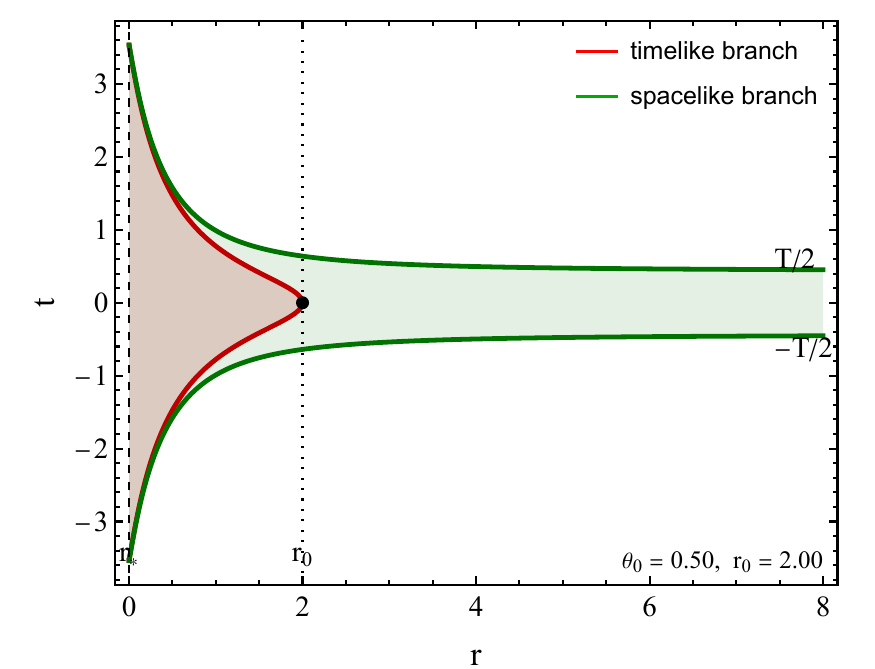}
		\caption{}
		\label{fig:complexity-exact-tr-loc4}
	\end{subfigure}
	\caption{
		Exact $t$-$r$ branch profiles before fixed-boundary-interval selection.
		The timelike branch ends at $r_0$, while the spacelike branch reaches the
		cutoff. The panels show how the branch geometry changes when $r_0$ and
		$\theta_0$ are varied.
	}
	\label{fig:complexity-exact-tr-branch-structure}
\end{figure}

Figure~\ref{fig:complexity-exact-tr-branch-structure} illustrates the two
roles of the branch labels. The angular label controls the effective
geometry through $K_y(r,\theta_0)$ and $G(r,\theta_0)$, while the turning
point controls where the timelike branch closes. These profiles are not yet
candidate saddles for the same boundary interval. They are branch-family
data. The fixed-boundary comparison is performed only after solving \eqref{eq:complexity-localized-fixedT}.

We next examine the time map. In the large-$r$ regime, the relation between
$r_0$ and $T$ is single-valued. In the exact black-pole geometry this need
not hold. At fixed $\theta_0$, the curve $T(r_0,\theta_0)$ can increase,
reach a maximum and then decrease. The same boundary interval can then be
represented by two radial branches.

For each angular label we define
\begin{equation}
	r_{0,\rm peak}(\theta_0)
	=
	\underset{r_0}{\operatorname{arg\,max}}\,
	T(r_0,\theta_0),
	\qquad
	T_{\rm max}(\theta_0)
	=
	T(r_{0,\rm peak}(\theta_0),\theta_0).
	\label{eq:complexity-exact-r0peak-Tmax}
\end{equation}
The smaller-$r_0$ and larger-$r_0$ branches are the two sides of this
maximum:
\begin{equation}
	r_0<r_{0,\rm peak}(\theta_0)
	\quad
	\hbox{smaller-}r_0\hbox{ branch},
	\qquad
	r_0>r_{0,\rm peak}(\theta_0)
	\quad
	\hbox{larger-}r_0\hbox{ branch}.
	\label{eq:complexity-small-large-branches}
\end{equation}
These names only specify the side of the peak. They do not by themselves
determine which branch is selected.

\begin{figure}[H]
	\centering
	\begin{subfigure}{0.48\textwidth}
		\centering
		\includegraphics[width=\textwidth]{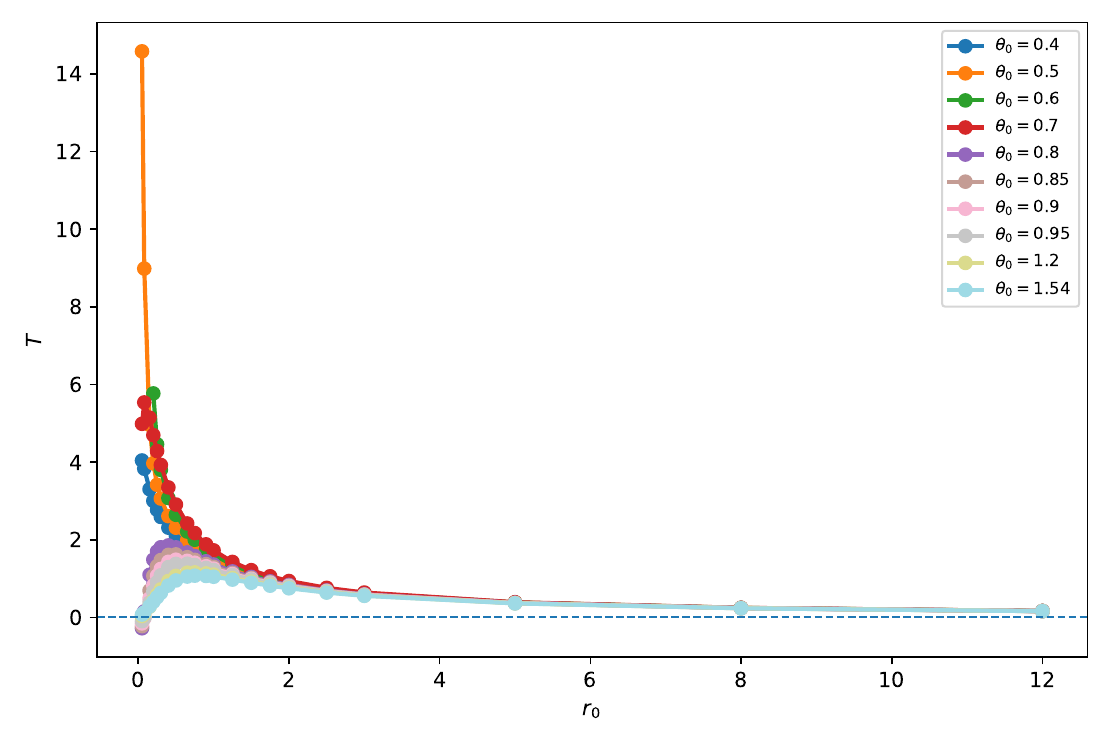}
		\caption{}
		\label{fig:complexity-temporal-families}
	\end{subfigure}
	\hfill
	\begin{subfigure}{0.48\textwidth}
		\centering
		\includegraphics[width=\textwidth]{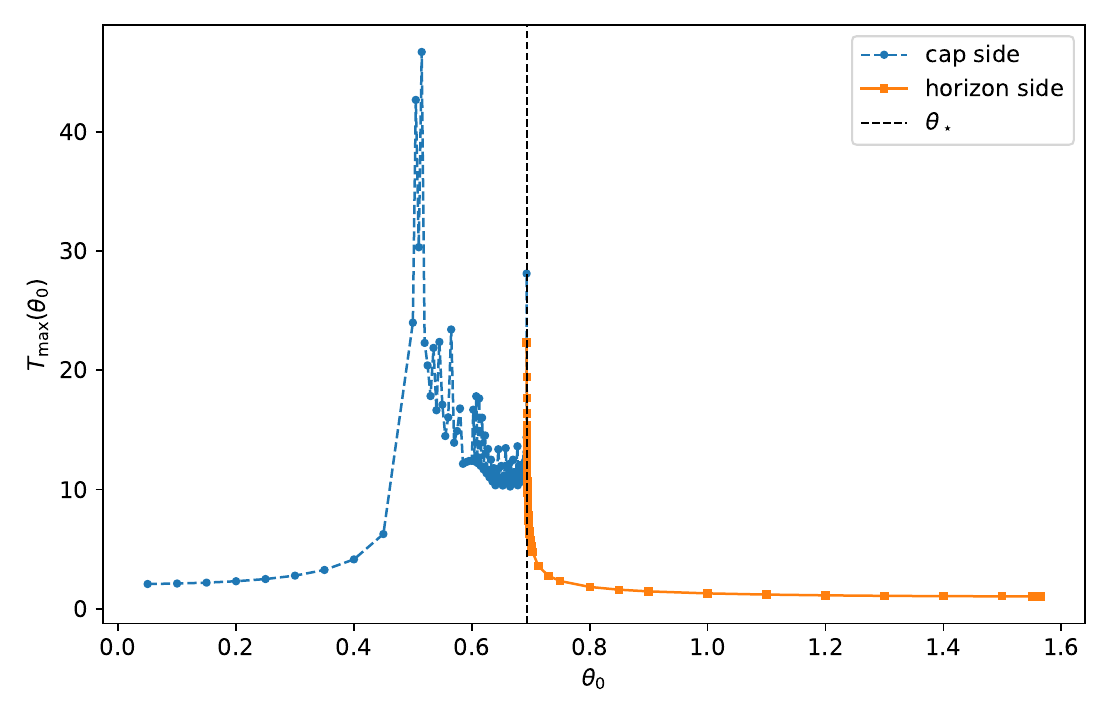}
		\caption{}
		\label{fig:complexity-Tmax-theta}
	\end{subfigure}
	\caption{
		Exact time map for the complexity choice $x_E=0.6$. Left:
		$T(r_0,\theta_0)$ for representative angular labels. Right:
		$T_{\rm max}(\theta_0)$ as a function of the angular label. The
		enhancement near $\theta_\star$ shows that angular families close to the
		cap/horizon transition region can support larger boundary intervals.
	}
	\label{fig:complexity-exact-temporal-structure}
\end{figure}

Figure~\ref{fig:complexity-exact-temporal-structure} shows why the
fixed-boundary-interval step is necessary. For a given target interval, an
angular family may have two radial roots. Both roots must be kept until the
finite volumes are compared. The right panel gives the angular
accessibility condition
\begin{equation}
	T_{\rm target}\leq T_{\rm max}(\theta_0).
	\label{eq:complexity-time-accessibility}
\end{equation}
Only angular labels satisfying this inequality can contribute at the chosen
target interval. The volume selection is made after this admissible branch
family has been constructed.

We now evaluate the finite lifted complexity on the exact branch family
before imposing a common boundary interval. For each $(r_0,\theta_0)$, the
boundary interval is computed from eq.~\eqref{eq:exact-tee-boundary-interval}
and the finite renormalized volume from eq.~\eqref{eq:complexity-Clift-final}.

\begin{figure}[H]
	\centering
	\begin{subfigure}{0.48\textwidth}
		\centering
		\includegraphics[width=\textwidth]{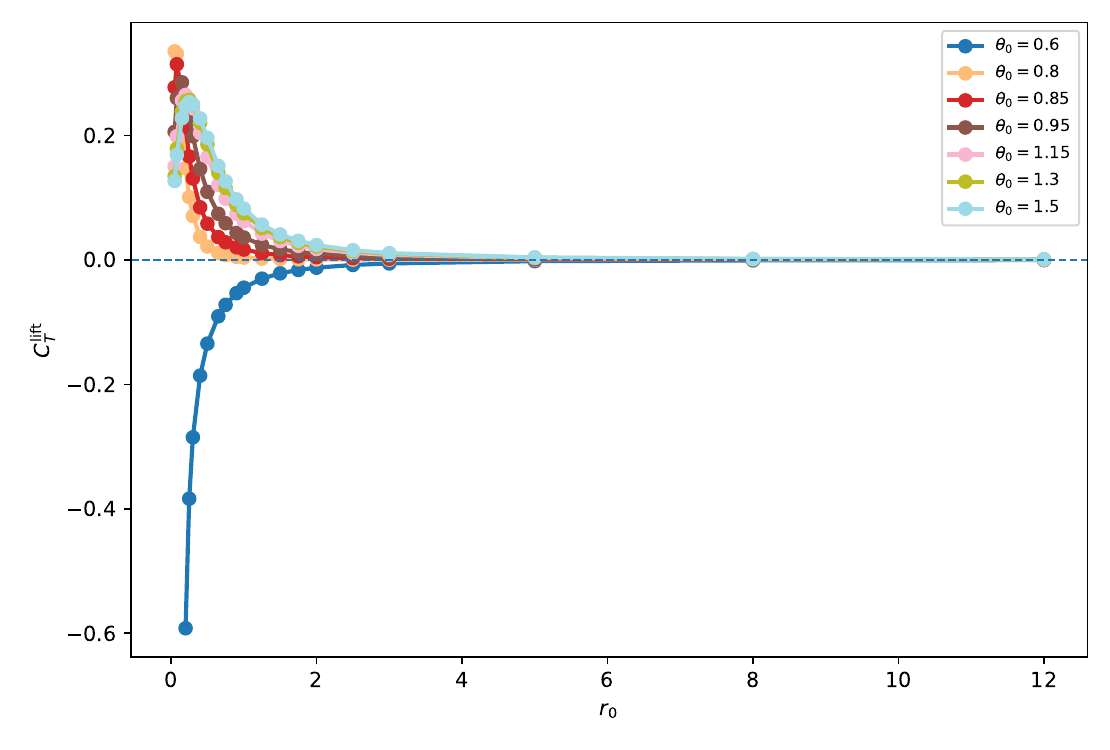}
		\caption{}
		\label{fig:complexity-unselected-r0}
	\end{subfigure}
	\hfill
	\begin{subfigure}{0.48\textwidth}
		\centering
		\includegraphics[width=\textwidth]{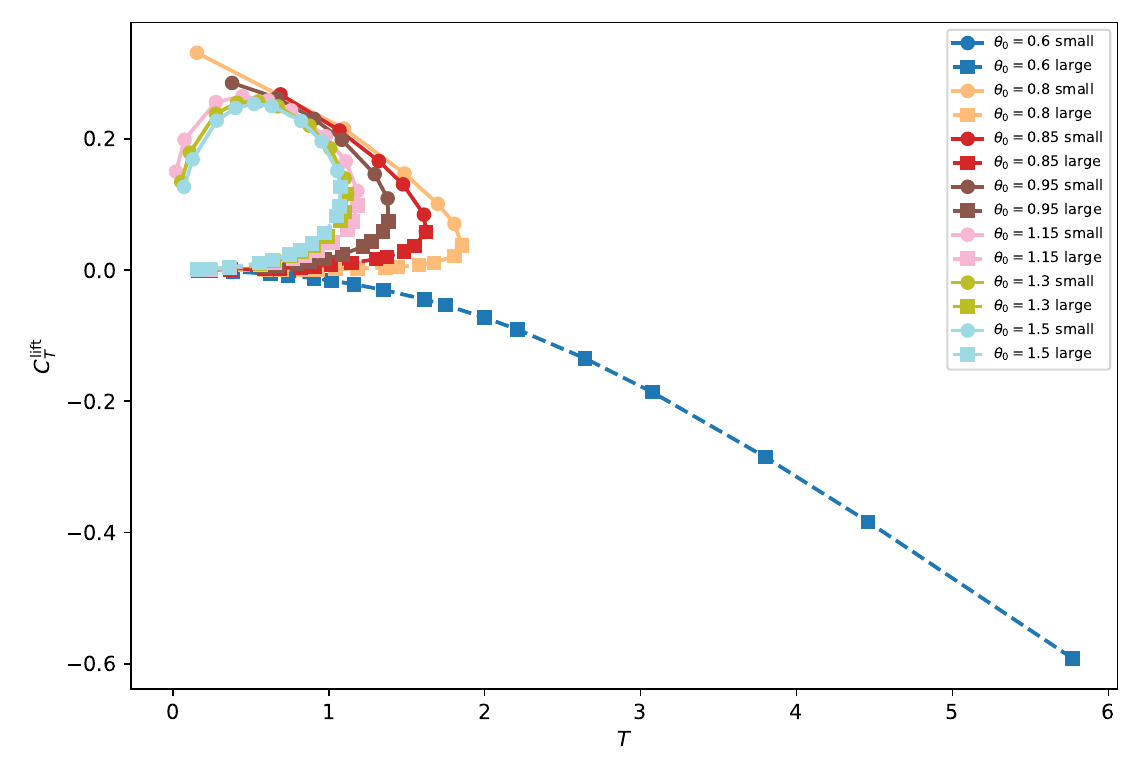}
		\caption{}
		\label{fig:complexity-unselected-T}
	\end{subfigure}
	\caption{
		Finite renormalized timelike complexity on the exact branch family before
		fixed-boundary-interval selection. Left: $C_T^{\rm lift}$ as a function
		of $r_0$ at fixed angular label. Right: the same branch-family values
		displayed against the corresponding boundary interval
		$T(r_0,\theta_0)$.
	}
	\label{fig:complexity-exact-unselected-complexity}
\end{figure}

Figure~\ref{fig:complexity-exact-unselected-complexity} shows the
unselected finite volume. For branches with large $r_0$, the result follows
the large-$r$ short-interval behaviour. When the turning point moves
inward, the branch kernels become sensitive to the exact functions
$K_y(r,\theta_0)$ and $G(r,\theta_0)$. The finite volume then develops a
clear angular and radial branch dependence. The lifted density
$\bar\mu(r)$ already contains the integration over the physical internal
angle $\theta$, so the remaining angular dependence in these curves enters
through the branch geometry.

The plot against $T$ makes the main point explicit. Before selection,
$C_T^{\rm lift}$ is not a single-valued function of the boundary interval.
The same $T$ can arise from different radial branches and angular labels,
and these branches can give different finite volumes. Therefore the curves
in figure~\ref{fig:complexity-exact-unselected-complexity} are not yet the
physical selected complexity.

The conclusion of this subsection is that the exact black-pole complexity
problem is genuinely two-parameter before selection. The angular label
chooses the effective branch family, while the turning point fixes the
radial extent of the timelike branch. Because the exact time map is
non-monotonic, the finite lifted complexity becomes multi-valued as a
function of $T$ before the fixed-boundary-interval condition is imposed.
The next subsection performs that selection.

\subsection{Fixed-boundary-interval minimisation for timelike complexity}
\label{sec:complexity-fixed-boundary-minimisation}

We now perform the fixed-boundary-interval minimisation for timelike
subregion complexity. The exact branch family was described in
section~\ref{sec:complexity-exact-branch-structure}. For each pair
$(r_0,\theta_0)$, the Lorentzian branches determine both the boundary
interval $T(r_0,\theta_0)$ and the finite renormalized volume
$C_T^{\rm lift}(r_0,\theta_0)$. These volumes cannot be compared at
arbitrary points in the $(r_0,\theta_0)$ plane, because different points
usually correspond to different boundary intervals.

The comparison is therefore made in two steps. First, for a prescribed
target interval $T_{\rm target}$, we impose
\begin{equation}
	T(r_0,\theta_0)=T_{\rm target}.
	\label{eq:complexity-fixed-boundary-selection}
\end{equation}
This equation is solved for every admissible angular label and for every
radial branch. Second, the finite renormalized volumes are compared only
inside this fixed-boundary-interval family. This is the same logic used in
the timelike-entanglement analysis of
section~\ref{sec:tee-fixed-boundary-interval}. The difference is the
quantity being minimized: timelike entanglement minimizes the real part of a
complex lifted area, while timelike complexity minimizes the real finite
volume $C_T^{\rm lift}$.

In the exact localized geometry, the temporal family
$T(r_0,\theta_0)$ can be non-monotonic. Hence
eq.~\eqref{eq:complexity-fixed-boundary-selection} can have two radial roots
at the same angular label. We denote these roots by
\begin{equation}
	r_0=r_{0,b}(T_{\rm target},\theta_0),
	\qquad
	b\in\{\mathrm{smaller},\mathrm{larger}\}.
	\label{eq:complexity-fixed-roots}
\end{equation}
The labels ``smaller'' and ``larger'' refer to the two sides of the maximum
of $T(r_0,\theta_0)$. The smaller-$r_0$ branch lies before the peak of the
temporal family, while the larger-$r_0$ branch lies after the peak. This is
only a branch classification; it is not a statement about the absolute size
of the selected turning point.

On every fixed-boundary branch we evaluate
\begin{equation}
	C_{T,b}^{\rm lift}(T_{\rm target},\theta_0)
	=
	C_T^{\rm lift}
	\left(
	r_{0,b}(T_{\rm target},\theta_0),
	\theta_0
	\right).
	\label{eq:complexity-fixed-T-branch-C}
\end{equation}
The selected finite complexity is
\begin{equation}
	C_T^{\rm selected}(T_{\rm target})
	=
	\min_{\theta_0,b}
	C_{T,b}^{\rm lift}(T_{\rm target},\theta_0),
	\label{eq:complexity-selected-C}
\end{equation}
where the minimisation is restricted to branches that exist at
$T_{\rm target}$. In particular, the angular label must satisfy
\begin{equation}
	T_{\rm target}\leq T_{\rm max}(\theta_0).
	\label{eq:complexity-admissibility-fixedT}
\end{equation}
The selected branch is described by
\begin{equation}
	\left(
	\theta_0^\ast(T_{\rm target}),
	r_0^\ast(T_{\rm target}),
	b^\ast(T_{\rm target})
	\right).
	\label{eq:complexity-selected-branch-data}
\end{equation}

\begin{figure}[H]
	\centering
	\begin{subfigure}[t]{0.25\textwidth}
		\centering
		\includegraphics[width=\linewidth]{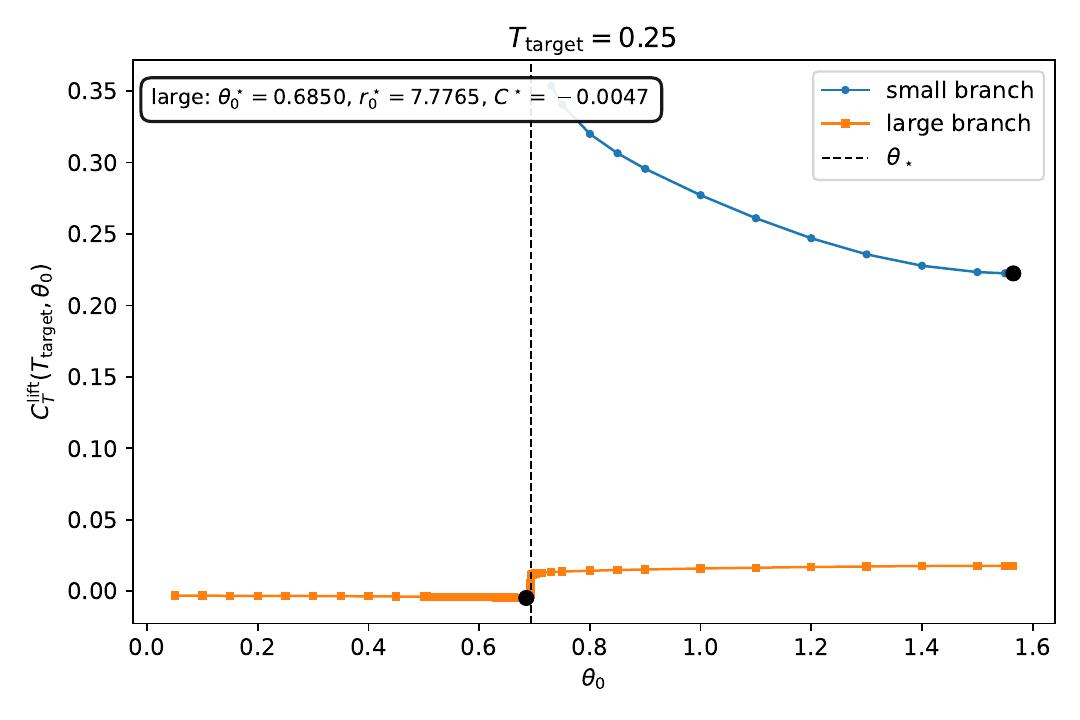}
	\end{subfigure}
	\hfill
	\begin{subfigure}[t]{0.25\textwidth}
		\centering
		\includegraphics[width=\linewidth]{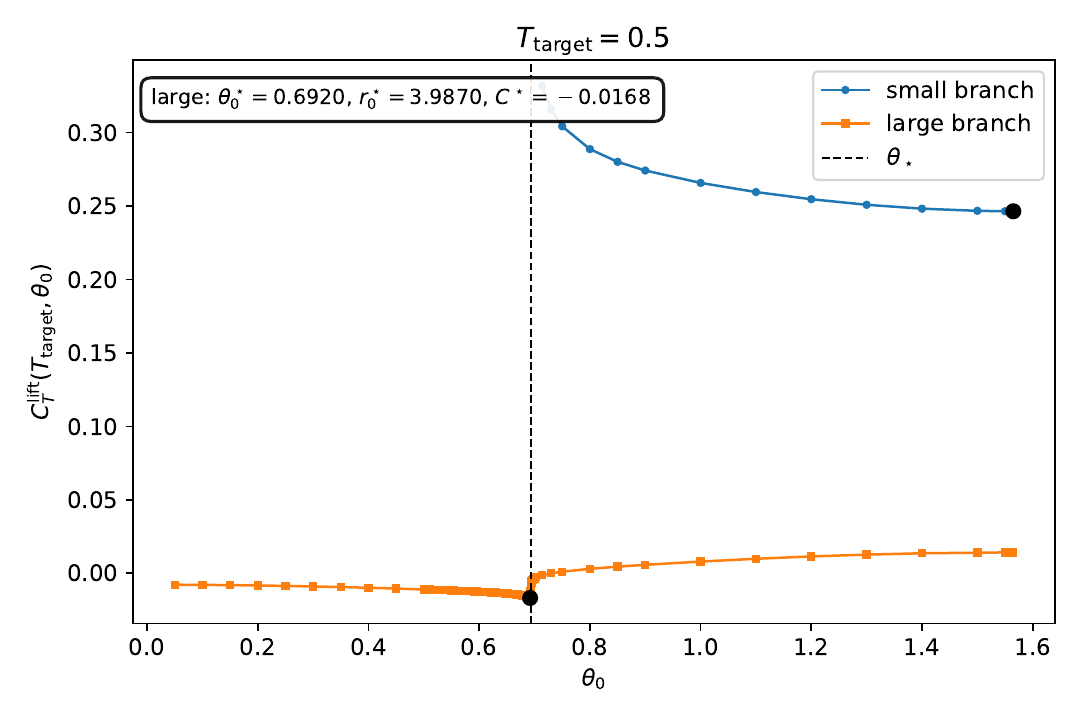}
	\end{subfigure}
	\hfill
	\begin{subfigure}[t]{0.25\textwidth}
		\centering
		\includegraphics[width=\linewidth]{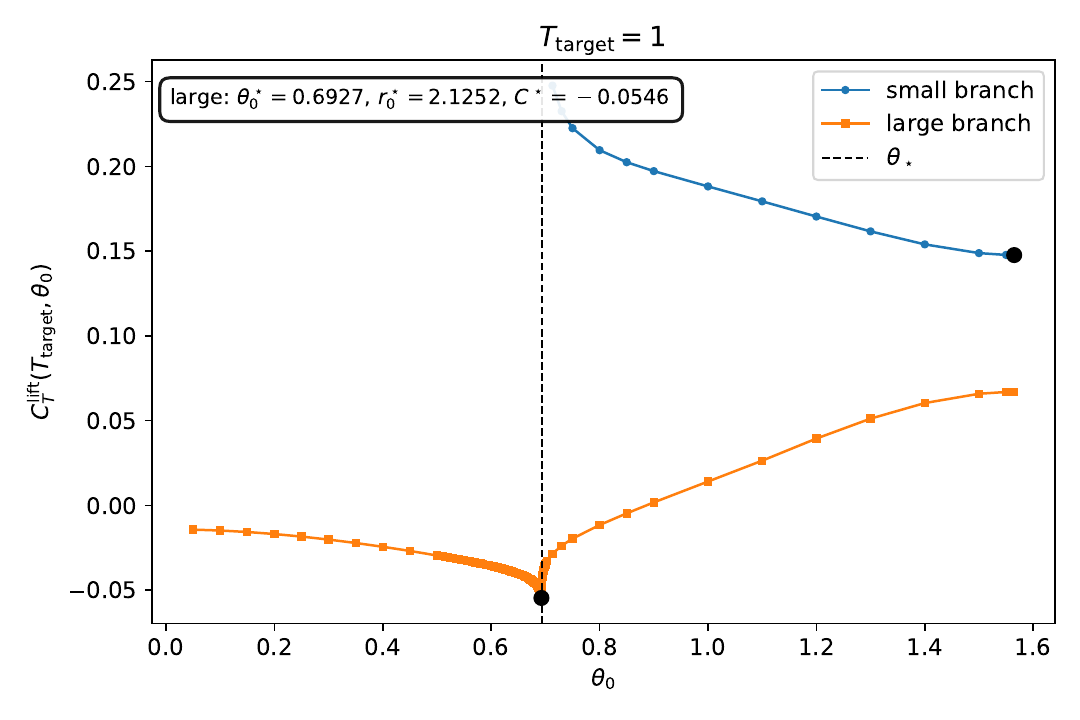}
	\end{subfigure}
	
	\medskip
	
	\begin{subfigure}[t]{0.25\textwidth}
		\centering
		\includegraphics[width=\linewidth]{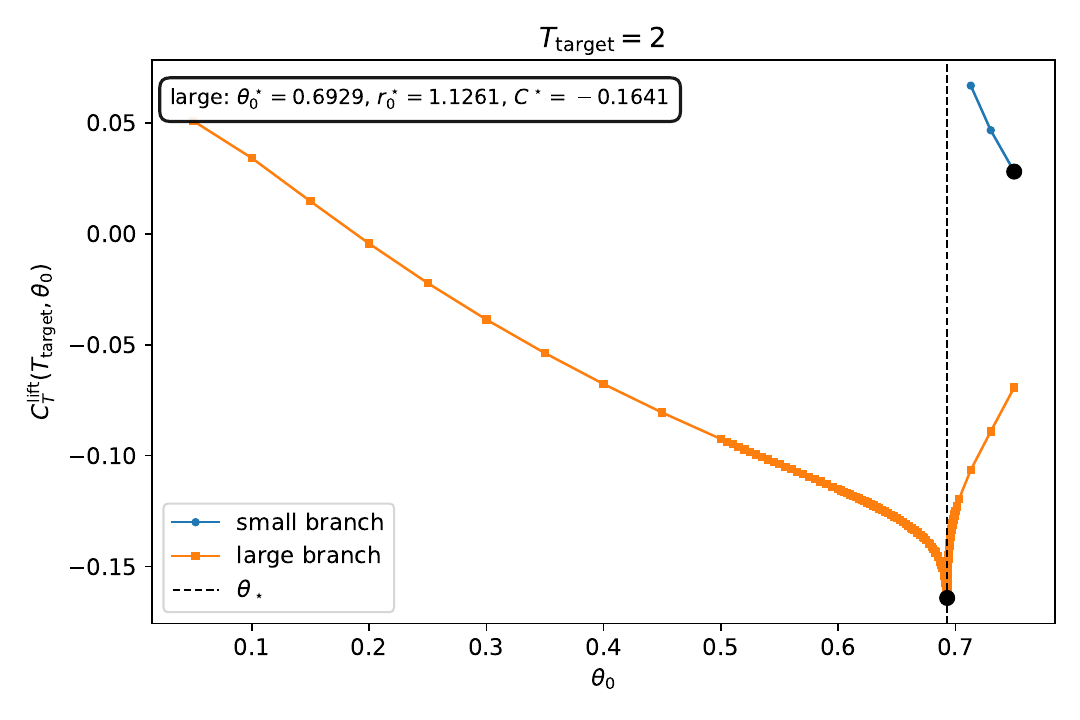}
	\end{subfigure}
	\hfill
	\begin{subfigure}[t]{0.25\textwidth}
		\centering
		\includegraphics[width=\linewidth]{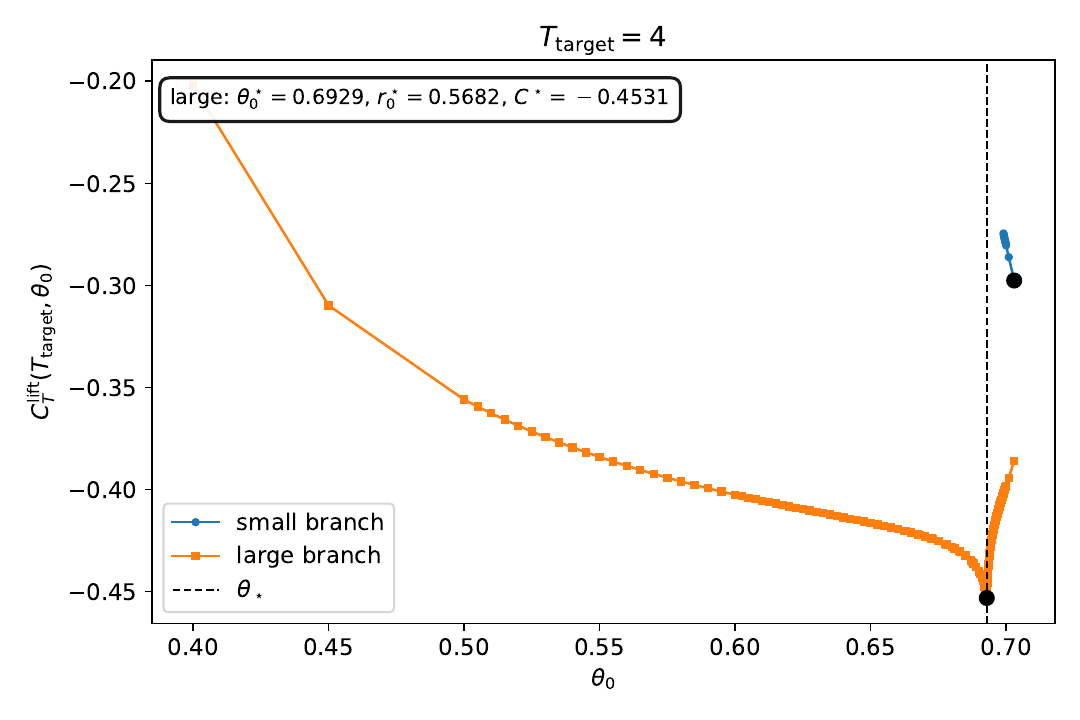}
	\end{subfigure}
	\hfill
	\begin{subfigure}[t]{0.25\textwidth}
		\centering
		\includegraphics[width=\linewidth]{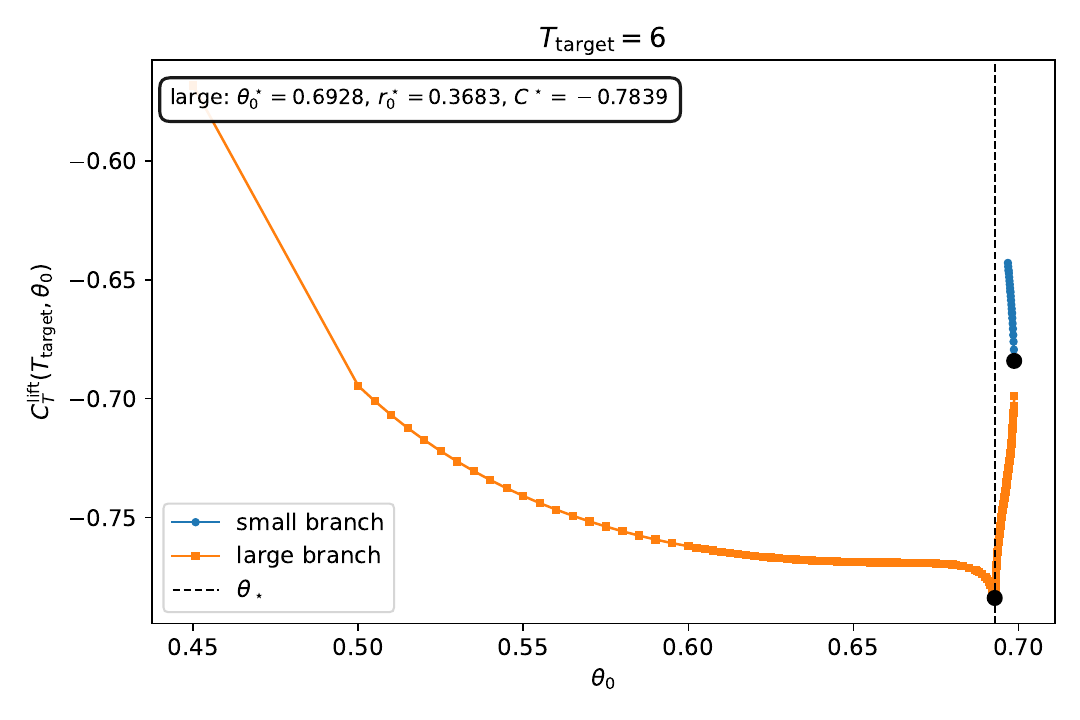}
	\end{subfigure}
	
	\medskip
	
	\begin{subfigure}[t]{0.25\textwidth}
		\centering
		\includegraphics[width=\linewidth]{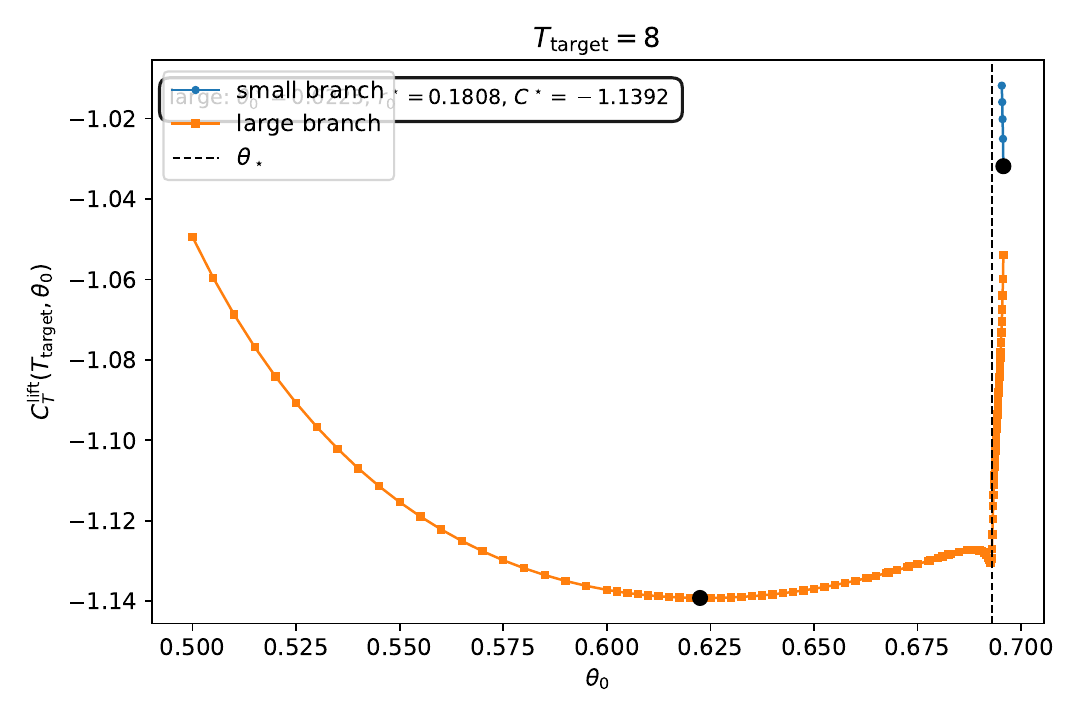}
	\end{subfigure}
	\hfill
	\begin{subfigure}[t]{0.25\textwidth}
		\centering
		\includegraphics[width=\linewidth]{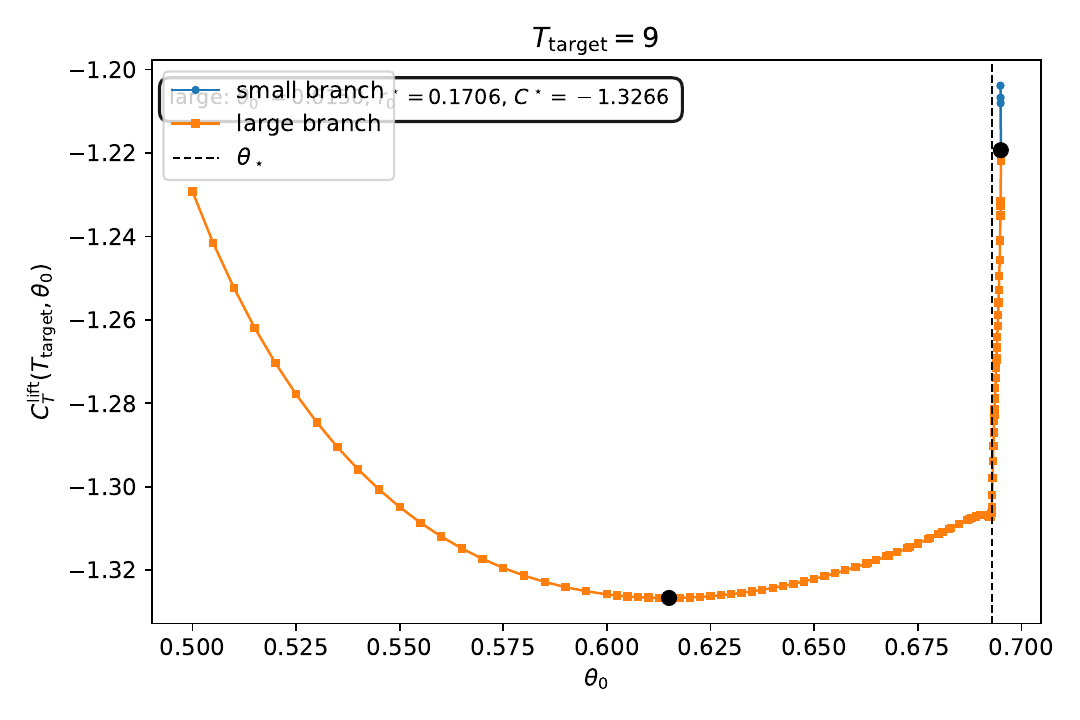}
	\end{subfigure}
	\hfill
	\begin{subfigure}[t]{0.25\textwidth}
		\centering
		\includegraphics[width=\linewidth]{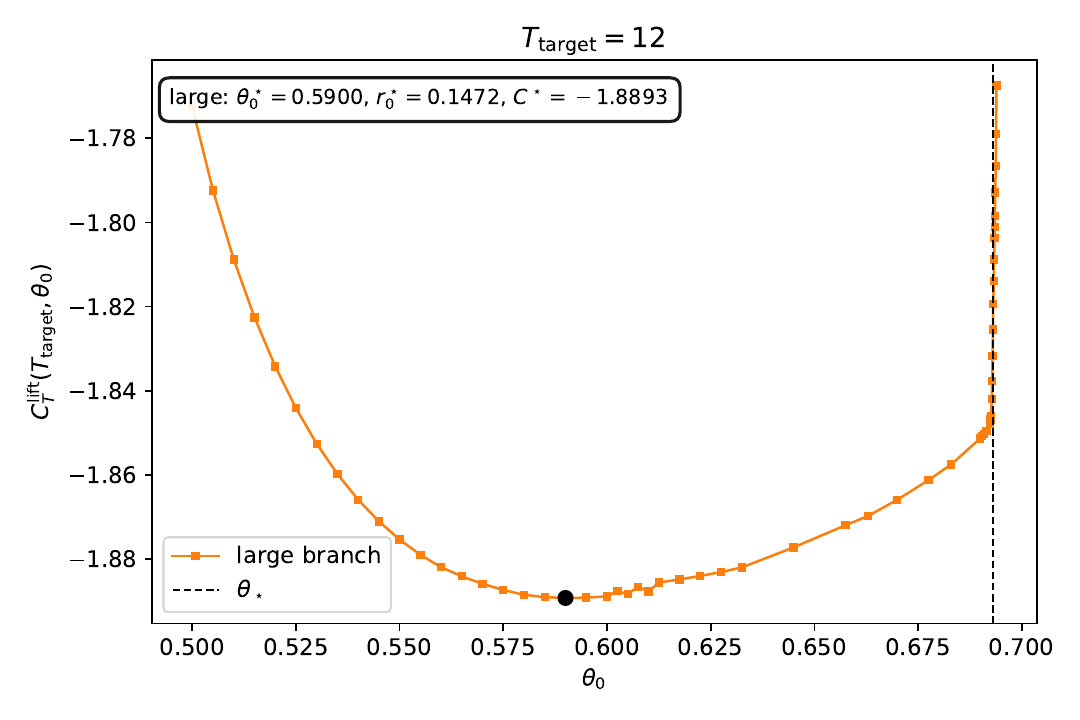}
	\end{subfigure}
	\caption{
		Fixed-boundary-interval minimisation of the exact black-pole timelike
		complexity for $x_E=0.6$. Each curve is obtained after imposing
		$T(r_0,\theta_0)=T_{\rm target}$. The smaller-$r_0$ and larger-$r_0$
		branches are compared only at the same target interval. The black marker
		denotes the minimum of the finite renormalized volume. Empty angular
		ranges indicate that no root of the time equation exists.
	}
	\label{fig:complexity-fixed-T-minimisation}
\end{figure}

Figure~\ref{fig:complexity-fixed-T-minimisation} shows the fixed-boundary
minimisation. The time equation first decides which angular labels and
radial branches are available at a given $T_{\rm target}$. For small target
intervals, many angular labels solve the time equation. As the target
interval increases, the condition
$T_{\rm target}\leq T_{\rm max}(\theta_0)$ removes more angular labels, and
the admissible range is concentrated near the cap/horizon transition
region. This restriction comes from the Lorentzian time map itself, before
the finite volume is minimized.

The local reason for this angular restriction is explained in
appendix~\ref{app:transition-region}. Near the transition angle, the
crossover scale behaves as
\begin{equation}
	r_c(\theta_0)\simeq \ell|\theta_0-\theta_\star|.
\end{equation}
When $\theta_0$ approaches $\theta_\star$, this scale becomes small and the
transition radial interval
\begin{equation}
	r_c\ll r\ll r_{\rm loc}
\end{equation}
becomes long. In this interval,
\begin{equation}
	K_{\rm tim}\simeq \frac{C_{\rm tim}}{r},
	\qquad
	K_{\rm sp}\simeq \frac{C_{\rm sp}}{r},
	\qquad
	C_{\rm tim}>C_{\rm sp}.
\end{equation}
Therefore $I_{\rm tim}-I_{\rm sp}$ receives a positive logarithmic
contribution. This is why angular labels close to $\theta_\star$ can
support larger target intervals as shown in Figure \ref{fig:complexity-exact-temporal-structure}.

The cap-side interpretation is subtle but important. The cap-side deep core
does not itself generate the logarithmic enhancement: there the branch
kernels remain finite. However, when $\theta_0$ approaches $\theta_\star$
from the cap side, the crossover scale $r_c$ becomes small and a long
transition-dominated interval opens before the deep cap region is reached.
This transition interval gives the same type of logarithmic enhancement as
on the horizon side. Thus the cap deep core is not the source of the large
time, but the cap side near $\theta_\star$ contains the transition region
that supports large accessible boundary intervals.

After the admissible branches have been found, the finite volume selects the
saddle. At fixed $T_{\rm target}$, the term
$T_{\rm target}{\cal M}_\infty$ in
eq.~\eqref{eq:complexity-Clift-final} is common to all candidates. The
branch dependence of the minimisation is therefore carried by
\begin{equation}
	2I_{\rm sp}^{V}(T_{\rm target},\theta_0)
	-
	2I_{\rm tim}^{V}(T_{\rm target},\theta_0).
	\label{eq:complexity-angular-volume-comparison}
\end{equation}
Thus an angular family that can support a large target interval is not
automatically selected. The selected branch is determined by the finite
renormalized volume after the boundary interval has been fixed.

The minimisation shows that the selected angular label remains close to the
cap/horizon transition region in the displayed range. At intermediate
target intervals it lies slightly on the cap side of the transition angle,
\begin{equation}
	\theta_0^\ast(T_{\rm target})
	\simeq
	\theta_\star^{-}.
	\label{eq:complexity-transition-selected}
\end{equation}
As the target interval is increased, the minimum remains near the transition
region and can approach it from the cap side. It should not be interpreted
as a saddle moving into the deep cap core. This behaviour is related to the
timelike-entanglement selection because both observables use the same time
map, but the final minimum is different because the quantity minimized here
is a finite volume rather than the real part of a lifted area.

It is useful to separate the two radial branches before taking the final
minimum. Define
\begin{subequations}
	\begin{align}
		C_{\rm smaller}^{\ast}(T_{\rm target})
		&=
		\min_{\theta_0}
		C_{T,\rm smaller}^{\rm lift}(T_{\rm target},\theta_0),
		\label{eq:complexity-Csmaller-star}
		\\
		C_{\rm larger}^{\ast}(T_{\rm target})
		&=
		\min_{\theta_0}
		C_{T,\rm larger}^{\rm lift}(T_{\rm target},\theta_0).
		\label{eq:complexity-Clarger-star}
	\end{align}
\end{subequations}
The selected value is then
\begin{equation}
	C_T^{\rm selected}(T_{\rm target})
	=
	\min
	\left[
	C_{\rm smaller}^{\ast}(T_{\rm target}),
	C_{\rm larger}^{\ast}(T_{\rm target})
	\right].
	\label{eq:complexity-selected-branch-resolved}
\end{equation}

\begin{figure}[H]
	\centering
	\includegraphics[width=0.78\textwidth]{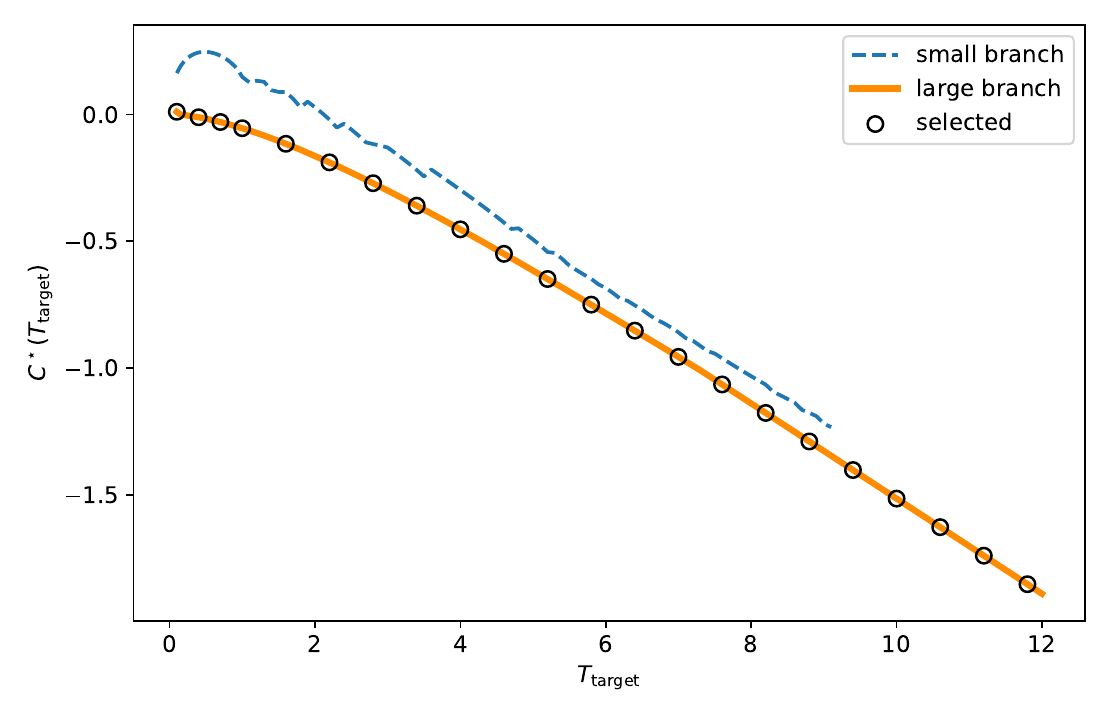}
	\caption{
		Branch-resolved selected timelike complexity. The smaller-$r_0$ and
		larger-$r_0$ branches are first minimized over the angular label
		separately. The selected curve is the lower of the two branch values. In
		the displayed range, the selected saddle lies on the larger-$r_0$ branch.
	}
	\label{fig:complexity-branch-resolved-selected}
\end{figure}

Figure~\ref{fig:complexity-branch-resolved-selected} shows that the selected
saddle remains on the larger-$r_0$ branch throughout the displayed range.
This statement is relative to the peak of the temporal family. Even when
$r_0^\ast$ becomes numerically smaller at larger $T_{\rm target}$, the
selected branch is still the larger branch if
\begin{equation}
	r_0^\ast(T_{\rm target})
	\geq
	r_{0,\rm peak}
	\left(
	\theta_0^\ast(T_{\rm target})
	\right).
	\label{eq:complexity-post-peak-condition}
\end{equation}
Thus the selected saddle can move inward while remaining on the larger
branch.

The behaviour of $C_T^{\rm selected}$ should be read within the
timelike-subregion prescription used here. The localized black-pole
solution is described in the exterior patch relevant for our branch
construction; the horizon interior is not part of the data used in this
finite-volume observable. This is different from the usual global BTZ
complexity intuition, where the main nontrivial volume growth is associated
with the black-hole interior. In the BTZ exterior patch, after the
corresponding subtraction, there is no independent exterior
contribution to the complexity. In the localized black pole, by contrast, the exterior patch
itself carries nontrivial structure through $K_y(r,\theta)$ and
$G(r,\theta)$. It is therefore meaningful that the renormalized
subregion volume receives a nonzero contribution from this exterior
localized geometry.

This finite quantity is
\begin{equation}
	C_T^{\rm lift}
	\propto
	T{\cal M}_\infty
	+
	2I_{\rm sp}^{V}
	-
	2I_{\rm tim}^{V}.
\end{equation}
The spacelike and timelike weighted volumes are positive separately, but
they enter the renormalized observable with opposite signs after the
asymptotic subtraction. As $T_{\rm target}$ is increased, the selected
values of $r_0^\ast$ and $\theta_0^\ast$ change, and the two branch volumes
are evaluated over different radial parts of the exact localized geometry.
A decrease or slow variation of $C_T^{\rm selected}$ should therefore be
understood as the behaviour of this finite exterior-patch branch-volume
difference, not as a claim about a global black-hole complexity.

\begin{figure}[H]
	\centering
	\includegraphics[width=\textwidth]{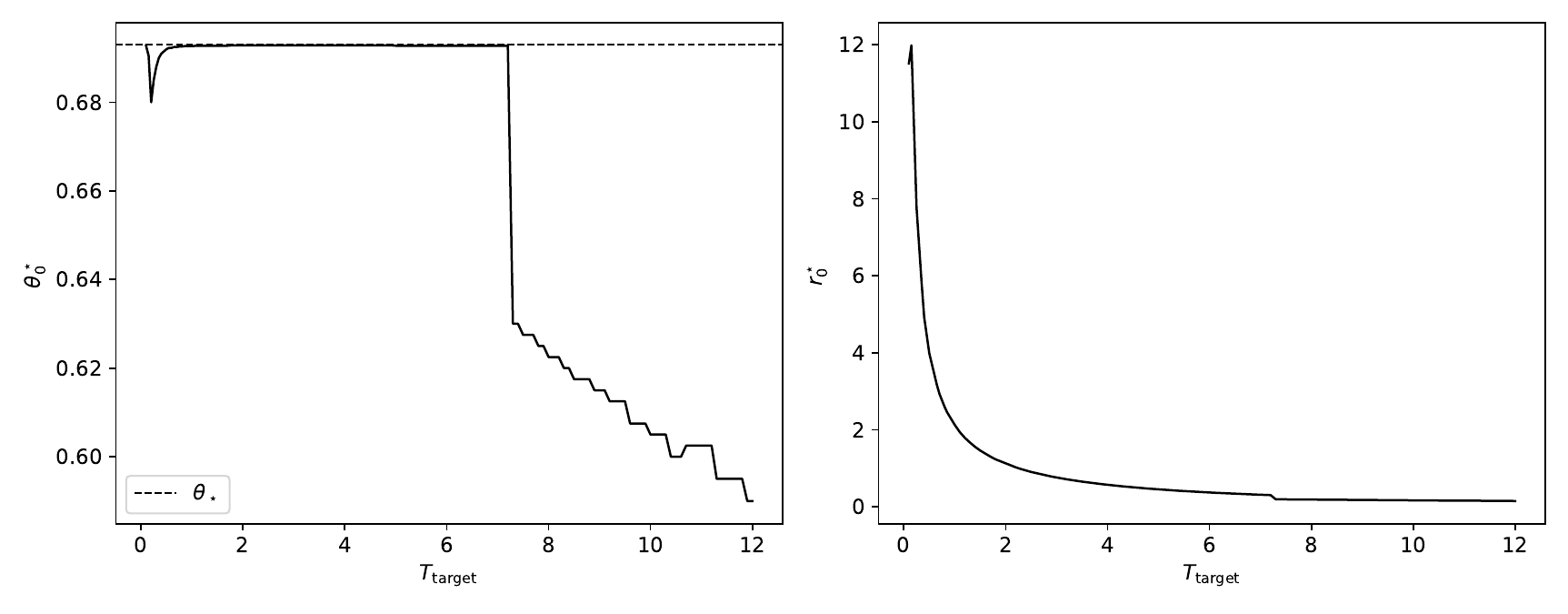}
	\caption{
		Selected saddle location as a function of the target boundary interval.
		The figure shows $\theta_0^\ast(T_{\rm target})$ and
		$r_0^\ast(T_{\rm target})$. The selected angular label remains close to
		the cap/horizon transition region, approaching it from the cap side in the
		displayed range, while the selected turning point decreases.
	}
	\label{fig:complexity-selected-location}
\end{figure}

The selected angular label and selected turning point are shown in
figure~\ref{fig:complexity-selected-location}.

Figure~\ref{fig:complexity-selected-location} shows the correlated angular
and radial motion of the selected saddle. As $T_{\rm target}$ increases,
the selected angular label remains near the transition region, while the
turning point moves inward. The selected saddle therefore becomes more
sensitive to the exact localized geometry, especially to the radial region
where $K_y(r,\theta)$ and $G(r,\theta)$ differ from their large-$r$ forms.
This inward motion should not be confused with a switch to the
smaller-$r_0$ branch; the branch assignment is determined by
eq.~\eqref{eq:complexity-post-peak-condition}.

Finally, we reconstruct the effective $t-r$ profiles of the selected
complexity saddles. For each target interval, the selected values
$\theta_0^\ast$, $r_0^\ast$ and $b^\ast$ are first determined. The profiles
then show where the selected Lorentzian branches lie in the effective
geometry.

\begin{figure}[H]
	\centering
	\begin{subfigure}[t]{0.25\textwidth}
		\centering
		\includegraphics[width=\linewidth]{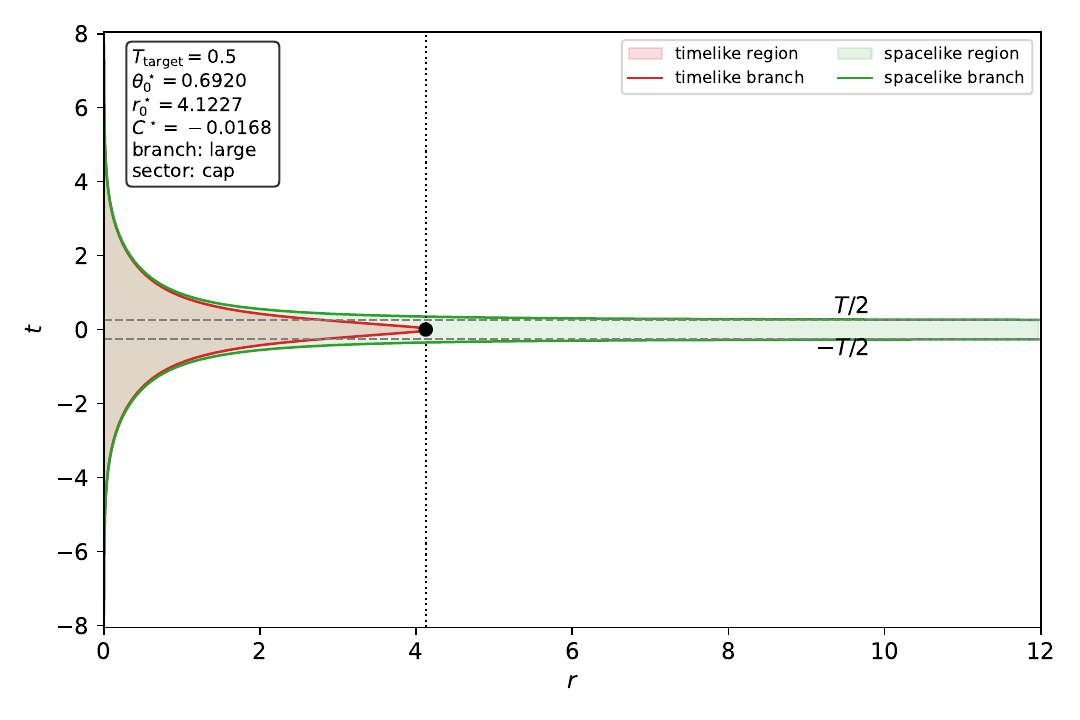}
	\end{subfigure}
	\hfill
	\begin{subfigure}[t]{0.25\textwidth}
		\centering
		\includegraphics[width=\linewidth]{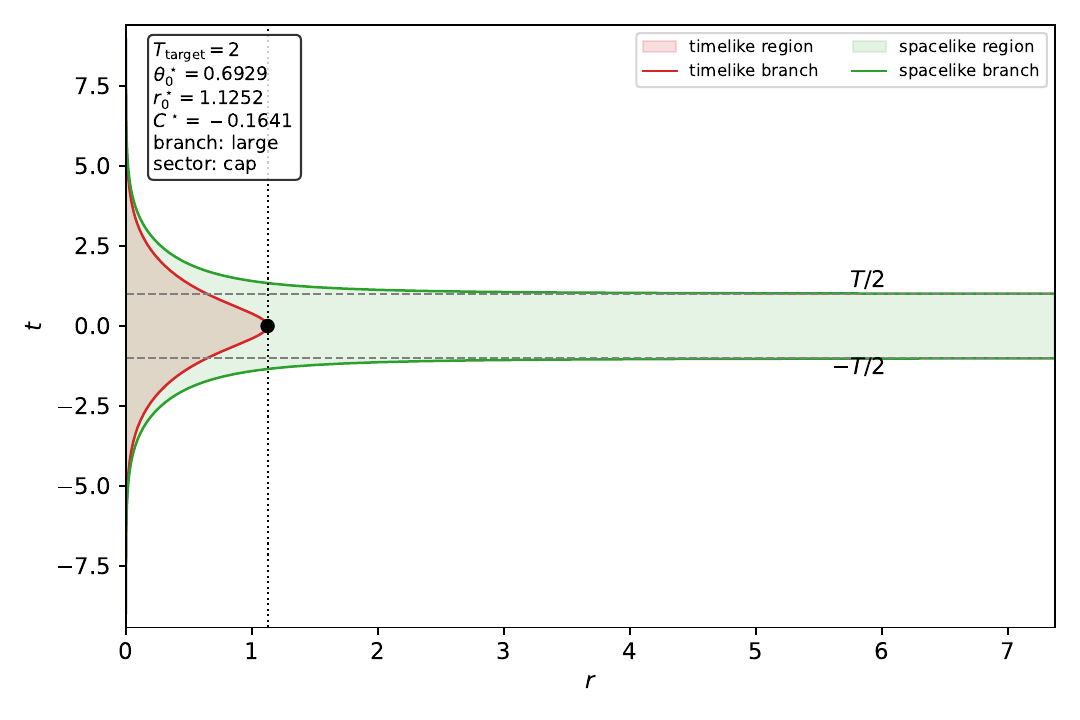}
	\end{subfigure}
	\hfill
	\begin{subfigure}[t]{0.25\textwidth}
		\centering
		\includegraphics[width=\linewidth]{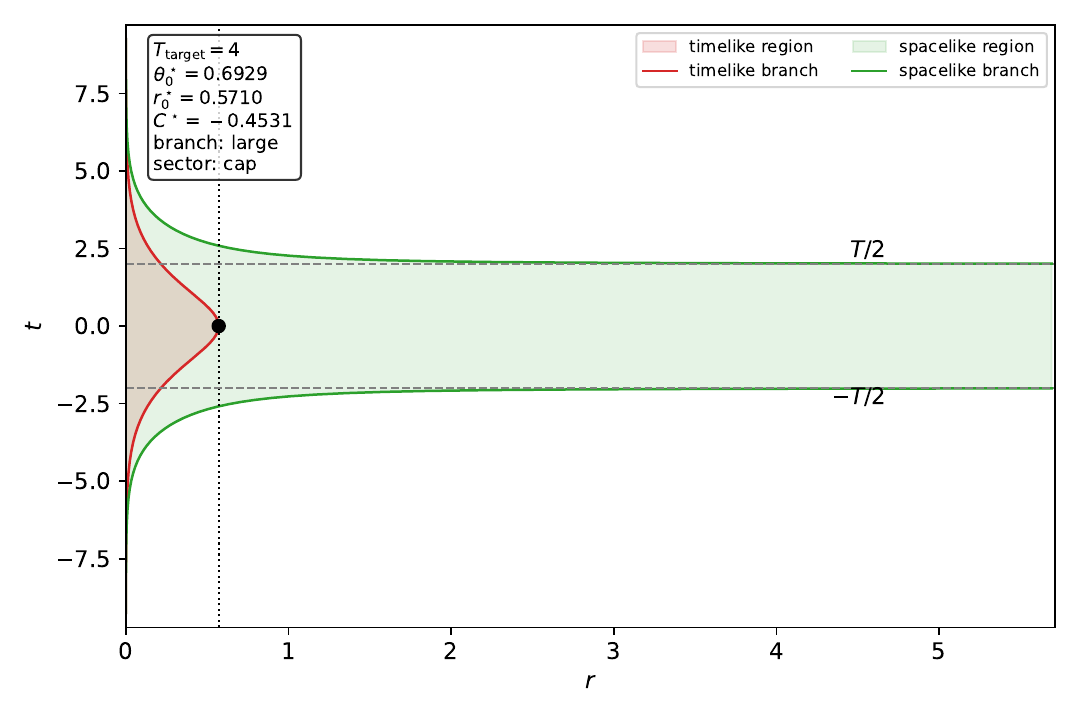}
	\end{subfigure}
	
	\medskip
	
	\begin{subfigure}[t]{0.25\textwidth}
		\centering
		\includegraphics[width=\linewidth]{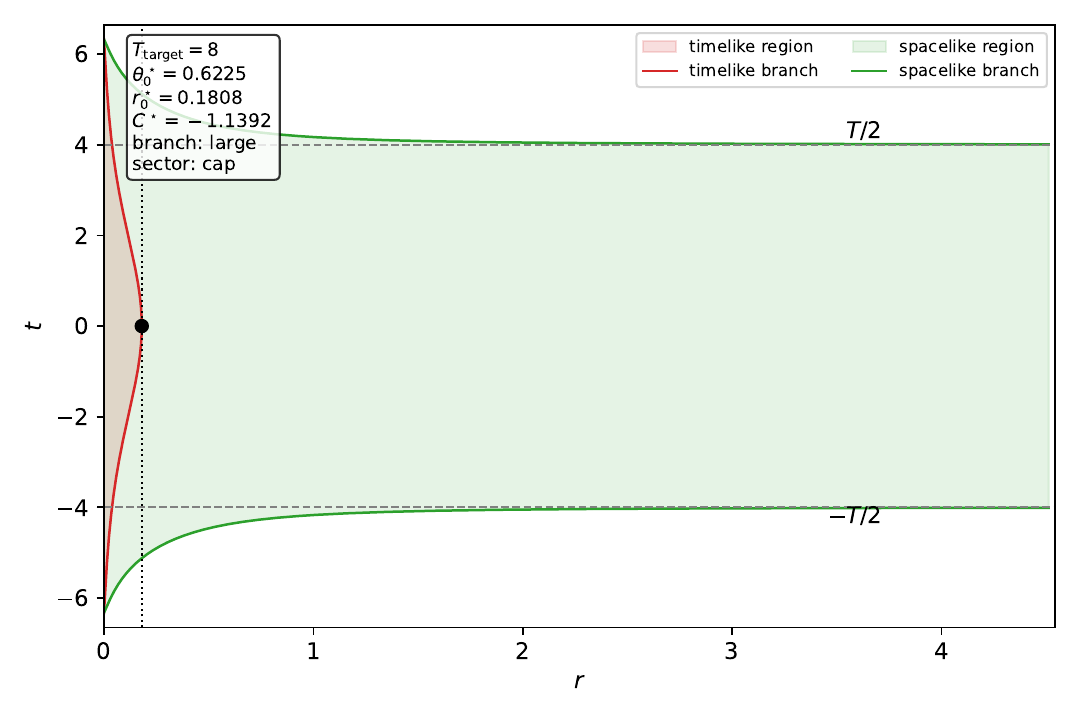}
	\end{subfigure}
	\hfill
	\begin{subfigure}[t]{0.25\textwidth}
		\centering
		\includegraphics[width=\linewidth]{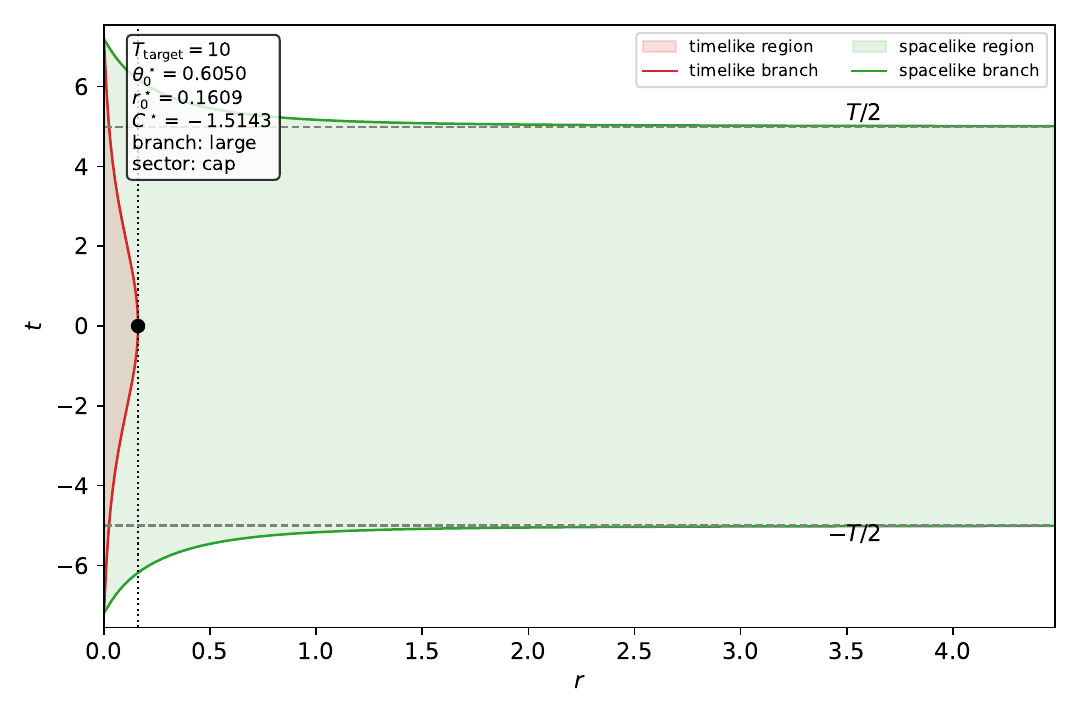}
	\end{subfigure}
	\hfill
	\begin{subfigure}[t]{0.25\textwidth}
		\centering
		\includegraphics[width=\linewidth]{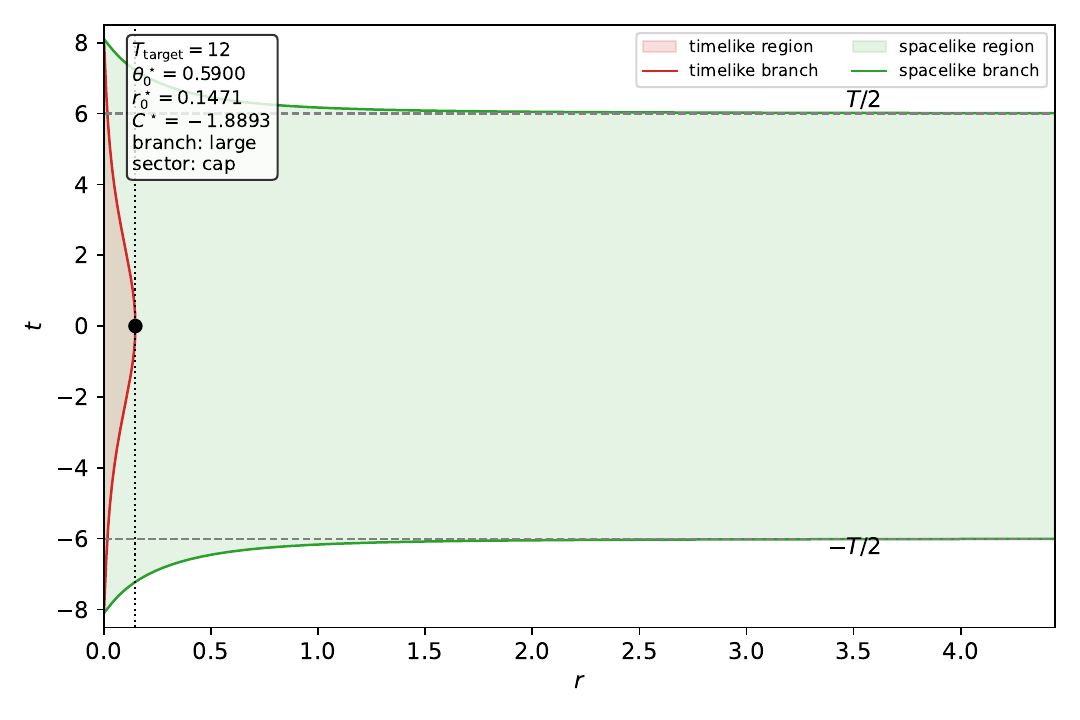}
	\end{subfigure}
	\caption{
		Post-minimisation $t-r$ profiles of the selected timelike-complexity
		saddles. Each profile is reconstructed after selecting
		$\theta_0^\ast(T_{\rm target})$, $r_0^\ast(T_{\rm target})$ and the
		radial branch. The spacelike branch reaches the cutoff at
		$t=\pm T_{\rm target}/2$, while the timelike branch closes at the selected
		turning point.
	}
	\label{fig:complexity-selected-tr-profiles}
\end{figure}

Figure~\ref{fig:complexity-selected-tr-profiles} shows how the selected
Lorentzian branches move as the target boundary interval is increased. For
shorter target intervals, the selected profile remains closer to the
large-$r$ regime, where the lifted volume is close to the asymptotic
reference behaviour. For larger target intervals, the selected turning
point moves inward and the branches enter the localized region more
strongly. In this regime the finite volume in
eq.~\eqref{eq:complexity-Clift-final} becomes sensitive to the exact
black-pole functions $K_y(r,\theta)$ and $G(r,\theta)$ through both the
branch kernels and the lifted radial weight. The post-selection profiles
therefore give the geometric origin of the selected complexity curve: the
change in $C_T^{\rm selected}$ follows from the inward motion of the
selected branch and from the signed competition between the spacelike and
timelike weighted volumes.

The fixed-boundary-interval minimisation gives the following physical
picture. The transition angle controls which angular families can support
larger target intervals, as explained in appendix~\ref{app:transition-region}.
The cap-side deep core does not produce the large-time enhancement, but the
cap side near $\theta_\star$ contains the transition interval that does.
After the target interval is fixed, the finite renormalized volume selects
one branch among the admissible families. In the displayed range, the
selected saddle remains on the larger-$r_0$ branch, while its turning point
moves inward and its angular label stays near the cap/horizon transition
region. The nonzero exterior-patch finite volume is a localized-geometry
effect of the exact functions $K_y(r,\theta)$ and $G(r,\theta)$ and is
absent in both large-$r$ regime and BTZ description.

\subsection{Summary of the timelike-complexity analysis}
\label{sec:complexity-physical-interpretation}

We now summarize the timelike-complexity result. The observable studied in
this section is the finite renormalized branch volume
$C_T^{\rm lift}$ defined in eq.~\eqref{eq:complexity-Clift-final}. It is
computed on the Lorentzian region bounded by the same timelike and spacelike
branches used in the timelike-entanglement analysis. The first step is not
a volume minimisation. One must first impose the fixed-boundary-interval
condition. Only branches with the same $T_{\rm target}$ describe the same
boundary problem and can be compared.

The large-$r$ calculation gives the short-boundary-interval check. In this
regime the turning point is far from the localized core, the leading angular
dependence of $K_y(r,\theta)$ and $G(r,\theta)$ drops out, and the time map
is single-valued. The lifted density reduces to its asymptotic form, and
the finite renormalized volume vanishes in the short-interval limit, as in
eq.~\eqref{eq:complexity-large-r-small-T}. This shows that very short
boundary intervals probe only the asymptotic region and do not see the
cap/horizon structure.

The exact black-pole geometry changes this behaviour. The branch kernels
depend on the angular label $\theta_0$ through $K_y(r,\theta_0)$ and
$G(r,\theta_0)$, while the lifted density $\bar\mu(r)$ contains the integral
over the physical internal angle $\theta$. The finite volume therefore
contains localized-geometry information in two ways: through the reduced
branch profile and through the ten-dimensional radial weight.

The first effect of the exact geometry is the non-monotonic time map. At
fixed $\theta_0$, the same target interval can be reached by two radial
branches. This is why $C_T^{\rm lift}$ cannot be minimized directly over
the two-parameter family $(r_0,\theta_0)$. The time equation must be solved
first, and the finite volumes must be compared only after the common
boundary interval has been imposed.

The second effect is the angular accessibility condition. A branch can
contribute at a given target interval only if
eq.~\eqref{eq:complexity-time-accessibility} is satisfied. As
$T_{\rm target}$ increases, the admissible angular labels are restricted
toward the cap/horizon transition region. The local origin of this
restriction is explained in appendix~\ref{app:transition-region}. Near
$\theta_\star$, the crossover scale $r_c(\theta_0)$ becomes small, and the
transition radial interval gives a positive logarithmic contribution to
$I_{\rm tim}-I_{\rm sp}$. This mechanism works from both sides of
$\theta_\star$: the cap-side deep core does not generate the large time, but
the cap side near $\theta_\star$ still contains the transition interval that
does.

After the admissible branch family has been constructed, the finite volume
selects the saddle. In the numerical range studied here, with $x_E=0.6$,
the selected saddle remains on the larger-$r_0$ branch, where the branch
label is defined relative to the peak of the temporal family. This does not
mean that the selected turning point remains in the asymptotic region. As
$T_{\rm target}$ increases, the selected value $r_0^\ast$ decreases. The
selected branch therefore moves inward and becomes more sensitive to the
exact localized geometry.

The selected angular label remains close to the cap/horizon transition
region in the displayed range, approaching it from the cap side. This
should not be interpreted as motion into the deep cap core. The relevant
local region is the transition radial interval near $\theta_\star$. This is
why the angular motion of the selected complexity saddle is consistent with
the local analysis in appendix~\ref{app:transition-region}.

The behaviour of $C_T^{\rm selected}$ should be read within the
timelike-subregion prescription used here. The localized black-pole
solution is described in the exterior patch relevant for the Lorentzian
branch construction; the horizon interior is not part of the data used in
this finite-volume observable. This differs from the usual global BTZ
intuition, where the main volume growth is tied to the black-hole interior.
For the localized black pole, the exterior patch already contains
nontrivial structure through $K_y(r,\theta)$ and $G(r,\theta)$. The
renormalized subregion volume can therefore receive a finite contribution
from this localized exterior geometry.

The finite observable is a signed branch-volume combination: the spacelike
weighted volume and the timelike weighted volume are positive separately,
but they enter with opposite signs after the asymptotic subtraction in
eq.~\eqref{eq:complexity-Clift-final}. As the selected branch moves inward,
these two weighted contributions are evaluated over different radial parts
of the exact geometry. A decrease or slow variation of the selected finite
volume is therefore a property of this renormalized branch-volume
difference.

This also clarifies the relation to timelike entanglement. Both observables
use the same Lorentzian branches and the same fixed-boundary-interval
logic. Timelike entanglement evaluates a lifted area and can be complex,
with its imaginary part controlled by the sign of the lifted area functions.
Timelike complexity evaluates a lifted volume and remains real. Its
nontrivial behaviour comes from the finite spacelike-minus-timelike volume
combination.

The final picture is therefore clear. The large-$r$ regime gives a
short-boundary-interval check with no angular selection. The exact
black-pole geometry introduces a non-monotonic time map, an angular
accessibility condition controlled by the transition region, and a selected
branch that moves inward as the target interval grows. Timelike complexity
is a real Lorentzian volume probe of the same localized cap/horizon
structure seen by timelike entanglement, but it encodes that structure
through a different bulk functional.

\section{Discussion and conclusions}
\label{sec:discussion-conclusion}

In this work we studied two Lorentzian observables in the localized
black-pole geometry: timelike entanglement entropy and timelike subregion
complexity. Both observables are built from the same Lorentzian branch
geometry, consisting of a spacelike branch reaching the asymptotic boundary
and a timelike branch ending at a turning point. They differ in the bulk
quantity assigned to this branch geometry. Timelike entanglement is
computed from a lifted area and can be complex, while timelike subregion
complexity is computed from a lifted volume and is real. This common branch
structure allows a direct comparison between the two probes.

A key feature of the localized black pole is that the AdS$_3$ part of the
metric depends on the internal sphere through the functions
$K_y(r,\theta)$ and $G(r,\theta)$. This makes the localized prescription
different from the BTZ case. The reduced branch profile is obtained at a
fixed angular label $\theta_0$, but the ten-dimensional area or volume is
evaluated by integrating over the physical internal angle $\theta$. Thus
$\theta_0$ selects the effective Lorentzian branch family, while $\theta$
enters the lifted bulk functional. This distinction is essential for both
observables.

The large-$r$ regime provides a useful short-boundary-interval check. In
this regime the leading angular dependence of $K_y$ and $G$ drops out. The
time map is single-valued, no angular selection is present, and the
cap/horizon transition region is not probed. For timelike entanglement, the
large-$r$ calculation gives the expected logarithmic real part after UV
subtraction and a finite leading imaginary part. For timelike complexity,
the finite renormalized volume vanishes in the short-interval limit. These
results confirm the consistency of the two prescriptions in the asymptotic
region, but they do not capture the localized structure of the black pole.

The exact black-pole geometry changes the branch problem in an essential
way. Once the full functions $K_y(r,\theta)$ and $G(r,\theta)$ are restored,
the time map can become non-monotonic. At fixed angular label, the same
boundary interval can be reached by more than one radial branch. Therefore
the saddle problem cannot be formulated as an unrestricted minimization over
$(r_0,\theta_0)$. One must first impose the fixed-boundary-interval
condition and then compare only those branches which describe the same
boundary interval. This step is common to timelike entanglement and
timelike complexity.

The appendix explains the local origin of the angular restriction that
appears at larger boundary intervals. Near the transition angle
$\theta_\star$, the crossover scale $r_c(\theta_0)$ becomes small. This
opens a transition radial interval in which the timelike branch kernel has a
larger logarithmic coefficient than the spacelike one. As a result,
$I_{\rm tim}-I_{\rm sp}$ receives a positive logarithmic contribution, and
larger boundary intervals can be supported near $\theta_\star$. This
mechanism works from both sides of the transition angle. The cap-side deep
core does not itself generate the large time, but the cap side near
$\theta_\star$ still contains the transition interval that does.

For timelike entanglement, the selected surface is obtained by minimizing
the real part of the renormalized lifted area at fixed boundary interval.
The imaginary part is then evaluated on the same selected surface. The real
part changes slowly once the selected branches are restricted close to the
transition region. This behaviour is a property of the selected
fixed-boundary-interval surfaces, not of every branch in the family. The
imaginary part carries different information: it is determined by the sign
structure of the lifted area functions over the internal sphere. For this
reason it can vary differently from the real part, and even the lifted
spacelike branch can contribute to the imaginary area.

For timelike subregion complexity, the selected observable is the finite
renormalized branch volume at fixed boundary interval. The local volume
density is positive, so the complexity observable is real. Its nontrivial
behaviour comes from the renormalized branch-volume combination: the
spacelike and timelike weighted volumes enter with opposite signs after the
asymptotic subtraction. The selected curve should therefore be interpreted
as a finite subregion volume associated with the selected Lorentzian
branches. It is not a global complexity of the complete black-pole
spacetime. This distinction is important because the black-pole calculation
uses the exterior patch relevant for the localized solution and for the
branch construction. Unlike the usual BTZ intuition, where the main volume
growth is tied to the black-hole interior, the localized exterior geometry
already contains nontrivial structure through $K_y$ and $G$. The nonzero
finite volume found here is therefore a localized-geometry effect of this
exterior branch construction.

The exact numerical results show the same broad pattern for both
observables. As the target boundary interval is increased, the admissible
angular labels move toward the cap/horizon transition region. The selected
turning point also moves inward, so the selected branch becomes sensitive to
the part of the geometry where the exact black-pole functions differ from
their large-$r$ form. In the complexity calculation, the selected branch
remains on the larger-$r_0$ branch in the sense defined relative to the peak
of the time map, even though the numerical value of the selected turning
point decreases. This distinction is important for interpreting the
post-selection branch data.

The comparison between the two observables is instructive. Timelike
entanglement and timelike complexity use the same Lorentzian branch
kinematics and the same fixed-boundary-interval logic. Their difference is
the lifted bulk functional. Timelike entanglement probes the complex
structure of the lifted area, especially through its imaginary part.
Timelike complexity probes the real finite volume of the same branch region,
with sensitivity to the radial volume weight and to the signed
spacelike-minus-timelike combination. Thus the two observables are
complementary probes of the localized black-pole geometry.

The main conclusion is that the exact localized geometry produces effects
that are absent both in BTZ and in the leading large-$r$ regime. These
include a non-monotonic time map, angular restrictions on admissible
branches, branch folding before fixed-boundary-interval selection, and
selected branches that move inward as the boundary interval grows. All
these effects are controlled by the exact functions $K_y(r,\theta)$ and
$G(r,\theta)$ and by the cap/horizon transition region. Timelike
Lorentzian observables therefore provide a useful way to probe localized
geometric structure beyond the asymptotic approximation.

Let us also comment on the robustness of these conclusions.  The detailed
numerical values of the selected angular label, turning point and finite
area or volume depend on the energy parameter $x_E$, because $x_E$ fixes
$\tau$, the length scales $\ell_1,\ell_2$ and the transition angle
$\theta_\star$.  Changing $x_E$ therefore shifts the location of the
cap/horizon transition region and changes the range of boundary intervals
that can be supported.  The qualitative mechanism, however, is local and
does not rely on the particular numerical choice of $x_E$.  Near
$\theta_\star$, the crossover scale
$r_c(\theta_0)\simeq \ell|\theta_0-\theta_\star|$ becomes small, and the
transition interval gives a positive logarithmic contribution to
$I_{\rm tim}-I_{\rm sp}$.  Thus the angular restriction toward
$\theta_\star$, the appearance of multiple radial branches, and the inward
motion of the selected saddles are expected to persist under moderate
changes of the energy or temperature, with the numerical location of the
transition region shifted accordingly.

The cutoff dependence is also under control.  The spacelike branch contains
the universal asymptotic divergence, which is removed by the analytic UV
subtraction, while the timelike branch has finite radial extent.  In the
complexity calculation the large-$r$ tail of the lifted density is included
explicitly, so increasing $R_{\rm max}$ changes only the residual numerical
error after subtraction.  The small-$r$ regulator controls the endpoint of
the localized core.  On the horizon side, the leading logarithmic endpoint
terms in the timelike and spacelike time integrals cancel in the boundary
interval, and on the cap side the kernels are finite.  Therefore the main
features reported here are not artifacts of the cutoffs, but follow from
the exact localized functions $K_y(r,\theta)$ and $G(r,\theta)$ and from the
cap/horizon transition structure.

A few concrete extensions remain open. The first is to apply the same
Lorentzian branch prescription to other localized AdS$_3\times S^3\times T^4$
solutions, such as the black belt, the black bi-pole and the more general
localized solutions constructed numerically. This would test whether the
transition-region mechanism found here is specific to the black pole or is a
common feature of localized horizons on an internal sphere.

A second direction is to compare the present timelike observables with the
spatial probes used in the localized-black-hole literature. Minimal RT
surfaces are mostly sensitive to the asymptotic region, while non-minimal
extremal surfaces can enter the entanglement shadow and probe the localized
core. The timelike branches studied here provide a different Lorentzian
route into the same ten-dimensional structure. Understanding the relation
between these probes could clarify which parts of the localized geometry are
seen by spatial entanglement, timelike entanglement and timelike complexity.

A third direction is to formulate a genuinely global complexity problem for
localized black holes. The timelike subregion complexity studied in this
paper is a finite renormalized branch-volume observable associated with a
finite boundary interval and with the exterior patch of the localized
solution. It should be kept distinct from the usual interior-volume
question in eternal black holes. A separate construction would be needed to
include the hidden interior region and to compare directly with the standard
BTZ complexity intuition.

Finally, the boundary interpretation of the angular selection deserves a
more direct understanding. In the bulk, the selected branches are guided
toward the cap/horizon transition region because the localized functions
$K_y(r,\theta)$ and $G(r,\theta)$ reshape the Lorentzian time map. This is
a genuine localized-geometry effect: it is absent in the BTZ uplift and in
the leading large-$r$ regime. On the CFT side, the natural question is
which probes can detect the breaking of the internal $S^3$ symmetry and the
non-uniform distribution of the horizon over the compact space. Timelike
entanglement and timelike complexity suggest that Lorentzian observables are
well suited for this purpose, because their selected branches are driven
toward precisely the region where the localized geometry departs most
strongly from BTZ. Making this connection sharper would help identify the
boundary signatures of the black pole beyond the universal asymptotic data.

\appendix

\section{Local transition region and cap/horizon large-time enhancement}
\label{app:transition-region}

In the main text, both timelike entanglement and timelike subregion
complexity use the same reduced Lorentzian branch geometry. For each
angular label $\theta_0$ and turning point $r_0$, the branch kernels define
the boundary interval $T(r_0,\theta_0)$. The fixed-boundary-interval
selection is therefore controlled first by the time map, before one compares
either the lifted area in the entanglement calculation or the finite lifted
volume in the complexity calculation.

This appendix explains the local origin of the enhancement of
$T_{\max}(\theta_0)$ near the cap/horizon transition angle $\theta_\star$.
The key point is simple: the large boundary interval is not produced by the
deepest cap-side or horizon-side core alone. It is produced by a transition
radial interval which becomes long when $\theta_0$ approaches
$\theta_\star$. Since the length of this interval is controlled by
$|\theta_0-\theta_\star|$, the enhancement can occur on both sides of
$\theta_\star$.

At fixed angular label $\theta_0$, the reduced metric is
\begin{equation}
	ds^2_{2,\theta_0}
	=
	-F(r;\theta_0)\,dt^2
	+
	H(r;\theta_0)\,dr^2.
	\label{eq:app-reduced-metric}
\end{equation}
We use the metric functions
$F(r;\theta_0)$ and $H(r;\theta_0)$ defined in
eqs.~\eqref{eq:reduced-two-dimensional-metric}--\eqref{eq:H-definition},
the turning-point value $F_0=F(r_0;\theta_0)$, and the branch kernels
$K_{\rm tim}$ and $K_{\rm sp}$ defined in
eqs.~\eqref{eq:exact-tee-Ktim} and \eqref{eq:exact-tee-Ksp}. The exact black-pole functions can be written as
\begin{align}
	K_y(r,\theta)
	&=
	1+
	\frac{2\ell_2^2}{
		D(r,\theta)+r^2+\ell_1^2\sin^2\theta
		-\ell_2^2\cos^2\theta
	},\nonumber\\ G(r,\theta)
	&=
	\frac{
		\left[D(r,\theta)+r^2\right]^2
		-
		\left[
		\ell_1^2\sin^2\theta-\ell_2^2\cos^2\theta
		\right]^2
	}{
		4r^2D(r,\theta)
	}.
	\label{eq:app-G-exact}
\end{align}
Here
\begin{equation}
	D(r,\theta)
	=
	\frac12
	\left[
	\left[
	(2r^2+\ell^2)\cos2\theta+\ell_2^2-\ell_1^2
	\right]^2
	+
	4r^2(r^2+\ell^2)\sin^2 2\theta
	\right]^{1/2}.
	\label{eq:app-D-exact}
\end{equation}
We now introduce the local angular variable which controls the transition.
Let
\begin{equation}
	a=\ell_1^2,
	\qquad
	b=\ell_2^2,
	\qquad
	L=\ell^2=a+b.
	\label{eq:app-abL}
\end{equation}
The combination that appears in both $K_y$ and $G$ is
\begin{equation}
	Y(\theta)
	=
	a\sin^2\theta-b\cos^2\theta
	=
	\ell_1^2\sin^2\theta-\ell_2^2\cos^2\theta.
	\label{eq:app-Y}
\end{equation}
In terms of $Y(\theta)$,
\begin{align}
	K_y(r,\theta)
	&=
	1+
	\frac{2b}{D(r,\theta)+r^2+Y(\theta)},\qquad G(r,\theta)
	=
	\frac{
		\left[D(r,\theta)+r^2\right]^2-Y(\theta)^2
	}{
		4r^2D(r,\theta)
	}.
	\label{eq:app-G-Y}
\end{align}
The physical transition is seen in $K_y$ and $G$. The variables $D$ and $Y$
are useful because they identify where the leading behaviour of these
functions changes. The transition angle is defined by
\begin{equation}
	\cos^2\theta_\star=\frac{a}{L},
	\qquad
	\sin^2\theta_\star=\frac{b}{L}.
	\label{eq:app-thetastar}
\end{equation}
Equivalently,
\begin{equation}
	Y(\theta_\star)=0.
	\label{eq:app-Y-star}
\end{equation}
The sign of $Y$ distinguishes the two angular sectors:
\begin{align}
	Y(\theta_0)<0
	&\quad \Longleftrightarrow \quad
	\theta_0<\theta_\star,
	\qquad
	\text{cap-side sector},\nonumber
	\\
	Y(\theta_0)>0
	&\quad \Longleftrightarrow \quad
	\theta_0>\theta_\star,
	\qquad
	\text{horizon-side sector}.
	\label{eq:app-Y-horizon}
\end{align}
Near $\theta_\star$,
\begin{equation}
	Y(\theta_0)
	\simeq
	2\ell_1\ell_2(\theta_0-\theta_\star).
	\label{eq:app-Y-near}
\end{equation}
Thus $Y(\theta_0)$ measures the signed angular distance from the transition
angle.

To find the local radial scale, we rewrite $D^2$. Using
\begin{equation}
	L\cos2\theta+b-a=-2Y(\theta),
\end{equation}
one obtains
\begin{align}
	D^2
	&=
	\left[
	r^2\cos2\theta-Y(\theta)
	\right]^2
	+
	r^2(r^2+L)\sin^2 2\theta,\nonumber\\&=
	Y^2
	+
	r^2
	\left[
	L\sin^2 2\theta
	-
	2Y\cos2\theta
	\right]
	+
	r^4.
	\label{eq:app-D2-expanded}
\end{align}
Define
\begin{equation}
	\alpha(\theta)^2
	=
	L\sin^2 2\theta
	-
	2Y(\theta)\cos2\theta.
	\label{eq:app-alpha-theta}
\end{equation}
Then
\begin{equation}
	D^2
	=
	Y(\theta)^2
	+
	\alpha(\theta)^2r^2
	+
	r^4.
	\label{eq:app-D2-alpha}
\end{equation}
Close to $\theta_\star$,
\begin{equation}
	\alpha(\theta_0)^2\simeq\alpha_\star^2,
	\qquad
	\alpha_\star
	=
	\frac{2\ell_1\ell_2}{\ell}.
	\label{eq:app-alpha-star}
\end{equation}
In the localized core region,
\begin{equation}
	r\ll r_{\rm loc},
	\qquad
	r_{\rm loc}\sim O(\ell,\ell_1,\ell_2),
	\label{eq:app-rloc}
\end{equation}
the $r^4$ term in eq.~\eqref{eq:app-D2-alpha} is subleading. Therefore
\begin{equation}
	D(r,\theta_0)
	\simeq
	\left[
	Y(\theta_0)^2+\alpha_\star^2r^2
	\right]^{1/2}.
	\label{eq:app-D-local}
\end{equation}
This expression is positive. Hence the crossover scale is not a zero of $D$. It is the radial scale at which the angular contribution and the radial contribution become comparable:
\begin{equation}
	Y(\theta_0)^2
	\sim
	\alpha_\star^2r_c^2.
\end{equation}
Thus
\begin{equation}
	r_c(\theta_0)
	=
	\frac{|Y(\theta_0)|}{\alpha_\star}
	\simeq
	\ell\,|\theta_0-\theta_\star|.
	\label{eq:app-rc-final}
\end{equation}
The dependence on the absolute value is crucial. The scale $r_c$ becomes
small as $\theta_0$ approaches $\theta_\star$ from either side. This opens a
transition radial interval on both the cap side and the horizon side.

We now examine the local regions. In the deep-core region,
\begin{equation}
	r\ll r_c,
	\label{eq:app-deep-core}
\end{equation}
we get
\begin{equation}
	D(r,\theta_0)\simeq |Y(\theta_0)|.
\end{equation}
The leading behaviour depends on the sign of $Y$.

On the cap side, $Y<0$. Writing $Y=-S$, with $S>0$, we have
\begin{equation}
	D
	\simeq
	S+\frac{\alpha_\star^2r^2}{2S}.
\end{equation}
Then
\begin{equation}
	D+r^2+Y
	\simeq
	r^2
	\left(
	1+\frac{\alpha_\star^2}{2S}
	\right),
\end{equation}
so
\begin{equation}
	K_y\sim \frac{1}{r^2}.
	\label{eq:app-Ky-cap-core}
\end{equation}
The function $G$ remains finite,
\begin{equation}
	G\sim O(1).
	\label{eq:app-G-cap-core}
\end{equation}
Consequently,
\begin{equation}
	F=Qr^2K_yG\sim O(1),
	\qquad
	H=\frac{Q^2G^2}{r^2+\ell^2}\sim O(1),
\end{equation}
and the branch kernels are finite:
\begin{equation}
	K_{\rm tim}\sim O(1),
	\qquad
	K_{\rm sp}\sim O(1).
\end{equation}
Thus the cap-side deep core does not produce a logarithmic enhancement of
the boundary interval.

On the horizon side, $Y>0$. In the deep-core region,
\begin{equation}
	D
	\simeq
	Y+\frac{\alpha_\star^2r^2}{2Y}.
\end{equation}
Near $\theta_\star$, this gives
\begin{equation}
	K_y\simeq \frac{b}{Y},
	\qquad
	G\simeq \frac{\alpha_\star^2}{4Y}.
	\label{eq:app-KyG-horizon-core}
\end{equation}
Therefore
\begin{align}
	F(r;\theta_0)
	\simeq
	f_2(\theta_0)r^2,\qquad
	H(r;\theta_0)
	\simeq
	h_0(\theta_0),
	\label{eq:app-H-horizon-core}
\end{align}
where
\begin{align}
	f_2(\theta_0)
	=
	\frac{Qb\alpha_\star^2}{4Y^2}
	=
	\frac{Q\ell_1^2\ell_2^4}{\ell^2Y^2},\qquad
	h_0(\theta_0)
	=
	\frac{Q^2\alpha_\star^4}{16\ell^2Y^2}
	=
	\frac{Q^2\ell_1^4\ell_2^4}{\ell^6Y^2}.
	\label{eq:app-h0-horizon}
\end{align}
Since
\begin{equation}
	\sqrt{\frac{h_0}{f_2}}
	=
	\frac{\sqrt Q\,\ell_1}{\ell^2}
	\equiv c_0,
	\label{eq:app-c0}
\end{equation}
and $F\ll F_0$ near $r=0$, both kernels behave as
\begin{equation}
	K_{\rm tim}
	\simeq
	K_{\rm sp}
	\simeq
	\frac{c_0}{r},
	\qquad
	c_0=
	\frac{\sqrt Q\,\ell_1}{\ell^2}.
	\label{eq:app-deep-core-kernels}
\end{equation}
The individual integrals therefore contain the same logarithm:
\begin{align}
	I_{\rm tim}^{\rm core}
	\simeq
	c_0\log\frac{r_c}{\epsilon},\qquad
	I_{\rm sp}^{\rm core}
	\simeq
	c_0\log\frac{r_c}{\epsilon}.
\end{align}
Their leading contribution cancels in the boundary interval,
\begin{equation}
	I_{\rm tim}^{\rm core}
	-
	I_{\rm sp}^{\rm core}
	\simeq
	0.
	\label{eq:app-core-cancellation}
\end{equation}
Thus the deepest horizon-side core also does not give the non-cancelling
large-time contribution.

The enhancement comes from the transition interval
\begin{equation}
	r_c\ll r\ll r_{\rm loc}.
	\label{eq:app-transition-domain}
\end{equation}
This interval exists on either side of $\theta_\star$ whenever
$\theta_0$ is close enough to the transition angle. In this region,
\begin{equation}
	D(r,\theta_0)\simeq \alpha_\star r,
	\qquad
	D\gg |Y|,
	\qquad
	D\gg r^2.
	\label{eq:app-D-transition}
\end{equation}
Equation~\eqref{eq:app-G-Y} then give
\begin{align}
	K_y(r,\theta_0)
	\simeq
	\frac{2\ell_2^2}{\alpha_\star r},\qquad
	G(r,\theta_0)
	\simeq
	\frac{\alpha_\star}{4r}.
	\label{eq:app-G-transition}
\end{align}
Thus, in the transition interval,
\begin{equation}
	K_y\sim \frac{1}{r},
	\qquad
	G\sim \frac{1}{r}.
	\label{eq:app-KyG-transition-scaling}
\end{equation}
This is the local scaling of the physical black-pole functions. The
variables $D$ and $Y$ only identify where this scaling begins.

The corresponding reduced metric functions are
\begin{align}
	F(r;\theta_0)
	&=
	Qr^2K_yG
	\simeq
	\frac{Q\ell_2^2}{2}
	\equiv F_p,
	\nonumber
	\\
	H(r;\theta_0)
	&=
	\frac{Q^2G^2}{r^2+\ell^2}
	\simeq
	\frac{h_p}{r^2},
	\label{eq:app-Hp}
\end{align}
with
\begin{equation}
	h_p
	=
	\frac{Q^2\alpha_\star^2}{16\ell^2}
	=
	\frac{Q^2\ell_1^2\ell_2^2}{4\ell^4}.
	\label{eq:app-hp}
\end{equation}
Substituting these forms into the branch kernels gives
\begin{align}
	K_{\rm tim}
	\simeq
	\frac{C_{\rm tim}}{r},\qquad
	K_{\rm sp}
	\simeq
	\frac{C_{\rm sp}}{r},
	\label{eq:app-Ksp-transition}
\end{align}
where
\begin{align}
	C_{\rm tim}
	=
	\left[
	\frac{
		F_0h_p
	}{
		F_p(F_0-F_p)
	}
	\right]^{1/2},\qquad
	C_{\rm sp}
	=
	\left[
	\frac{
		F_0h_p
	}{
		F_p(F_0+F_p)
	}
	\right]^{1/2}.
	\label{eq:app-Csp}
\end{align}
The timelike transition branch requires $F_0>F_p$. Since $F_p>0$,
\begin{equation}
	F_0-F_p<F_0+F_p,
\end{equation}
and hence
\begin{equation}
	C_{\rm tim}>C_{\rm sp}.
	\label{eq:app-Ctim-greater-Csp}
\end{equation}
The transition interval therefore gives a positive logarithmic contribution
to $I_{\rm tim}-I_{\rm sp}$:
\begin{align}
	I_{\rm tim}-I_{\rm sp}
	&\simeq
	\int_{r_c}^{r_{\rm loc}}
	dr\,
	\left(
	K_{\rm tim}-K_{\rm sp}
	\right)
	\nonumber\\
	&\simeq
	(C_{\rm tim}-C_{\rm sp})
	\int_{r_c}^{r_{\rm loc}}
	\frac{dr}{r}
	\nonumber\\
	&=
	(C_{\rm tim}-C_{\rm sp})
	\log\frac{r_{\rm loc}}{r_c}.
	\label{eq:app-log-enhancement}
\end{align}
Using
\begin{equation}
	r_c\simeq \ell|\theta_0-\theta_\star|,
	\qquad
	r_{\rm loc}\sim \ell,
\end{equation}
one obtains
\begin{equation}
	I_{\rm tim}-I_{\rm sp}
	\sim
	(C_{\rm tim}-C_{\rm sp})
	\log
	\frac{1}{|\theta_0-\theta_\star|}
	+
	O(1).
	\label{eq:app-Itim-Isp-log}
\end{equation}
Since $T=2(I_{\rm tim}-I_{\rm sp})$,
\begin{equation}
	T(r_0,\theta_0)
	\sim
	2(C_{\rm tim}-C_{\rm sp})
	\log
	\frac{1}{|\theta_0-\theta_\star|}
	+
	O(1).
	\label{eq:app-T-log}
\end{equation}
This formula is the local reason why large values of
$T_{\max}(\theta_0)$ appear near $\theta_\star$. The logarithm depends on
the distance from the transition angle, not on the side from which the
transition angle is approached.

This also clarifies the cap-side behaviour. The large boundary time is not
generated by the deep cap core, where the branch kernels are finite. It
comes from the transition interval $r_c\ll r\ll r_{\rm loc}$, which opens as
$\theta_0$ approaches $\theta_\star$ from either side. Thus the enhancement
of $T_{\max}(\theta_0)$ is controlled by $|\theta_0-\theta_\star|$, not by
the deep-core sector. The same time-map mechanism restricts the admissible
branch family in both observables; only after the fixed-boundary-interval
family is constructed do the two calculations differ, with timelike
entanglement evaluating a lifted area and timelike complexity evaluating a
lifted volume.

\acknowledgments
This work was supported by National Natural Science Foundation of China under Grant No.  12075059, as well as by the start-up fund of USTC. The authors are thankful to Chanyong Park for comments on the manuscript.

\end{document}